\documentclass[12pt, a4paper]{article}
\usepackage[utf8]{inputenc}
\usepackage{dcolumn,lscape}%,setspace}
\usepackage{amsmath,longtable,multicol,dcolumn,tabularx,graphicx,amssymb}
\usepackage{exscale,amsthm,multirow,rotating}
\usepackage{natbib}
\usepackage{bbm,soul}
\usepackage{tikz,dsfont,float}

\usepackage{hyperref}

\hypersetup{
    colorlinks=true,
    linkcolor=blue,
    citecolor=blue,
    filecolor=magenta,      
    urlcolor=cyan,
    pdftitle={review paper},
    pdfpagemode=FullScreen,
}

\usepackage{subfigure}
\newcommand{\xvec}{\boldsymbol}
\newcommand{\xmat}{\mathbf}

\newtheorem{algorithm}{Algorithm}

\newcommand{\blue}{\textcolor{blue}}

\newcommand{\mf}{\mathbf}
\newcommand{\bs}{\boldsymbol}

\newcommand{\h}{\xvec{h}}
\newcommand{\y}{\xvec{Y}}
\newcommand{\W}{\xmat{W}}
\newcommand{\x}{\xvec{x}}
\newcommand{\z}{\xvec{z}}
\newcommand{\X}{\xmat{X}}

\newcommand{\I}{\mathbf{I}}
\newcommand{\U}{\mathbf{U}}
\newcommand{\A}{\mathbf{A}}
\newcommand{\Ss}{\mathbf{S}}
\newcommand{\V}{\mathbf{V}}
\newcommand{\J}{\mathbf{J}}
\newcommand{\0}{\xvec{0}}
\newcommand{\K}{\mathbf{K}}
\newcommand{\Pp}{\mathbf{P}}
\newcommand{\B}{\mathbf{B}}
\newcommand{\Y}{\xvec{Y}}

\newcommand{\Z}{\xmat{Z}}
\renewcommand{\ln}{\log}
\newcommand{\rh}{\xvec{\rho}}
\newcommand{\Om}{\boldsymbol{\Omega}}
\newcommand{\Si}{\xmat{\Sigma}}

\DeclareMathOperator*{\E}{E}
\DeclareMathOperator*{\Diag}{diag}
\DeclareMathOperator{\e}{\varepsilon}

\begin{document}

\def\spacingset#1{\renewcommand{\baselinestretch}%
	{#1}\small\normalsize} \spacingset{1}

%%%%%%%%%%%%%%%%%%%%%%%%%%%%%%%%%%%%%%%%%%%%%%%%%%%%%%%%%%%%%%%%%%%%%%%%%%%%%%

\title{\bf Spatial and Spatiotemporal Volatility Models: A Review}
\author{Philipp Otto\footnote{University of Glasgow, School of Mathematics and Statistics, Email: philipp.otto@ikg.uni-hannover.de} \and Osman Do\u{g}an\footnote{Department of Economics, Istanbul Technical University, Istanbul, Türkiye, Email: osmandogan@itu.edu.tr} \and Süleyman Ta\c{s}p\i nar\footnote{Department of Economics, The City University of New York, Queens College, New York, USA, Email: staspinar@qc.cuny.edu} \and Wolfgang Schmid\footnote{Department of Statistics, European University Viadrina, Frankfurt, Germany, Email: schmid@europa-uni.de} \and Anil K. Bera\footnote{Department of Economics, University of Illinois at Urbana-Champaign, Illinois, USA, Email: abera@illinois.edu}}
\maketitle

\begin{abstract}
	\noindent 
	Spatial and spatiotemporal volatility models are a class of models designed to capture spatial dependence in the volatility of spatial and spatiotemporal data. Spatial dependence in the volatility may arise due to spatial spillovers among locations; that is, if two locations are in close proximity, they can exhibit similar volatilities. In this paper, we aim to provide a comprehensive review of the recent literature on spatial and spatiotemporal volatility models. We first briefly review time series volatility models and their multivariate extensions to motivate their spatial and spatiotemporal counterparts. We then review various spatial and spatiotemporal volatility specifications proposed in the literature along with their underlying motivations and estimation strategies. Through this analysis, we effectively compare all models and provide practical recommendations for their appropriate usage. We highlight possible extensions and conclude by outlining directions for future research.	
	
%	In this paper, we aim to provide a comprehensive review of the recent literature on spatial and spatiotemporal volatility models---a class of models designed to capture volatility clusters in spatial and spatiotemporal data. Spatial volatility clusters may arise due to spatial spillovers among locations; that is, if two locations are in close proximity, they can exhibit similar volatilities. We first briefly review time series volatility models and their multivariate extensions to motivate their spatial and spatiotemporal counterparts. We then review various spatial and spatiotemporal volatility specifications proposed in the literature along with their underlying motivations and estimation strategies. Through this analysis, we effectively compare all models and provide practical recommendations for their appropriate usage. We highlight possible extensions and conclude by outlining directions for future research.	
	
 \end{abstract}

\noindent%

% \newpage
\spacingset{1.45} % DON'T change the spacing!

\newpage

\tableofcontents

% \newpage

\section{Introduction and motivation}

When observations of a random process have a natural ordering, such as the temporal ordering for time series or geographical locations for spatial processes, they are typically dependent. For time series, observations close together in time will be more closely related than observations further apart. Similarly, for spatial processes, \cite{fisher1935design} stated that ``patches in close proximity are commonly more alike, as judged by the yield of crops, than those which are further apart'', or more commonly known as Tobler's first law of Geography: ``Everything is related to everything else, but near things are more related than distant things'' \citep{Tobler70}. The similarity may be reflected in the mean or trend behaviour, which motivates autoregressive processes and integrated processes, but also in the process variation or volatility, motivating autoregressive heteroscedasticity models or stochastic volatility models. Spatial and spatiotemporal data models must fulfil this key property: instant and direct dependence due to spatial or temporal proximity.

Various types of models have been introduced in time series analysis and spatial analysis to describe the dependence over time and space. The most popular approaches are based on linear structures where, e.g., the value of the time series (spatial process) at a certain time point (point in space) is a linear combination of preceding or neighbouring observations and possible regressors. Such types of processes, e.g. SARIMA models, have been analysed in time series analysis in detail \citep[see, e.g.,][]{brockwell2009time,brockwell2002introduction}. Additionally, over the last few decades, the field of econometrics has seen the development of numerous linear spatial models, which have proven to be highly practical and useful \citep[see, e.g.,][]{anselin1988spatial, Lee:2004, Lesage:2009, KP:2010, elhorst2010applied, Elhorst:2014}. 

Unfortunately, these linear approaches are only of limited use to describe a dependence behaviour in the process' volatility. This point was discussed in detail by \cite{Engle82,Bollerslev86} for temporal processes. They introduced a new type of nonlinear temporal process, which has turned out to be extremely useful for modelling returns of financial data. They are based on a multiplicative decomposition and are, thus, nonlinear in the white noise process. Their main advantage is that they can describe a time-varying behaviour of the conditional variance, the so-called volatility. In general, there are different definitions of volatility \citep[see][for a discussion of different volatility measures]{andersen2010parametric}. For instance, for ARCH and GARCH models, the volatility term coincides with the conditional variance of the response variable. For other volatility time series models, such as stochastic volatility models, the conditional variance is non-explicit \cite[see][]{francq2019garch}. The time-varying volatility makes these nonlinear models quite attractive in practice since the risk behaviour of the process may change over time. Linear time series, e.g., ARMA processes, do not possess this property; they have a constant volatility. For an overview of nonlinear time series models, we refer the interested reader to \cite{turkman2016non} or \cite{fan2003nonlinear}.

In time series analysis, generalised autoregressive conditional heteroscedasticity (GARCH) and stochastic volatility models are widely used to model time-varying volatility \citep[see also][for a thorough overview]{Francq11,francq2019garch}. However, many real-world phenomena are observed at multiple geo-referenced locations and, thus, exhibit spatial or spatiotemporal dependence, which means that the volatility of neighbouring series may influence the volatility of a series. Spatial and spatiotemporal volatility models have been recently developed to account for this spatial dependence. After first being mentioned ``as a byproduct'' in \citet{Bera04}, \cite{Otto16_arxiv} introduced spatial ARCH models (published in \citealt{Otto18_spARCH,otto2019stochastic}), and \citet{Sato17} introduced spatial log-ARCH models. Interestingly, both models were developed simultaneously and independently of each other to be able to represent the ARCH-like dependencies in the variance of spatial processes. Moreover, \citet{Robinson:2009} and \citet{tacspinar2021bayesian} introduced spatial stochastic volatility models that include an additional stochastic term in the volatility process. 

Since then, various nonlinear spatial and spatiotemporal volatility models with spatial and spatiotemporal dependence in the (conditional) variance have been proposed. However, there has been little comparison among these models (e.g., \citet{OttoSchmid19_arxiv_unified} compared some models in a unified framework). In this paper, we aim to provide a comprehensive review of the recent literature on spatial and spatiotemporal volatility models. As these models resemble their time series counterparts, we begin by briefly reviewing important aspects of discrete time series volatility models. Subsequently, we describe spatial and spatiotemporal volatility models considered in the literature. Besides motivating each model, we describe important estimation strategies that are applicable in the spatiotemporal context, e.g., quasi-maximum-likelihood (QML) approach, generalised methods-of-moments (GMM) approach, and the Bayesian MCMC sampling schemes. We describe possible extensions and provide directions for future research. 

The remainder of the paper is structured as follows. In Section \ref{sec:timeseries}, we start with discrete time series volatility models and their multivariate extensions, which are also applicable to spatiotemporal data. In Section~\ref{sec:spatial}, we discuss spatial GARCH properties (i.e., instantaneous GARCH-type interactions across the cross-sectional or spatial domain) and compare different specifications that have been proposed in the literature. In Section \ref{sec:spatiotemporal}, our focus is on spatiotemporal volatility models allowing for these instantaneous spatial GARCH-type interactions in addition to the temporal GARCH structure. Further extensions and related models are discussed in the ensuing Section \ref{sec:furthermodels}. Finally, in Section \ref{sec:conclusion}, we conclude the review with an outlook on future research and a discussion of open problems. Some of the technical details are collected in an appendix.

\section{Time series volatility models}\label{sec:timeseries}

\begin{figure}
    \centering
    \includegraphics[width=0.7\textwidth]{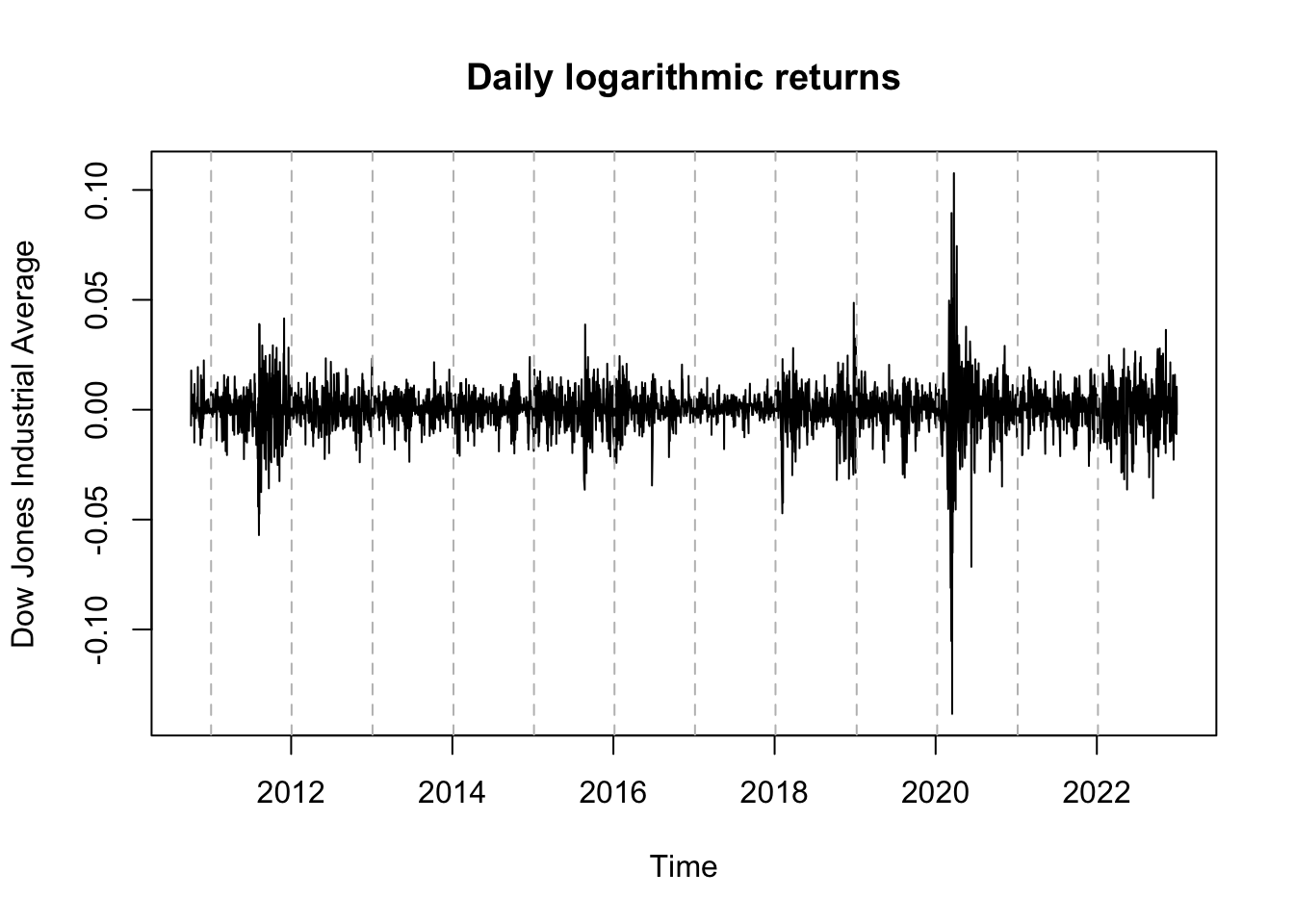}\\
    \includegraphics[width=0.7\textwidth]{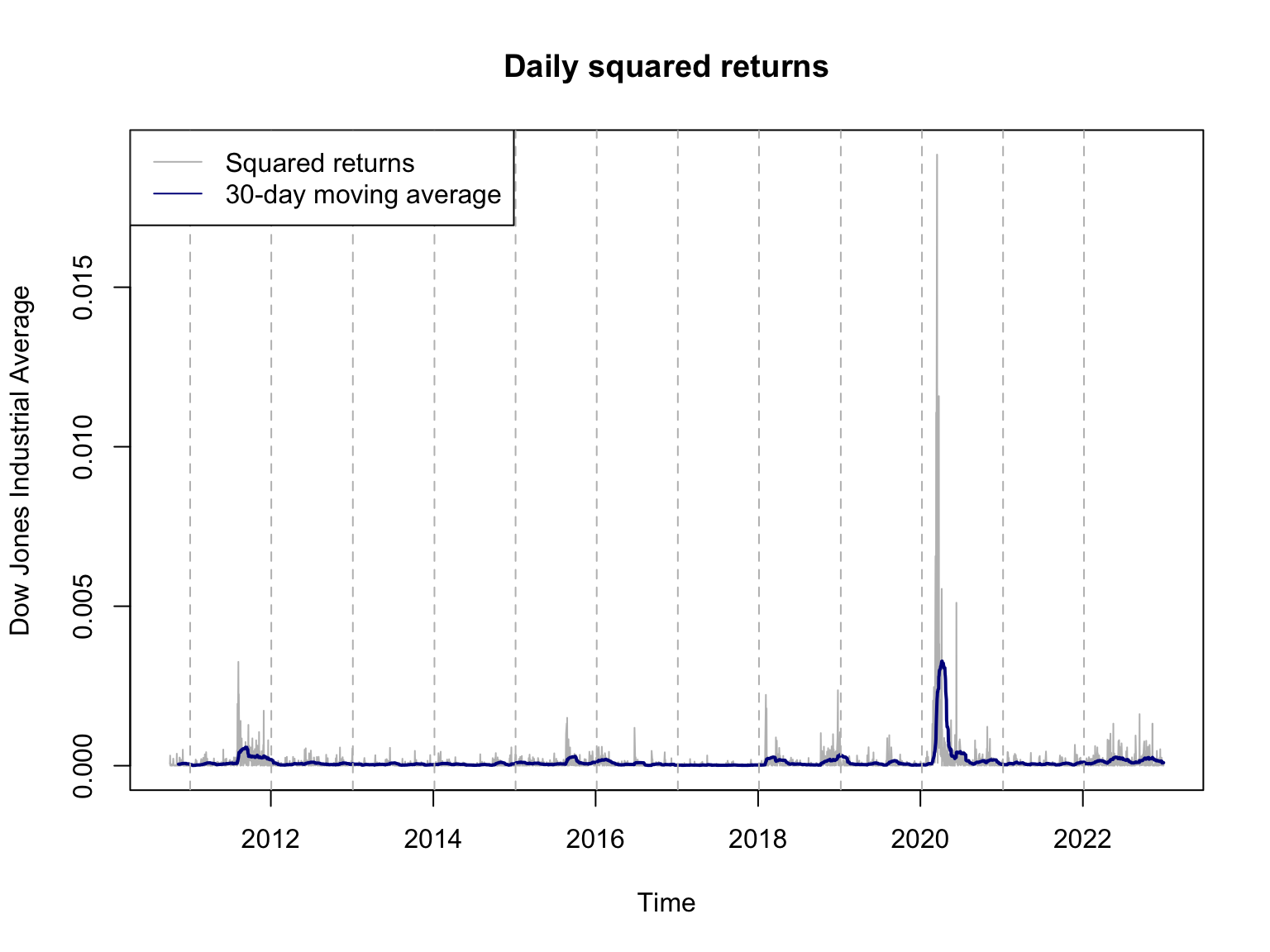}
    \caption{Example of a univariate time series. Top: Daily logarithmic returns of the Dow Jones Index from 2011 to 2023. Bottom: Squared returns and the 30-day moving averages as volatility measures, indicating temporally varying volatility levels/clusters.}
    \label{fig:univariate_ts}
\end{figure}

This section will briefly describe two classes of time series volatility models: (i) ARCH and GARCH-type models and (ii) stochastic volatility models. These models provide the basis for all spatial and spatiotemporal ARCH/GARCH and stochastic volatility models developed in the literature. It is well known that the volatility of many financial time series, such as asset returns and exchange rates, changes  over time. For example, Figure \ref{fig:univariate_ts} displays a univariate time series of daily logarithmic returns of the Dow Jones Index from 2011 to 2023. The bottom plot shows measures of volatility using absolute returns and a 30-day moving average, which indicates temporal variations in volatility levels. These variations can be useful for identifying patterns or trends in the data and informing investment strategies or risk management decisions.

Both classes of models aim to describe the dependence in the conditional variance, i.e., volatility, which should be separated from the mean behaviour of the random process, such as temporal trends, seasonality, or autoregressive dependence. For ARCH and GARCH models, the volatility is a function of the squares of previous observations and past conditional variances (i.e., volatilities). In the case of stochastic volatility models, the volatility is modelled through a latent stochastic process, also depending on the temporally lagged squared observations. In Sections~\ref{sArch} and \ref{sSV}, we briefly describe important univariate and multivariate versions suggested in the literature for both classes of models.  Throughout this section, we consider the discrete time series $\{ \xvec{Y}_t \in \mathbb{R}^r : t \in \mathbb{Z}\}$, i.e., an $r$-dimensional random process $\xvec{Y}_t$ with equidistantly ordered temporal observations.

\subsection{ARCH and GARCH models}\label{sArch}

\subsubsection{Univariate ARCH and GARCH models}

In the univariate case, $r = 1$, the random process $Y_t$ is given by
\begin{equation}\label{eq1}
    Y_t = h_t^{1/2} \varepsilon_t \, ,
\end{equation}
where $h_t^{1/2}$ is a scaling factor and $\{ \varepsilon_t \}$ is a sequence of independent and identically distributed (i.i.d.) variables with mean $0$ and variance $1$. Moreover, $h_t$ depends on the past realisations of $Y_t$, and it coincides with the conditional variance of $Y_t$. Thus, it can be interpreted as the volatility of the series. This idea traces back to the seminal work of \cite{Engle82}. To be precise, the volatility process is defined as
\begin{equation}
    h_t  =  \alpha_0 + \sum_{i = 1}^{p} \alpha_i Y_{t-i}^2,
\end{equation}
for an ARCH($p$) process with unknown parameters $\alpha_0, \ldots, \alpha_p$, which have to be estimated. To obtain a positive conditional variance, the constant term $\alpha_0$ is assumed to be positive and $\alpha_i \geq 0$ for all $i = 1, \ldots, p$. In practice, large lag orders $p$ are often needed to capture the volatility dynamics, e.g., those of inflation indices \citep{engle1983estimates}.  A GARCH process extends this model by allowing for autoregression and moving average components in the conditional variance equation, reducing the required lag orders and, thus, the computational burdens. The basic GARCH($p$,$q$) model is given by
\begin{equation}
    h_t  =  \alpha_0 + \sum_{i = 1}^{p} \alpha_i Y_{t-i}^2 + \sum_{i = 1}^{q} \beta_i h_{t-i},
\end{equation}
with $\beta_i \geq 0$, $i = 1, \ldots, q$, being additional parameters to be estimated \citep[see][]{Bollerslev86}. A GARCH process is weakly stationary if $\sum_{i=1}^p \alpha_i + \sum_{j=1}^q \beta_j < 1$. For a detailed discussion of the statistical properties of ARCH and GARCH models and the parameter estimation, we refer the interested reader to the textbook of \cite{francq2019garch}.

ARCH and GARCH models are also particularly useful as error processes of other regression and time series models to describe time-dependent and dynamic model uncertainties. Since they have an expectation of zero, they can be directly applied to any other mean model, as \cite{Engle82} already demonstrated for an ARCH regression model with a linear regression component. Later, \cite{weiss1984arma,weiss1986asymptotic,weiss1986arch} considered ARCH models for the errors of autoregressive moving average processes. These models exhibit an autoregressive dependence in both the mean and the conditional variance.

Since then, several extensions and adaptations of GARCH models have been proposed. Each of these models has its own strengths and weaknesses, and choosing the best model for a particular data set depends on the nature of the data, the research question, and other factors. First, to avoid the non-negative constraints of the model coefficients, \cite{Geweke86,Pantula86,Milhoj87} proposed a logarithmic expression of the (log-)volatility equation, i.e., 
\begin{equation}
    \log h_t  = \alpha_0 + \sum_{i = 1}^{p} \alpha_i \log Y_{t-i}^2 + \sum_{i = 1}^{q} \beta_i \log h_{t-i} \, .
\end{equation}
Consequently, the process exhibits multiplicative dynamics in the volatility, while the standard ARCH and GARCH models have additive volatility dynamics. Moreover, this logarithmic GARCH (log-GARCH) model allows for a direct transformation of the log-squared observations to an autoregressive moving average process of orders $p$ and $q$. Interestingly, the log-GARCH model is invariant to power log-GARCH models where $\log h_t$ is replaced by $\log h_t^\delta$ with the power $\delta > 0$, which acts as a scaling factor of the logarithmic terms \citep{sucarrat2019log}. Since $h_t = \text{exp}(\alpha_0 + \sum_{i = 1}^{p} \alpha_i \log Y_{t-i}^2 + \sum_{i = 1}^{q} \beta_i \log h_{t-i})$, the conditional variance is always positive for real-valued coefficients $\{\alpha_i:i=1,\ldots,p\}$ and $\{\beta_i:i=1,\ldots,q\}$. This is particularly interesting when exogenous covariates influence volatility. That is, 
\begin{equation}
    \log h_t  = \alpha_0 + \sum_{i = 1}^{p} \alpha_i \log Y_{t-i}^2 + \sum_{i = 1}^{q} \beta_i \log h_{t-i}  + \sum_{j=1}^{s}\delta_j x_{j,t-1}, 
\end{equation}
with $\delta_j$ being the corresponding regression coefficients of the $j$-th covariate $x_{j,t}$ \citep{sucarrat2019log}. In the GARCH counterpart, the regressors are assumed to be almost surely positive with non-negative coefficients \citep{francq2019qml}. 

Second, to replicate the so-called leverage effect, which is often observed in financial  return data, the model has been extended to Exponential GARCH models (EGARCH, \citealt{Nelson91}) that allow for asymmetric dependence by including the sign of the residuals in the log-volatility equation:
\begin{equation}
    	\log h_t  =  \alpha_0 + \sum_{i = 1}^{q} \beta_i g(\varepsilon_{t-i}) \quad \text{with} \quad g(\varepsilon_{t}) = \theta \varepsilon_t + \gamma(|\varepsilon_t| - E(|\varepsilon_t|)) \, .
\end{equation}

Third, \cite{higgins1992class} proposed a non-linear ARCH (NARCH) model, which nests the linear ARCH model of \cite{Engle82} as a special case and converges to the log-ARCH model in the limiting case. More precisely, the volatility equation of the NARCH model is given by
\begin{equation}
    	h_t  =  \left[ \phi_0(\alpha_0)^{\delta} + \sum_{i = 1}^{p} \phi_i(\varepsilon_{t-i}^2)^\delta\right]^{1/\delta} \, 
\end{equation}
with $\phi_i \geq 0$ for $i = 0,1,\ldots, p$ and $\delta > 0$. This model converges to a log-ARCH model for $\delta$ approaching zero.  

Apart from these models, several other extensions have been proposed, which should only briefly be mentioned, e.g., threshold GARCH models (TGARCH, \citealt{zakoian1994threshold,glosten1993relation}, adding a threshold term to the GARCH equation for different volatility dynamics below and above the threshold), Glosten, Jaganathan and Runkle GARCH models (GJR-GARCH, asymmetric volatility by including both positive and negative residuals in the GARCH equation), or fractionally integrated GARCH models (FIGARCH, long memory in the volatility process by using fractional integration in the conditional variance equation). For a more detailed overview of univariate time-series ARCH and GARCH models, we refer the interested reader to the survey paper of \cite{bera1993arch} and the textbook of \cite{francq2019garch}.

% https://mpra.ub.uni-muenchen.de/100386/1/MPRA_paper_100386.pdf

\subsubsection{Multivariate ARCH and GARCH models} \label{sMARCH}

\begin{figure}
    \centering
    \includegraphics[width=0.7\textwidth]{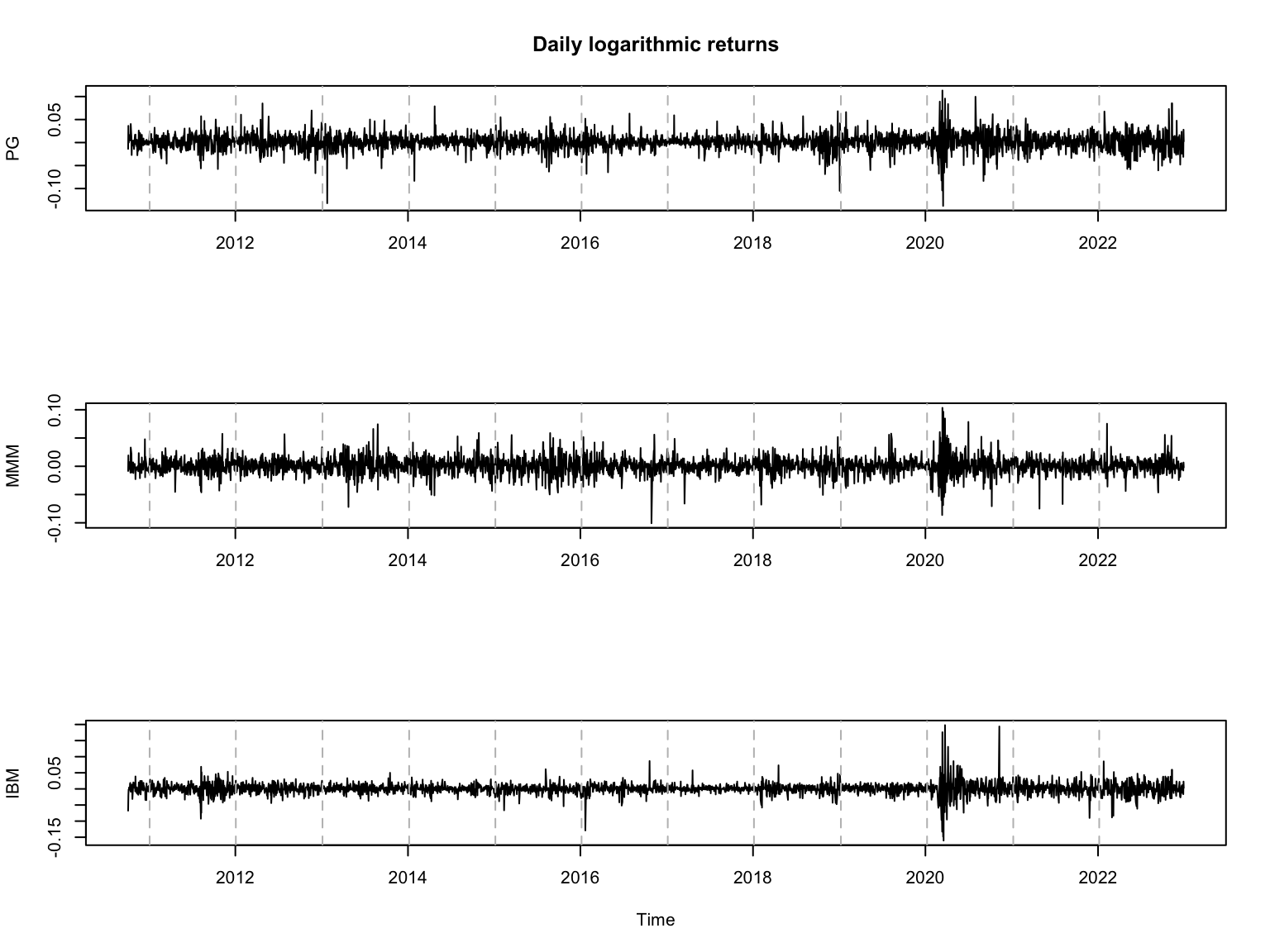}\\
    \includegraphics[width=0.7\textwidth]{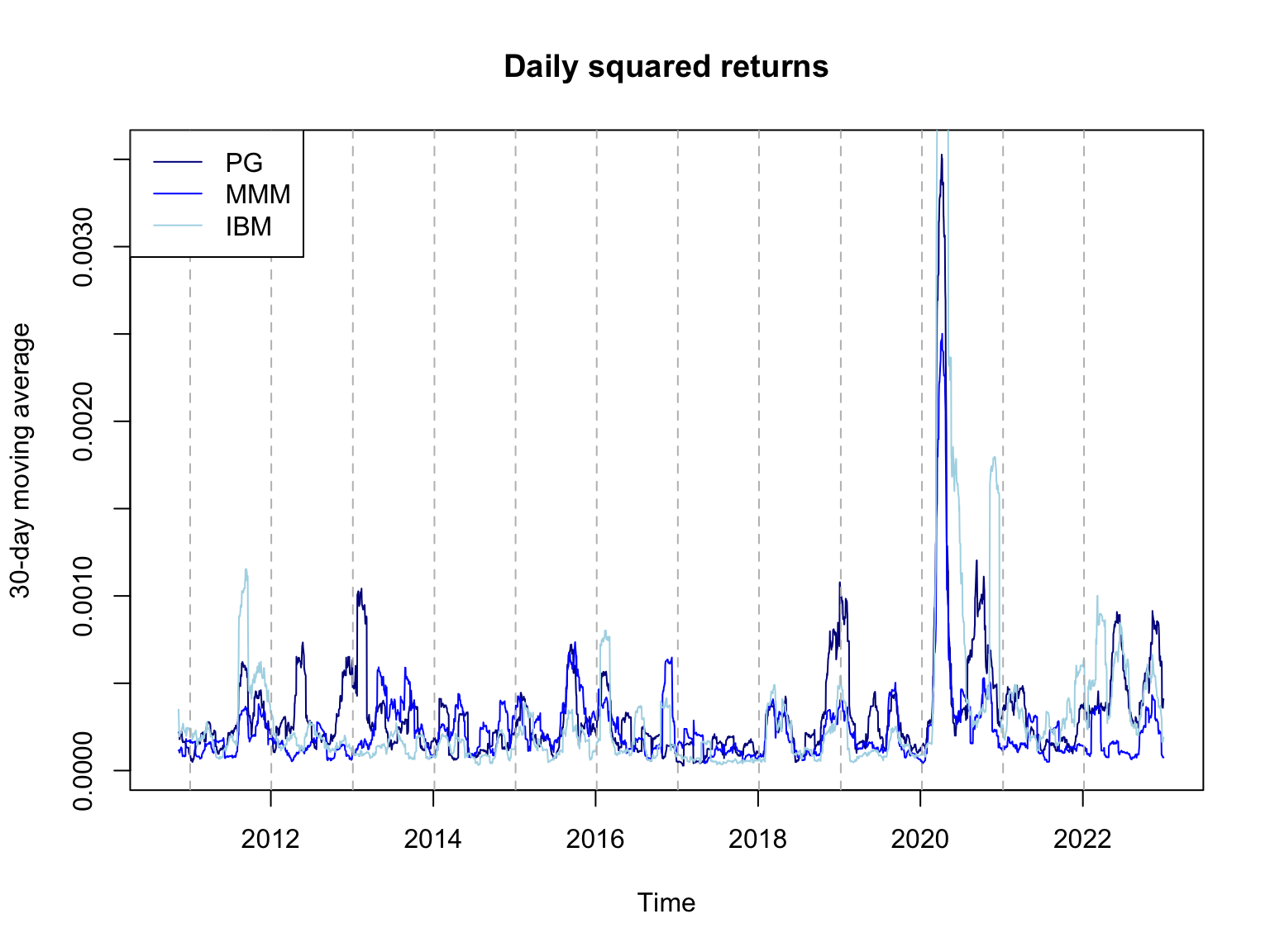}
    \caption{Example of a multivariate time series. Top: daily logarithmic returns of three selected stocks in the Dow Jones Index, Procter \& Gamble (PG), 3M (MMM), and International Business Machines (IBM). Bottom: 30-days moving averages of the squared daily returns showing the co-movement of the volatility.}
    \label{fig:multivariate_ts}
\end{figure}

In general, financial asset returns tend to move together over time. Figure \ref{fig:multivariate_ts} shows exemplarily a multivariate time series consisting of daily logarithmic returns of three selected stocks in the Dow Jones Index, Procter \& Gamble (PG), 3M (MMM), and International Business Machines (IBM). The top plot displays the raw data of the daily returns over time, and the bottom plot shows the 30-day moving averages of the squared daily returns, indicating the co-movement of volatility among the three stocks. This can provide insights into the correlations and dependencies between the stocks, which can be useful for portfolio and risk management. Thus, a multivariate framework is well-suited for modelling the time-varying conditional covariance matrix for all returns. Several multivariate GARCH (MGARCH) models have been proposed to account for such dependence in the volatilities.

Recall that $\xvec{Y}_t$  is an $r$-dimensional vector of returns at time point $t$. It is assumed that
\begin{equation}
    {\xvec Y}_t = {\xmat H}_t^{1/2} \xvec{\varepsilon}_t \, ,
\end{equation} 
where $\{ \xvec{\varepsilon}_t \}$ is a sequence of i.i.d. $r$-dimensional random variables with $\E( \xvec{\varepsilon}_t ) = {\xvec 0}$ and $\text{Var}(\xvec{\varepsilon}_t) = \xmat{I}_r$. The square root has to be understood in the sense of the Cholesky factorization, i.e. $\xmat{H}_t^{1/2}$ is the unique symmetric and positive symmetric matrix with $\xmat{H}_t^{1/2} (\xmat{H}_t^{1/2})^\prime = \xmat{H}_t$. The matrix $\xmat{H}_t$ is allowed to depend on certain parameters, and it is assumed to be a measurable function with respect to the $\sigma$-algebra generated by $\xvec{Y}_v, v < t$. Thus,
\begin{equation}
\text{Var}(\xvec{Y}_t | \xvec{Y}_v, v <t) = \xmat{H}_t ,
\end{equation} 
i.e., $\xmat{H}_t$ is the volatility matrix of $\xvec{Y}_t$.

Most of the published papers on this topic appeared at the end of the 80s and the 90s of the 20th century. In principle, all these models only differ in modelling the matrix $\xmat{H}_t$. Here we want to sketch some of the most relevant approaches briefly.

The first paper on MGARCH models is due to \cite{bollerslev1988capital}. They introduced the so-called vector GARCH  model, briefly VEC model. It is based on the idea of transforming the matrix $\xmat{H}_t$ to a vector by using the vech-operator \citep[cf.][]{harville1998matrix}. For a square $r \times r$ matrix $\xmat{B}$, the operator $\text{vech}(\xmat{B})$ is defined as the $r(r+1)/2$-dimensional vector obtained by stacking the columns of the lower triangular part of $\xmat{B}$. Let $\xvec{h}_t = \text{vech}(\xmat{H}_t)$. Then, it holds for a VEC-GARCH(1,1) process that 
\begin{equation}
\xvec{h}_t = \xvec{\omega} + \xmat{A} \xvec{\eta}_{t-1} + \xmat{B} \xvec{h}_{t-1} ,
\end{equation} 
where $\xvec{\eta}_{t-1} = \text{vech}(\xvec{\varepsilon}_{t-1} \xvec{\varepsilon}_{t-1}^\prime)$, $\xmat{A}$ and $\xmat{B}$ are assumed to be $r(r+1)/2 \times r(r+1)/2$ parameter matrices, and $\xvec{\omega}$ is a $r(r+1)/2$ parameter vector. The total number of parameters of this model is $r(r+1)(r(r+1)+1)/2$. For $r=2$ it is $21$, for $r=3$ it is $78$, and for $r=4$ it is already $210$. Thus, the number of parameters increases very fast as $r$ increases. This is the reason why in practice, the model is only used for small values of $r$, e.g., $r=2$ or $r=3$. 

Another possibility to reduce the number of parameters is to assume that $\xmat{A}$ and $\xmat{B}$ are diagonal matrices \citep{bollerslev1988capital}. This model is denoted as the diagonal vector GARCH model, briefly DVEC-GARCH(1,1) model. Using the Hadamard product \citep[cf.][]{harville1998matrix}, it can be written as
\begin{equation}
\xmat{H}_t = \xmat{\Omega}^\# + \xmat{A}^\# \odot (\xvec{\varepsilon}_{t-1} \xvec{\varepsilon}_{t-1}^\prime) + \xmat{B}^\# \odot \xmat{H}_{t-1}, 
\end{equation}
where $\xmat{A}^\#$, $\xmat{B}^\#$, and $\xmat{\Omega}^\#$ are $r \times r$-matrices implied by $\xmat{A} = \text{diag}(\text{vech}(\xmat{A}^\#))$, $\xmat{B} = \text{diag}(\text{vech}(\xmat{B}^\#))$, and $\xmat{\Omega} = \text{diag}(\text{vech}(\xmat{\Omega}^\#))$. Consequently,  $h_{ij,t}$ depends only on its own lag and on $\varepsilon_{i,{t-1}} \varepsilon_{j,{t-1}}$. This restriction dramatically simplifies the parameter estimation; however, it is still not suitable for large-scale systems. 

A further simplification of this model was discussed by \cite{ding2001large}. They choose $\xmat{A}^\#$ and $\xmat{B}^\#$ as matrices of rank one or as multiples of the matrix consisting purely on the element $1$. Further, \cite{riskmetrics1996jp} uses the exponentially weighted moving average model, which can be written as
\begin{equation}
\xvec{h}_t = (1-\lambda) \xvec{\eta}_{t-1} + \lambda \xvec{h}_{t-1} \, ,
\end{equation}
which corresponds to a scalar VEC-GARCH(1,1) model. Riskmetrics recommends choosing the factor $\lambda$ equal to $0.94$ for daily data and $0.97$ for monthly data \citep[critically discussed by][based on empirical evidence]{bollen2015should,gonzalez2007optimality}.

% see also arxiv preprint: Araneda, A. A. (2021). Asset volatility forecasting: The optimal decay parameter in the EWMA model. arXiv preprint arXiv:2105.14382.

Strong restrictions on the parameters are necessary to ensure the matrix $\xmat{H}_t$ to be positive definite. For that reason, \cite{engle1995multivariate} proposed another approach, the BEKK (Baba, Engle, Kraft, and Kroner) model. The BEKK-GARCH(1,1,$K$) process is given by \begin{equation}
\xmat{H}_t = \xmat{\Omega} + \sum_{i=1}^K \xmat{A}_i \xvec{\varepsilon}_{t-1} \xvec{\varepsilon}_{t-1}^\prime \xmat{A}_i^\prime + \sum_{i=1}^K \xmat{B}_i \xmat{H}_{t-1} \xmat{B}_i^\prime  \, ,
\end{equation}
where $\xmat{A}_i$, $\xmat{B}_i$, and $\xmat{\Omega}$ are $r \times r$ matrices and $\xmat{\Omega}$ is positive definite. The BEKK-GARCH model is a special case of the VEC-GARCH approach, but the converse is not true \citep{stelzer2008relation}. The number of parameters is again high. Similar to the VEC-GARCH and DVEC-GARCH, the application of the BEKK-GARCH model reduces to cases where $r$ is small. 

\cite{Bollerslev90} introduced a type of MGARCH model where the conditional correlations are constant (CCC model). For the CCC-GARCH(1,1) process, it holds that 
\begin{equation}
\xmat{H}_t = \xmat{D}_t \xmat{R} \xmat{D}_t = (\rho_{ij} \sqrt{h_{ii,t} h_{jj,t}} ) \, ,
\end{equation}
where $\xmat{R} = ( \rho_{ij} )$ and $\xmat{D}_t = \text{diag}(h_{11,t}^{1/2},..., h_{rr,t}^{1/2})$ and 
\begin{equation}
h_{ii,t} = \omega_i + \alpha_i \varepsilon_{i,t-1}^2 + \beta_i h_{ii,t-1}, \quad i=1,...,r  .
\end{equation}
Here, the number of parameters is $r(r+5)/2$.

A model with dynamic conditional correlations (DCC model) was proposed by \cite{Engle02} and \cite{tse2002multivariate}. For the DCC-GARCH(1,1) model of \cite{Engle02} it holds that
\begin{equation}
\xmat{H}_t = \xmat{D}_t \xmat{R}_t \xmat{D}_t = ( h_{ij,t} )
\end{equation}
with $\xmat{D}_t$ as above and
\begin{eqnarray}
\xmat{R}_t & = & \text{diag}(q_{11,t}^{-1/2},..., q_{rr,t}^{-1/2}) \xmat{Q}_t \text{diag}(q_{11,t}^{-1/2},..., q_{rr,t}^{-1/2}) , \\
\xmat{Q}_t & = & (1-\alpha-\beta) \bar{\xmat{Q}} + \alpha \xvec{u}_{t-1} \xvec{u}_{t-1}^\prime + \beta \xmat{Q}_{t-1}  = ( q_{ij,t} )
\end{eqnarray}
and $\xvec{u}_t = ( u_{i,t} )$ with $u_{i,t} = \varepsilon_{i,t}/\sqrt{h_{ii,t}}$. $\bar{\xmat{Q}}$ denotes the unconditional covariance matrix of $\xvec{u}_t$ and $\alpha$ and $\beta$ are non-negative numbers satisfying that $\alpha + \beta < 1$. 

Besides these models, many further proposals have been made. The above models seem to be the most applied ones. There are also other attempts to overcome the problem of dimensionality. Factor GARCH models are one step in that direction. Here the idea is that some factors drive the behaviour of the stock returns. Such an approach was discussed by, e.g., \cite{engle1990asset}, \cite{lin1992alternative} and \cite{bollerslev1993common}. 

We refer to the overview papers by \cite{bauwens2006multivariate} and \cite{silvennoinen2009multivariate}, as well as the book by \cite{francq2019garch}, where the presented multivariate models and many further ones are discussed in more detail.

\subsection{Stochastic volatility models}\label{sSV}

\subsubsection{Univariate stochastic volatility models}\label{suSV}

In stochastic volatility models, the volatility process is modelled through a latent stochastic process. Although it is difficult to determine the exact origin of these models as they arose from various research efforts addressing different issues, \citet{Taylor:1982, Taylor:1986} seem to be the first to consider a univariate discrete version that can be considered as an alternative to the ARCH process. To learn more about the origin and development of stochastic volatility models, refer to \citet{Eric:1996} and \citet{Shephard:2005}. A standard discrete time stochastic volatility model is specified in the following way:
\begin{align}
    &Y_t=e^{h_t/2}\varepsilon_t,\label{s1}\\
    &h_t-\mu_h=\phi(h_{t-1}-\mu_h)+u_t,\label{s2}
\end{align}
where $Y_t$ is the observed response variable, $\{h_t\}$ is the sequence of the unobserved log-volatility, assuming an AR(1) process with a mean parameter $\mu_h$ and an autoregressive parameter $|\phi|<1$. The model includes two independent disturbance terms denoted by $\varepsilon_t$ and $u_t$. The sequence $\{\varepsilon_t\}$ includes the independent random variables with an identical distribution with mean $0$ and variance $1$. The disturbance term $\{u_t\}$ in the log-volatility equation are independent and have identical distribution with mean $0$ and variance $\sigma^2_h$. In this specification, the sign of $Y_t$ is determined by that of $\varepsilon_t$, and the volatility clustering and fat tail properties observed in the marginal distribution of $Y_t$ are delivered by the time-varying log-volatility (see \citet{Eric:1996} on the statistical properties of stochastic volatility models). This standard model can also be obtained as a discrete-time approximation  to various diffusion processes in the continuous-time asset pricing literature \citep{Hull:1987, Wiggins:1987, Melino:1990, Scott:1989}.

The outcome and log-volatility equations can be modified to formulate alternative specifications. The log-volatility equation is specified as an AR(1) process and can be generalised to any ARMA process. For example, if $h_t$ follows an AR(p) process, then it will take the following form:
\begin{align}
    &h_t-\mu_h=\sum_{j=1}^p\phi_j(h_{t-j}-\mu_h)+u_t,
\end{align}
where $\phi_1,\hdots,\phi_p$ are unknown autoregressive parameters. 

An alternative specification can be obtained by assuming the presence of infrequent jumps in the outcome equation. Adding a jump component to the outcome equation can improve the fit of the observed time series of returns because the jump component may capture outliers as well as asymmetry in the return distribution \citep{Andersen:2002, Chib:2002}.  The stochastic volatility model with a jump component can be specified as
\begin{align}
    &Y_t=k_tq_t+e^{h_t/2}\varepsilon_t,\\
    &h_t-\mu_h=\phi(h_{t-1}-\mu_h)+u_t,
\end{align}
where $q_t$ is the jump random variable, and $k_t$ is the jump size random variable. The jump random variable is a Bernoulli random variable with success probability $P(q_t=1)=\kappa$, and the jump size is modelled as $\log(1+k_t)\sim N(-0.5\delta^2,\delta^2)$. In this model, $\kappa$ and $\delta$ are additional unknown parameters that we need to estimate along with $\mu_h$, $\phi$, and $\sigma^2_u$.

Another variant can be defined by allowing the volatility feedback in the outcome equation \citep{Koopman:2002}:
\begin{align}
    &Y_t=\alpha e^{h_t}+e^{h_t/2}\varepsilon_t,\\
    &h_t-\mu_h=\phi(h_{t-1}-\mu_h)+u_t,
\end{align}
where the scalar unknown parameter $\alpha$ gives the effect of volatility on the outcome variable. \citet{Chan:2017} extended this model by considering time-varying parameters:
\begin{align}
    &Y_t=\xvec{x_t}^{'}\xvec{\beta}_t+\alpha_t e^{h_t}+e^{h_t/2}\varepsilon_t,\\
    &h_t-\mu_h=\phi(h_{t-1}-\mu_h)+u_t,
\end{align}
where $\xvec{x_t}$ is the $k\times1$ vector of covariates with matching time-varying parameter vector $\xvec{\beta}_t$. This model generalises the model suggested in \citet{Koopman:2002} by allowing time-varying parameters $\xvec{\beta}_t$ and $\alpha_t$ in the outcome equation. Let $\xvec{\gamma}_t=(\alpha_t,\xvec{\beta}^{'}_t)^{'}$ be the $k\times1$ vector of time-varying parameters. \citet{Chan:2017} assumes an a random walk process for $\xvec{\gamma}_t$ such that $\xvec{\gamma}_t=\xvec{\gamma}_{t-1}+\xvec{\nu}_t$, where $\xvec{\nu}_t\sim N(\xvec{0},\xmat{\Gamma})$, and $\xmat{\Gamma}$ is the $(k+1)\times(k+1)$ covariance matrix.

\citet{Harvey:1996} consider a variant that allows for the so-called ``leverage effect'' via introducing correlation  in the disturbance terms of the outcome and the log-volatility equations. See also \citet{Eric:2004}, \citet{Yu:2005} and \citet{Omori:2007} on the different versions of this model. \citet{Shephard:2005} notes that \citet{Hull:1987} were the first to propose a continuous-time stochastic volatility model incorporating the leverage effect. This work, in turn, inspired the development of the EGARCH model proposed by Nelson (1991) \citep{Shephard:2005}. The variant proposed by \citet{Harvey:1996} can be specified as 
\begin{align}
    &Y_t=e^{h_t/2}\varepsilon_t,\label{eq15}\\
    &h_{t+1}-\mu_h=\phi(h_{t}-\mu_h)+u_{t},\label{eq16}\\
    &(\varepsilon_t,u_t)^{'}\sim N\left(\xvec{0},\begin{pmatrix}1&
    \varrho\sigma_u\\\varrho\sigma_u&\sigma^2\end{pmatrix}\right),\label{eq17}
\end{align}
where $\varrho$ is the correlation parameter. In this specification, \citet{Yu:2005} defines the leverage effect as the negative relationship between $E(h_t|Y_t)$ and $Y_t$, and derived the following equation:
\begin{align*}
    \text{E}(h_t|Y_t)=\frac{\mu_h(1+\phi-\phi^2)}{(1-\phi)}+\varrho\sigma_u\exp\left(-\frac{\sigma^4_u}{4(1-\phi^2)^2}+\frac{\sigma^2_u\mu_h}{(1-\phi)^2}\right)Y_t.
\end{align*}
This result suggests that this specification will exhibit the leverage effect whenever $\varrho<0$. 

Another variant can be obtained by assuming a scale mixture distribution for the outcome variable \citep{Eric:2004, Chib:2002}. This version takes the following form:
\begin{align}
    &Y_t=e^{h_t/2}\omega^{1/2}_t\varepsilon_t,\\
    &h_t-\mu_h=\phi(h_{t-1}-\mu_h)+u_t,
\end{align}
where $\{\omega_t\}$ is the sequence of latent variables that are independent and have identical distribution.  Under the assumptions that $\varepsilon_t\sim N(0,1)$ and $\omega_t\sim IG(\nu/2,\nu/2)$, where $IG$ denotes the inverse gamma distribution, it can be shown that  the marginal distribution of $\omega^{1/2}_t\varepsilon_t$ (unconditional on $\omega_t$) is the standard $t$ distribution with $\nu$ degrees of freedom \citep{Geweke:1993}. \citet{Omori:2007} consider the same model under the assumption that $\log(\omega_t)\sim N(-0.5\tau^2,\tau^2)$, where $\tau$ is a scalar unknown parameter with $\tau^2\sim\text{Gamma}(1,1)$. 

\citet{Chan:2013} introduces a class of models that includes both the moving average and stochastic volatility components that can nest a variety of specifications as special cases. This specification takes the following form:
\begin{align}
    &Y_t=\mu_t+\varepsilon_t,\\
    &\varepsilon_t=e^{h_t/2}v_t+\psi_1e^{h_{t-1}/2}v_{t-1}+\hdots+\psi_qe^{h_{t-q}/2}v_{t-q},\quad v_t\sim N(0,1),\\
    &h_t-\mu_h=\phi(h_{t-1}-\mu_h)+u_t,\quad u_t\sim N(0,\sigma^2_u),
\end{align}
where $\mu_t$ is the time-varying conditional mean process, $\psi_1,\hdots,\psi_q$ are the unknown moving average parameters, and the disturbance terms $v_t$ and $u_t$ are independent of each other for all leads and lags. Let $\xvec{\mu}=(\mu_1,\hdots,\mu_T)^{'}$, $\xvec{h}=(h_1,\hdots,h_T)^{'}$ and $\xvec{\psi}=(\psi_1,\hdots,\psi_T)^{'}$. Then, this specification gives 
\begin{align*}
\text{Var}(Y_t|\bs{\mu},\xvec{h},\xvec{\psi})=e^{h_t}+\psi^2_1e^{h_{t-1}}+\hdots+\psi^2_qe^{h_{t-q}},
\end{align*}
which indicates that the conditional variance of $Y_t$ is a moving average of $q+1$ most recent variances $e^{h_t},e^{h_{t-1}},\hdots,e^{h_{t-q}}$. Moreover, unlike the standard
stochastic volatility model in \eqref{s1} and \eqref{s2}, $Y_t$ is serially correlated even after conditioning on $\xvec{h}$. \citet{Chan:2013} shows that
\begin{align*}
\text{Cov}(Y_t,Y_{t-j}|\xvec{\mu},\xvec{\psi},\xvec{h})=
\begin{cases}
    \sum_{i=0}^{q-j}\psi_{i+j}\psi_ie^{h_{t-i}},\quad\text{for}\quad j=1,\hdots,q,\\
    0,\quad\text{for}\quad j>q,
\end{cases}
\end{align*}
where $\psi_0=1$. Thus, the conditional covariances are also time-varying because of the presence of the log-volatility $h_t$. As stated in \cite{Chan:2013}, popular specifications can be obtained from this model by choosing a suitable conditional mean process $\mu_t$. For example, some of these specifications are (i) an AR(p) model: $\mu_t=\beta_0+\beta_1Y_{t-1}+\hdots+\beta_pY_{t-p}$,
(ii) a linear regression model: $\mu_t=\xvec{x}^{'}_t\bs{\beta}$, where $\xvec{x}_t$ is the $k\times1$ vector of covariates,
(iii) an unobserved component model: $\mu_t=\tau_t$, where $\tau_t=\tau_{t-1}+\varepsilon^{\tau}_t$ and $\varepsilon^{\tau}_t\sim N(0,\sigma^2_{\tau})$, (iv) a time-varying regression model: $\mu_t=\xvec{x}^{'}_t\xvec{\beta}_t$, where $\xvec{\beta}_t=\xvec{\beta}_{t-1}+\xvec{\varepsilon}^{\xvec{\beta}}_t$ and $\xvec{\varepsilon}^{\xvec{\beta}}_t\sim N(\xvec{0},\xmat{\Sigma}_{\xvec{\beta}})$.

The likelihood function of a stochastic volatility model is a mixture over the distribution of $\mf{h}$ and therefore requires evaluation of a high-dimensional integral. For example, the likelihood function of the model in \eqref{s1}-\eqref{s2} can be defined as $f(\xvec{Y}|\xvec{\theta})=\int f(\xvec{Y}|\xvec{h},\xvec{\theta})f(\xvec{h}|\xvec{\theta})\text{d}\xvec{h}$, where $\xvec{Y}=(Y_1,\hdots, Y_T)^{'}$, $f(\xvec{h}|\xvec{\theta})$ is the prior distribution of $\xvec{h}$ determined by \eqref{s2}, and $\xvec{\theta}=(\mu_h,\phi,\sigma^2_h)^{'}$.\footnote{Note that $f(\xvec{Y}|\xvec{\theta})$ is also called the observed-data likelihood function or the integrated likelihood function. Two other functions that can be defined are (i) the conditional likelihood function $f(\mf{Y}|\mf{h},\bs{\theta})$, and (ii) the complete-data likelihood function given by $f(\xvec{Y},\xvec{h}|\xvec{\theta})=f(\xvec{Y}|\xvec{h})\times f(\xvec{h}|\xvec{\theta})$. The conditional and complete-data likelihood functions are readily available for all stochastic volatility models.} This feature indicates that the estimation based on the exact maximum likelihood is not readily available and poses difficulties for likelihood-based estimation procedures. In the literature, various estimation methods exist, including the generalised method of moments (GMM), the quasi-maximum likelihood method, spectral GMM based on the characteristic function, indirect inference methods, Monte Carlo maximum likelihood methods, and Markov chain Monte Carlo (MCMC) based methods. According to \citet{Asai:2006}, the choice of an estimation method can be based on the following properties: (i) efficiency, (ii) estimation of volatility, (iii) optimal filtering, smoothing, and forecasting methods, (iv) computational efficiency, and (v) applicability to flexible models. Among others, see \citet{JPR:1994}, \citet{Kim:1998}, and \citet{Broto:2004} on the estimation methods suggested in the literature for the univariate stochastic volatility models. 

\subsubsection{Multivariate stochastic volatility models}\label{smSV}
As in the case of multivariate ARCH and GARCH models described in Section~\ref{sMARCH}, the cross-dependence among the volatility of different financial assets leads  to modelling volatility in a multivariate framework, which can lead to efficient estimation. Let $\bs{h}_t=(h_{1t},\hdots,h_{rt})^{'}$ be the $r\times 1$ vector of log-volatility terms, and $\xmat{H}^{1/2}_t=\text{diag}\left(e^{h_{1t}/2},\hdots,e^{h_{rt}/2}\right)$ be the $r\times r$ diagonal matrix with the $i$th diagonal element $e^{h_{it}/2}$. \citet{Harvey:1994} consider the following multivariate version:
\begin{align}
&\xvec{Y}_t=\xmat{H}^{1/2}\xvec{\varepsilon}_t,\\
& \xvec{h}_t=\xvec{\mu}+\xvec{\phi}\circ\xvec{h}_{t-1}+\xvec{u}_t,\\
&(\xvec{\varepsilon}^{'}_t,\,\xvec{u}^{'}_t)^{'} \sim
N\left(
\begin{pmatrix}
    \xvec{0}\\
    \xvec{0}
\end{pmatrix}
,\,
\begin{pmatrix}
    \xmat{P}_{\xvec{\varepsilon}}&\bs{0}\\
    \xmat{0}&\bs{\Sigma}_{\xvec{u}}
\end{pmatrix}
\right),
\end{align}
where $\xvec{\mu}$ and  $\xvec{\phi}$ are the $r\times1$ vectors of unknown coefficients, $\circ$ denotes the Hadamard product, $\xmat{P}_{\xvec{\varepsilon}}=(\rho_{ij})$ is a positive definite correlation matrix with $\rho_{ii}=1$ and $|\rho_{ij}|<1$ for $i,j=1,\hdots,r$, and $\xmat{\Sigma}_{\xvec{u}}=(\sigma_{u,ij})$ is the $r\times r$ positive definite covariance matrix.  \citet{Harvey:1994} also consider the multivariate $t$ distribution for $\xvec{\varepsilon}_t$. The volatility process can be generalised to a VARMA structure in the following way:
\begin{align*}
    \xmat{\Phi}(L)\xvec{h}_t=\xvec{\mu}+\xmat{\Theta}(L)\xvec{u}_t,
\end{align*}
where $\xmat{\Phi}(L)=\left(\xmat{I}_r-\sum_{i=1}^p\xvec{\phi}_i\circ L^i\right)$ and $\xmat{\Theta}(L)=\left(\xmat{I}_r-\sum_{i=1}^p\xvec{\theta}_i\circ L^i\right)$, $\xmat{I}_r$ is the $r\times r$ identity matrix, $L$ is the lag operator, and $\{\xvec{\phi}_i\}$ and $\{\xvec{\theta}_i\}$ are vectors of parameters.

Note that if the off-diagonal elements of $\xmat{P}_{\xvec{\varepsilon}}$ and $\xmat{\Sigma}_{\xvec{u}}$ are zeros, then this multivariate model simply specifies a univariate standard stochastic volatility model for each component of $\xvec{Y}_t$. Although the non-zero off-diagonal elements of $\xmat{\Sigma}_{\bs{u}}$ introduce correlation across the volatility terms, the model does not allow for the leverage effect. To introduce the leverage effect, \citet{Asai:2006} consider the following model:
\begin{align}
&\xvec{Y}_t=\xmat{H}^{1/2}\xvec{\varepsilon}_t,\\
& \xvec{h}_t=\xvec{\mu}+\xvec{\phi}\circ\xvec{h}_{t-1}+\xvec{u}_t,\\
&(\xvec{\varepsilon}^{'}_t,\,\xvec{u}^{'}_t)^{'} \sim
N\left(
\begin{pmatrix}
    \xvec{0}\\
    \xvec{0}
\end{pmatrix}
,\,
\begin{pmatrix}
    \xmat{P}_{\xvec{\varepsilon}}&\xmat{L}\\
    \xmat{L}&\xmat{\Sigma}_{\xvec{u}}
\end{pmatrix}
\right),
\end{align}
where $\xmat{L}=\text{diag}\left(\lambda_1\sigma_{\xvec{u},11},\hdots,\lambda_r\sigma_{\xvec{u},rr}\right)$. Thus, the model exhibits the leverage effect when $\lambda_i<0$ for $i=1,\hdots,r$. 

Some other alternative versions that may not lead to the leverage effect as defined by \cite{Yu:2005} are also considered by \cite{Dani:1998} and \cite{Chan:2006}. See \cite{Yu:2006} for further details on the models that can deliver the leverage and asymmetric effects. There are also alternative specifications in the literature, including parsimonious specifications based on additive and multiplicative factors structures, time-varying correlation matrix models, matrix exponential models, Cholesky decomposition-based models, Wishart models, and range-based models. The details of these models and the estimation approaches considered in the literature are surveyed in \cite{Yu:2006}, \citet{Renate:2006}, and \citet{Chib:2009}.   

\section{Spatial volatility models}\label{sec:spatial}
\begin{figure}
    \centering
    \includegraphics[width=0.49\textwidth]{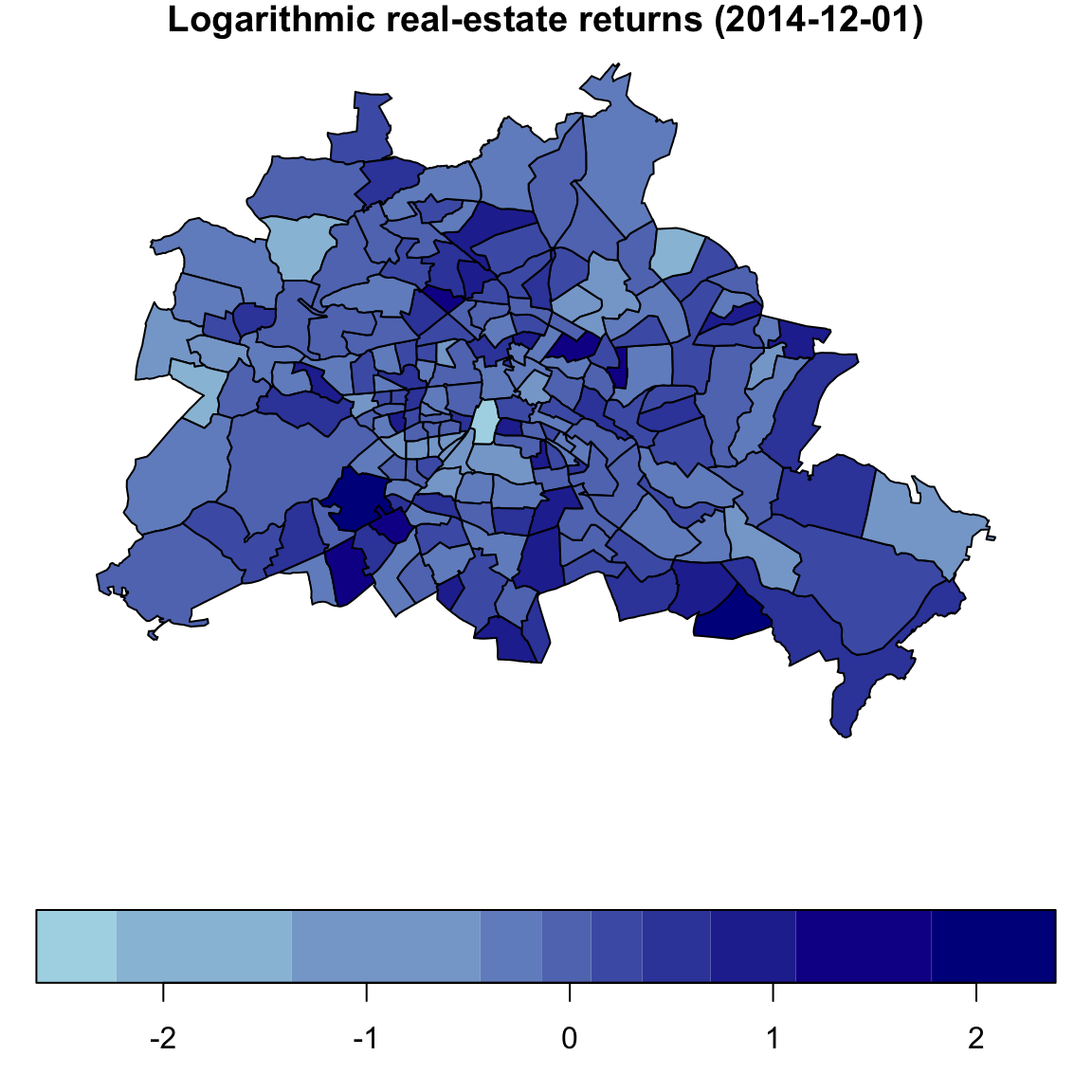}
    \includegraphics[width=0.49\textwidth]{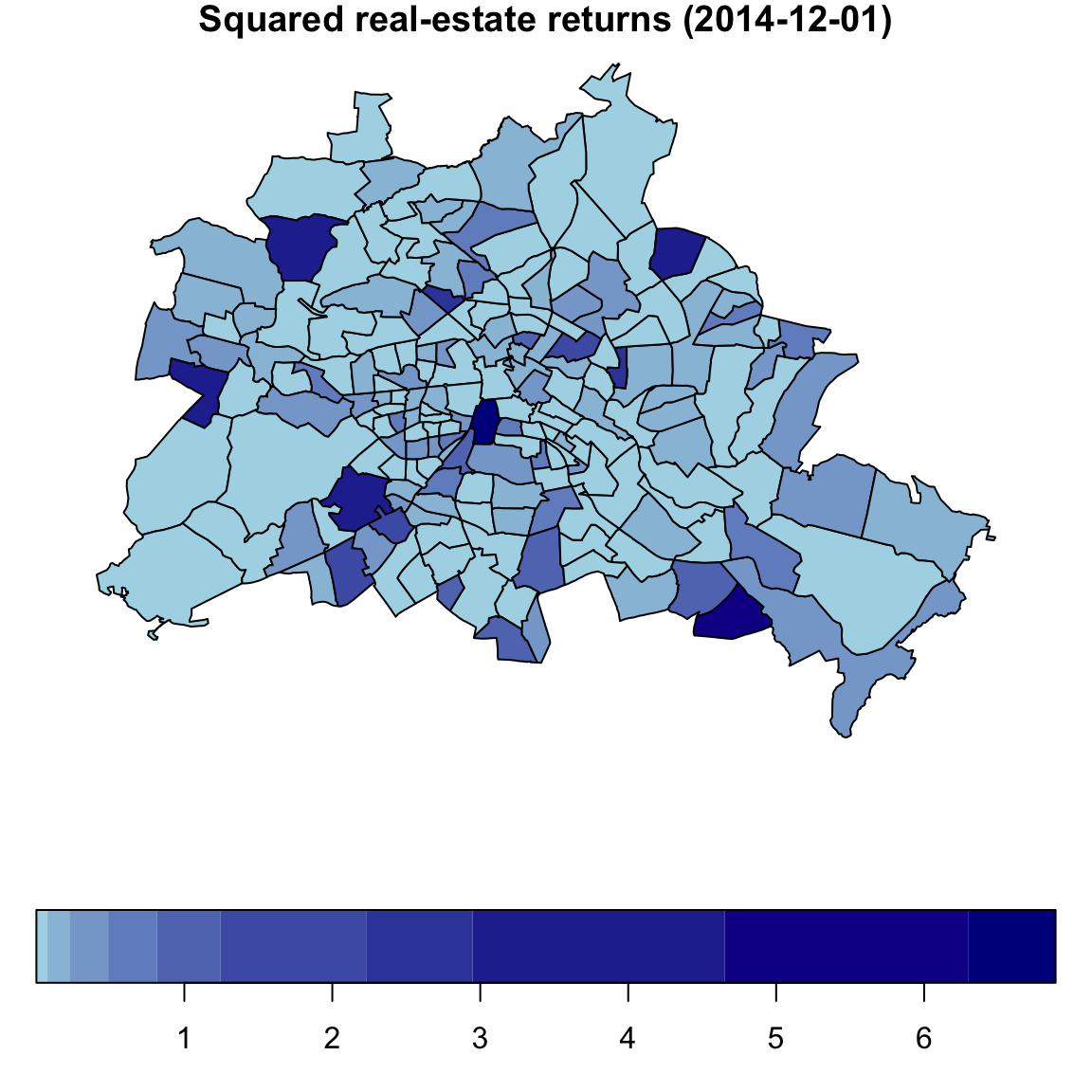}
    \caption{Example of a spatial process. Left: Monthly logarithmic real-estate returns (condominium prices) in all postcode regions of Berlin in December 2014. Left: Squared logarithmic real-estate returns in December 2014.}
    \label{fig:spatial_map}
\end{figure}

Now, suppose that the process is observed across space. In contrast to time series, where the index is a scalar value, the index of each observation is now at least two-dimensional. Consider the random process $\{ \xvec{Y}(\xvec{s}) \in \mathbb{R}^r : \xvec{s} \in D \subseteq \mathbb{R}^d, d > 1\}$, where $D$ is the spatial domain, typically a subset of $\mathbb{R}^d$ with $d > 1$ and positive volume. For instance, if $D \subseteq \mathbb{Z}^d$, the spatial domain would be called lattice (e.g., satellite image sequences for $d = 2$, or CT images for $d = 3$), while a continuous spatial process is present if $D \subseteq \mathbb{R}^d$ (e.g., soil samples, or species distributions). Another typical example is the case when $D$ is a discrete set $\{ \xvec{s}_1, \ldots, \xvec{s}_n\}$ of locations or polygons (e.g., air quality measurement stations, economic country/county-level data). Furthermore, $D$ could be considered to be a spherical space, $\mathbb{S}^d = \{ \xvec{x} \in \mathbb{R}^{d+1} : || \xvec{x} || = c\}$ with the radius $c$, which is particularly useful for modelling global data on the Earth. Figure \ref{fig:spatial_map} presents an example of a purely spatial process where the monthly log returns of condominium prices in Berlin are shown on the map. Each location, postcode region, is represented by a polygon precisely defined by geographical coordinates. 

\subsection{Spatial ARCH and GARCH models}\label{sec:spatial_arch_garch}

Spatial ARCH models were first mentioned ``as a byproduct'' in \citet[][World econometrics conference proceedings]{Bera04}. Later, \cite{Otto18_spARCH} and \cite{otto2019stochastic} introduced the first purely spatial ARCH model (jointly published on arXiv in \citealt{Otto16_arxiv}), which has an ARCH-like dependence structure in the conditional variances. The first spatial ARCH models were introduced for univariate processes and purely spatial domains, i.e., the process is observed only once for all locations in $D$. In other words, there is only one time point $t$, and we do not repeatedly observe the process over time. For the univariate case with $r = 1$ and $n$ spatial locations, a spatial ARCH model is defined by
\begin{equation}\label{Garchmain}
    Y(\xvec{s}_i) = h(\xvec{s}_i)^{1/2} \varepsilon(\xvec{s}_i) \qquad \text{for all $i = 1, \ldots, n$},
\end{equation}
where $h(s_i)$ is a scaling factor of the error $\varepsilon(\xvec{s}_i)$ of the $i$-th location, analogously to the time-series case. The errors are supposed to be i.i.d. across all spatial locations and have a zero mean and a constant variance of 1. Further, $h(\xvec{s}_i)$ depends on all adjacent realisations of the response variable $Y$, where the adjacency is defined by a so-called spatial weights matrix $\xmat{W} = (w_{ij})_{i,j = 1, \ldots, n}$. The weights $w_{ij}$ are non-zero if $h(\xvec{s}_i)$ might be influenced by $h(\xvec{s}_j)$, i.e., $\xvec{s}_i$ is in proximity to $\xvec{s}_j$. Typically, the definition of the weights matrix depends on the geographical space and the coordinates of all locations. For example, $w_{ij}$ can be chosen as the inverse-distance between $\xvec{s}_i$ and $\xvec{s}_j$, or $w_{ij}$ could be equal to $1/k$ for all $k$ nearest neighbours. Then, 
\begin{equation}\label{eq:h_spatialGARCH}
    h(\xvec{s}_i)  =  \alpha_0 + \alpha_1 \sum_{j = 1}^{n} w_{ij} Y(\xvec{s}_j)^2 \, .
\end{equation}
Due to this temporal simultaneity, the volatility $h(\xvec{s})$ at locations $\xvec{s}$ given all other locations is difficult to interpret because all other locations simultaneously depend on $Y(\xvec{s})$. Thus, the interpretation of $h(\xvec{s})$ is slightly different compared to time-series ARCH models, and $h(\xvec{s}_i)$ does not coincide with the conditional variance at location $\xvec{s}_i$ given all neighbouring observations. Thus, we refer to $\{ h(\xvec{s}_i) \}$ as the volatility process. For parameter estimation, \cite{Otto18_spARCH} proposed a quasi-maximum-likelihood estimator, which is computationally implemented in the \texttt{R}-package \texttt{spGARCH} \citep{Otto19_RJournal}.

Moreover, higher-order spatial dependence can be considered by
\begin{equation}
    h(\xvec{s}_i)  =  \alpha_0 + \sum_{k = 1}^{p} \sum_{j = 1}^{n} \alpha_k w_{k,ij} Y(\xvec{s}_j)^2 \, ,
\end{equation}
where $\{\xmat{W}_k = (w_{k,ij})_{i,j = 1, \ldots, n} : k = 1, \ldots, p \}$ is a set of suitable weight matrices, e.g., separating the influence for different directions (i.e., $k = 1$ corresponds to the northward direction, $k = 2$ to the eastward direction, and so on). In contrast to time-series models, higher-order spatial dependence is often directly included in the first spatial lag by choosing $w_{ij}$ to be positive also for larger lag-orders or distances between $\xvec{s}_i$ and $\xvec{s}_j$. For instance, typical choices of $w_{ij}$ are inverse-distance-based weights, i.e., $w_{ij} = d(\xvec{s}_i, \xvec{s}_j)^{-k}$ where $k$ controls weight decay across space and $d(\xvec{s}_i, \xvec{s}_j)$ is a suitable metric to measure the distance between $\xvec{s}_i$ and $\xvec{s}_j$. 

Alternatively, a spatial autoregressive term of the volatilities can also be included. In this way, a spatial GARCH model of order $(1,1)$ can be defined as 
\begin{equation}
    h(\xvec{s}_i)  =  \alpha_0 + \alpha_1 \sum_{j = 1}^{n} w_{1,ij} Y(\xvec{s}_j)^2 + \beta_1 \sum_{j = 1}^{n} w_{2,ij} h(\xvec{s}_j) ,
\end{equation}
which was introduced by \cite{otto2022general}. This model was first discussed within a unified framework proposed by \citet[][arXiv]{OttoSchmid19_arxiv_unified}.
By allowing for higher-order spatial lag terms, we obtain the following volatility equation of spatial GARCH models
\begin{equation}
    h(\xvec{s}_i)  =  \alpha_0 + \sum_{k = 1}^{p} \sum_{j = 1}^{n} \alpha_k w_{k,ij} Y(\xvec{s}_j)^2 + \sum_{s = 1}^{q} \sum_{j = 1}^{n} \beta_s w_{s,ij} h(\xvec{s}_j) \, .
\end{equation}

In contrast to time-series models that often employ the natural one-way ordering (i.e., past observations can only influence future observations, but not vice versa), spatial models must allow for two-sided influence. Thus, there is typically no causal order between the observations and further assumptions are needed for the existence of a real-valued process. Furthermore, if the spatial locations $\xvec{s}$ are one-dimensional (i.e., $d = 1$), the spatial ARCH models coincide with the time-series ARCH model by \cite{Engle82}, where $\xmat{W}$ acts like a backward-shift operator. More precisely, if the locations are ordered and equidistantly spaced, $\xmat{W}$  would be a sparse matrix with ones on the first subdiagonal for an ARCH(1) process. Similarly, an ARCH($p$) process can obtained if $\xmat{W}_k$ has ones on the $k$-th subdiagonal for $k = 1, \ldots, p$. Note that such models with a triangular weight matrix lead to directional spatial processes \citep[e.g.,][]{merk2021directional}. In these cases, a causal ordering of the locations exists.

% However, in spatial statistics, the higher-order spatial lags are typically included in one single weight matrix unless one is interested in separating different spatial spillover effects, e.g., asymmetric spillovers into different directions. It is worth noting that some widely applied weighting schemes, such as inverse-distance weights, already include all other locations. 

Below, we again focus on the special case of spatial ARCH(1) and GARCH(1,1) processes with two \blue{possibly different} weight matrices $\xmat{W}_1$ and $\xmat{W}_2$ for the ARCH and GARCH term, respectively. Like for time-series GARCH models, also spatial GARCH models require additional assumptions on the parameters and/or weight matrix to ensure the non-negativity of $h(\xvec{s}_i)$ for all $i = 1, \ldots, n$. To analyse this in more detail, let $\xvec{h} = (h(\xvec{s}_i))_{i = 1, \ldots, n}$ and $\xvec{Y}^{(2)} = (Y(\xvec{s}_i)^2)_{i = 1, \ldots, n}$ be the $n$-dimensional vectors of all $h(\xvec{s}_i)$ and the squares of $Y(\xvec{s}_i)$, respectively. Then, \eqref{eq:h_spatialGARCH} can be written in a matrix notation as  
\begin{equation}\label{eq:h_spatialGARCH_vector}
    \xvec{h}  =  \alpha_0 \xvec{1}_n + \alpha_1 \xmat{W}_1 \xvec{Y}^{(2)} \, .
\end{equation}
For a spatial GARCH model, the volatility is specified in a matrix notation as follows
\begin{equation}
    \xvec{h}  =  \alpha_0 \xvec{1}_n + \alpha_1 \xmat{W}_1 \xvec{Y}^{(2)} + \beta_1 \xmat{W}_2 \xvec{h} \, ,
\end{equation}
or in the reduced form, i.e.,
\begin{equation}
    \xvec{h}  =  \left(\xmat{I} - \beta_1 \xmat{W}_2\right)^{-1} \left(\alpha_0 \xvec{1}_n + \alpha_1 \xmat{W}_1 \xvec{Y}^{(2)}\right) \, .
\end{equation}
Furthermore, let $\xmat{A} = \text{diag}(\varepsilon(\xvec{s}_1)^2, \ldots, \varepsilon(\xvec{s}_n)^2)\xmat{W}_1$. For the existence of a real-valued process, i.e., $h(\xvec{s}_i) \geq 0$ for all $i = 1, \ldots, n$, we need to ensure that (i) the inverse $\xmat{S}(\beta_1) = \left(\xmat{I} - \beta_1 \xmat{W}_2\right)^{-1}$ exists, and (ii) the inverse of $\tilde{\xmat{S}}(\alpha_1) = (\xmat{I} - \alpha_1 \xmat{A}^2)^{-1}$ exists and all elements are non-negative. The first condition is a typical assumption in spatial econometrics, and several choices of the weights matrix and the corresponding parameter space have been discussed in the literature, e.g., if $\xmat{W}_2$ is a row standardised weights matrix and $\beta_1 \in [0,1]$ the inverse exists. Generally, the spectral radius of $\beta_1 \xmat{W}_2$ must be smaller than one. The second condition is more complicated to check in practice. On the one side, it can be guaranteed by limiting the squared error terms $\{\varepsilon(\xvec{s}_i)^2\}$, e.g. by assuming a truncated error distribution \citep[see also][]{otto2019stochastic}. On the other hand, it can be ensured by considering the characteristics of underlying spatial dynamics implied by $\xmat{W}_1$. In several specific cases, the condition is always fulfilled, e.g., in the case of directional processes in which $\xmat{W}_1$ can be expressed as a strictly triangular matrix \citep[cf.][]{merk2021directional}.

By contrast, in the logarithmic setting, weaker restrictions on the parameter space are needed for the non-negativity of $\xvec{h}$, but they also imply a different volatility structure. \cite{Sato17} introduced a spatial log-ARCH process, for which the  (log-)volatility equation is given by 
\begin{equation}
   \log h(\xvec{s}_i)  =  \alpha_0 + \alpha_1 \sum_{j = 1}^{n} w_{1,ij} \log Y(\xvec{s}_j)^2 \, .
\end{equation}
With $\log \xvec{h} = (\log h(\xvec{s}_1), \ldots, \log h(\xvec{s}_n))'$ and $\ln \xvec{Y}^{(2)} = (\log Y(\xvec{s}_1)^2, \ldots, \log Y(\xvec{s}_n)^2)'$ being the vectors of the element-wise logarithms of $\xvec{h}$ and the squared observation $\xvec{Y}^{(2)}$, the model can be expressed in a matrix notation as
\begin{equation}
   \log \xvec{h}  =  \alpha_0 \xvec{1}_n + \alpha_1 \xmat{W}_1 \log \xvec{Y}^{(2)} \, .
\end{equation}
Using the log-squared transformation \citep[cf.][]{Robinson:2009}, the model can be transformed into a spatial autoregressive model of the log-squared observations. The transformed errors $\{\log \varepsilon(\xvec{s}_i)^2\}$ of this spatial autoregressive representation follow a log $\chi_1^2$ distribution. Thus, they no longer have zero means and are heavily left-skewed. For the existence of the process, the regular conditions of spatial autoregressive models apply, i.e., the inverse $(\xmat{I} - \alpha_1 \xmat{W})^{-1}$ must exist.

As an extension of the spatial log-ARCH model, \cite{su2023statistical} proposed the following expression of the log volatility
\begin{equation}
   \log \xvec{h}  =   \alpha_0\xvec{1}_n + F(\alpha) \log \xvec{Y}^{(2)} \, 
\end{equation}
with $F(\alpha) = \xmat{I}_n - A(\alpha)$ and
\begin{equation}
   A(\alpha)  =  \xmat{I}_n - \sum_{l = 1}^{\infty} a_l(\alpha) \xmat{W}_1^l \, .
\end{equation}
The coefficients $a_l(\alpha)$ are real-valued deterministic functions with $A(0) = \xmat{I}_n$. In this general form, different kinds of spatial dependence structures can be considered, e.g., spatial autoregressive structures with $A(\alpha)  =  \xmat{I}_n - \alpha \xmat{W}_1$ (i.e., the log-ARCH model of \citealt{Sato17}), spatial moving average structures with $A(\alpha)  =  (\xmat{I}_n - \alpha \xmat{W}_1)^{-1}$, or matrix exponential structures with $A(\alpha)  =  e^{\alpha \xmat{W}_1}$. Besides, \cite{su2023statistical} consider exogenous regressors influencing the log volatility. For this reason, the constant term $\alpha_0\xvec{1}_n$ should be replaced by $\xmat{X}\xvec{\delta}$ with an $n \times (k+1)$-dimensional matrix $\xmat{X}$ and $\xvec{\delta} = (\delta_{0}, \ldots \delta_{k})'$ being a vector of linear regression coefficients.

Furthermore, the log-ARCH model can be generalised to a spatial log-GARCH model. For that reason, \cite{Takaki:2021} defined the log-volatility process as follows
\begin{eqnarray}
    \log h(\xvec{s}_i)  & = & \alpha_0 + \alpha_1 \sum_{j = 1}^{n} w_{1,ij} \log Y(\xvec{s}_j)^2 + \beta_1 \sum_{j = 1}^{n} w_{2,ij} \log h(\xvec{s}_j)\, , \qquad \text{or alternatively,}\label{eq:logGARCH} \\
     \log \xvec{h}      & = & \alpha_0 \xvec{1}_n + \alpha_1 \xmat{W}_1 \log \xvec{Y}^{(2)} + \beta_1 \xmat{W}_2 \log \xvec{h}  \, .   % + \xvec{x}(\xvec{s}_i)' \xvec{\beta} \, , 
\end{eqnarray}
Moreover, \cite{Takaki:2021} also allowed for regressive effects on the volatility, which can easily be included by adding $\xvec{x}(\xvec{s}_i)' \xvec{\delta}$ in \eqref{eq:logGARCH}, where $\xvec{x}(\xvec{s}_i) = (x_1(\xvec{s}_i),\hdots,x_k(\xvec{s}_i))^{'}$ is a vector of exogenous variables at location $\xvec{s}_i$.

\begin{figure}
    \centering  
     \subfigure[The $\log\chi^2_1$ and $N(-1.2704, \pi^2/2)$ densities]{
      \label{sd_N}	
      \includegraphics[width=3.1in, height=2.8in]{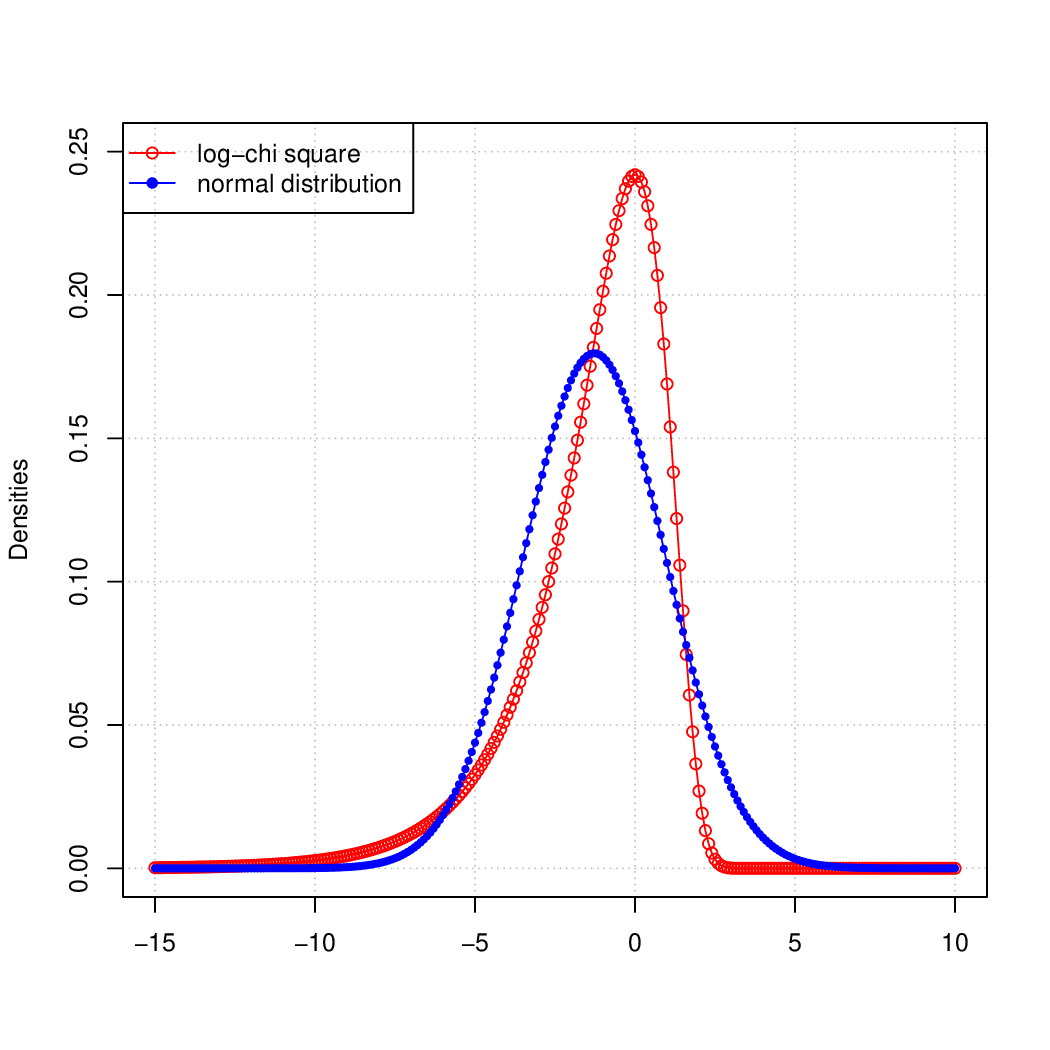}
      }\hspace{1em}
     \subfigure[The $\log\chi^2_1$ and Gaussian mixture densities]
    {\label{sd_G}
     \includegraphics[width=3.1in, height=2.8in]{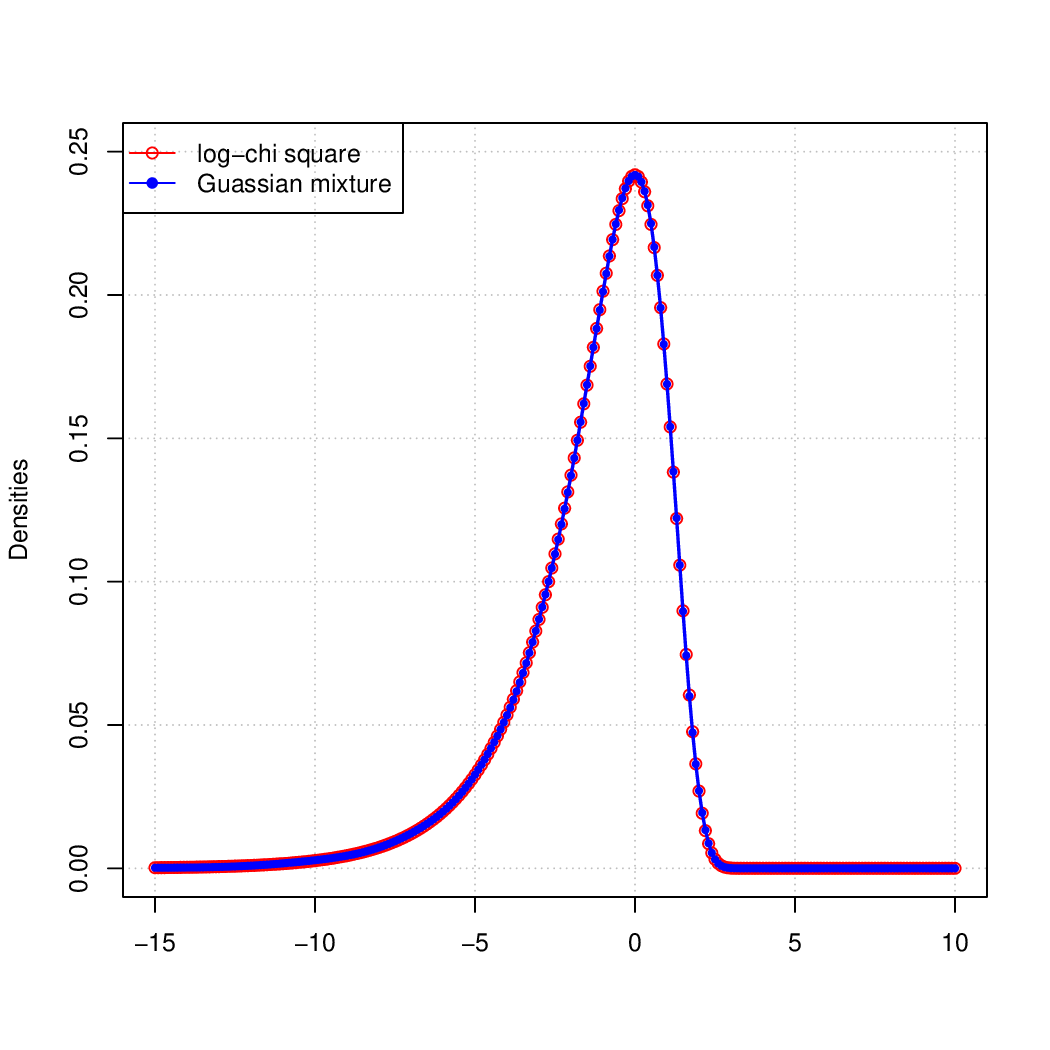}
    }
  \caption{The density of $\e(\xvec{s}_i)^{*}$ and its approximations}
\label{comp}		
\end{figure}

\cite{Dogan2023bayesian} consider a higher-order version of the spatial log-GARCH model suggested in \citet{Takaki:2021}, where the scaling factors in \eqref{Garchmain} follow
\begin{align}\label{eq56}
\log h(\xvec{s}_i) &= \sum_{r=1}^p\sum_{j=1}^n\alpha_{1r} w_{1r,ij}\log Y(\xvec{s}_j)^2 + \sum_{l=1}^q \sum_{j=1}^n\beta_{1l}w_{2l,ij}\log h(\xvec{s}_j) +\xvec{x}(\xvec{s}_i)^{'}\xvec{\delta}
\end{align}
for $i=1,2,\hdots,n$. Here, $p$ and $q$ are two finite positive integers, and $(w_{1r,ij})_{r=1}^p$ and $(w_{2l,ij})_{l=1}^q$ are non-stochastic weights matrices with zero diagonal elements. The corresponding $\{\alpha_{1r}\}_{r=1}^p$ and $\{\beta_{1l}\}_{l=1}^q$ are unknown scalar parameters. For the estimation, \cite{Dogan2023bayesian} transform \eqref{Garchmain} by taking the square of both sides and then taking its natural logarithm (i.e., the log-squared transformation). Then, the transformed outcome equation can be written as
\begin{align}
Y^{*}(\xvec{s}_i) = h^{*}(\xvec{s}_i) + \e^{*}(\xvec{s}_i),
\end{align}
where $Y^*(\xvec{s}_i) = \ln Y(\xvec{s}_i)^2$, $h^{*}(\xvec{s}_i) = \ln h(\xvec{s}_i)$, and $\e^*(\xvec{s}_i) = \ln\e(\xvec{s}_i)^2$. Note that $\e^*(\xvec{s}_i)$ has a $\log\chi^2_1$ distribution with the density
\begin{align} 
f(\e^*(\xvec{s}_i))=\frac{1}{\sqrt{2\pi}}\exp\left(-\frac{1}{2}(e^{\e^*(\xvec{s}_i)}-\e^*(\xvec{s}_i))\right),\quad-\infty<\e^*(\xvec{s}_i)<\infty,
\end{align}
with $\E(\e^*(\xvec{s}_i))\approx-1.2704$ and $\text{Var}(\e^*(\xvec{s}_i))=\pi^2/2\approx4.9348$. This density function is highly skewed with a long left tail, as visualised in Figure~\ref{comp}.

Let $\xvec{Y}^*=(Y^*(\xvec{s}_1),\hdots,Y^*(\xvec{s}_n))^{'}$,  $\xvec{h}^{*}=(h^{*}(\xvec{s}_1),\hdots, h^{*}(\xvec{s}_n))^{'}$ and $\xvec{\e}^*=(\e^*(\xvec{s}_1),\hdots,\e^*(\xvec{s}_n))^{'}$. Then, the higher-order spatial GARCH model in \cite{Dogan2023bayesian} can be written as
\begin{align}
&\y^*=\xvec{h}^{*}+\xvec{\e}^*,\label{space}\\
&\xvec{h}^{*} = \sum_{r=1}^p\alpha_{1r} \xmat{W}_{1r}\y^* + \sum_{l=1}^q\beta_{1l}\xmat{W}_{2l}\xvec{h}^{*} + \xmat{Z}\xvec{\delta},\label{state}
\end{align}
where $\xmat{W}_{1r}=(w_{1r,ij})$ and $\xmat{W}_{2l}=(w_{2l,ij})$ are the $n\times n$ weights matrices and $\xmat{Z}=(\z(\xvec{s}_1),\hdots,\z(\xvec{s}_n))^{'}$ is the $n\times k$ matrix of exogenous variables. Let $\xmat{S}(\bs{\beta}_1)=(\xmat{I}_n - \sum_{l=1}^q\beta_{1l}\xmat{W}_{2l})$, where $\xmat{I}_n$ is the $n\times n$ identity matrix and $\xvec{\beta}_1=(\beta_{11},\hdots,\beta_{1q})^{'}$. Also, let $\xvec{\alpha}_1=(\alpha_{11},\hdots,\alpha_{1p})^{'}$. Under the assumption that $\left\Vert \sum_{l=1}^1\beta_{1l}\xmat{W}_{2l}\right\Vert < 1$, $\xmat{S}(\xvec{\beta}_1)$ is invertible \citep{Horn:2013}. Then, the reduced form of equation \eqref{state} is given by  
\begin{align}
\xvec{h}^{*} &= \xmat{S}^{-1}(\xvec{\beta}_1)\left(\sum_{r=1}^p\alpha_{1r} \xmat{W}_{1r}\y^* + \xmat{Z}\xvec{\delta}\right).
\end{align}
Substituting this equation into \eqref{space} and rearranging yield
\begin{align}
\y^* &= \left(\sum_{r=1}^p\alpha_{1r} \xmat{W}_{1r} +  \sum_{l=1}^1\beta_{1l}\xmat{W}_{2l}\right)\y^* + \xmat{Z}\xvec{\delta} + \xmat{S}(\xvec{\beta}_1)\xvec{\e}^*
\end{align}
Let $\xmat{G}(\bs{\theta})= \left(\xmat{I}_n - \sum_{r=1}^p\alpha_{1r} \xmat{W}_{1r} -  \sum_{l=1}^1\beta_{1l}\xmat{W}_{2l}\right)$ and $\xvec{\theta}= (\xvec{\alpha}_1^{'},\xvec{\beta}_1^{'})^{'}$. Then, under the assumption that $\left\Vert \sum_{r=1}^p\alpha_{1r} \xmat{W}_{1r} +  \sum_{l=1}^q\beta_{1l}\xmat{W}_{2l}\right\Vert<1$, we obtain
\begin{align}\label{reducedYs}
\y^* &= \xmat{G}^{-1}(\xvec{\theta})\xmat{Z}\xvec{\delta} + \xmat{G}^{-1}(\xvec{\theta})\xmat{S}(\xvec{\beta}_1)\xvec{\e}^*.
\end{align}
When the weights matrices are row normalised, $\left\Vert \sum_{r=1}^p\alpha_{1r} \xmat{W}_{1r} +  \sum_{l=1}^q\beta_{1l}\xmat{W}_{2l}\right\Vert<1$ simplifies to $\sum_{r=1}^p|\alpha_{1r}| + \sum_{l=1}^q|\beta_{1l}| < 1$ by the triangle inequality.

\citet{Takaki:2021} propose a Gaussian pseudo maximum likelihood estimator for their spatial log-GARCH model by approximating the distribution of the transformed error terms with a normal distribution. They show that the resulting likelihood estimator attains the standard large sample properties. However, it is well known in the time series literature that the Gaussian pseudo maximum likelihood estimator obtained in this way might have poor finite sample properties because the normal approximation to the distribution of the log-squared error terms provides a poor approximation \citep{JPR:1994, Shephard:1994, Kim:1998, Koopman:1998}. 

\cite{Dogan2023bayesian} instead propose approximating the distribution of $\e^{*}(\xvec{s}_i)$ using a mixture of Gaussian distributions and develop a Bayesian estimation algorithm. They assume that the distribution of $\e^{*}(\xvec{s}_i)$ can be approximated by the following $10$-component Gaussian mixture distribution \citep{Omori:2007}: 
\begin{align}\label{10comp}
f(\e^*(\xvec{s}_i)) \approx \sum_{j=1}^{10} c_j\times \varphi(\e^*(\xvec{s}_i)|\mu_j,\,\sigma^2_j),
\end{align}
where $\varphi(\e^*(\xvec{s}_i)|\mu_j,\,\sigma^2_j)$ denotes the Gaussian density with mean $\mu_j$ and variance $\sigma^2_j$, and $c_j$ is the probability of $j$-th mixture component. The parameters of the $10$-component Gaussian mixture distribution are given in Table~\ref{mixt}. A comparison between the $10$-component Gaussian mixture distribution and the normal distribution in approximating the distribution of $\e^{*}(\xvec{s}_i)$ is illustrated in Figure~\ref{comp}. It is evident that the $10$-component Gaussian mixture distribution provides a very accurate approximation, whereas the normal distribution offers a poor approximation. 
\definecolor{LightCyan}{rgb}{0.88,1,1}
\begin{table}
\begin{center}
\caption{The $10$-component Gaussian mixture approximation for the $\log\chi^2_1$ density} 
\label{mixt} 
\setlength{\tabcolsep}{5pt} 
\renewcommand{\arraystretch}{1} 
\begin{tabular*}{0.65\textwidth}{@{\extracolsep{\fill} }ccrc} 
\hline\hline
Components& $c_j$ &\multicolumn{1}{c}{$\mu_j$}&$\sigma^2_j$ \\
\hline
1 &0.00609 &1.92677& 0.11265 \\ 
2&0.04775& 1.34744& 0.17788\\
3 &0.13057& 0.73504& 0.26768\\
4&0.20674 &0.02266& 0.40611\\
5 &0.22715&-0.85173 &0.62699\\
6&0.18842&-1.97278& 0.98583\\
7&0.12047&-3.46788& 1.57469\\
8 &0.05591&-5.55246& 2.54498\\
9&0.01575 &-8.68384 &4.16591\\
10&0.00115&-14.65000&7.33342\\
\hline\hline
\end{tabular*}
\end{center}
\end{table}

%%%%%%%%%%%%%%%%%%%%%%%%%%%%%%%%%%%%%%%%%%%%%

To complete the model specification, \cite{Dogan2023bayesian} assume the following independent prior distributions: $\alpha_{1r}\sim \text{Uniform}(-1,\,1)$ for $r=1,\hdots,p$, $\beta_{1l}\sim \text{Uniform}\left(-1,\,1\right)$ for $l=1,\hdots,q$, and $\xvec{\delta}\sim N(\xvec{\mu}_{\delta},\xmat{V}_{\delta})$, where $\text{Uniform}(-1,\,1)$ is the uniform distribution over the unit interval. A detailed description of the proposed Gibbs sampler can be found in Algorithm \ref{algo1} in the Appendix.

% ------------------------
\subsection{Spatial stochastic volatility models}\label{sec3.2}

%spatial econometric approach:

In this section, we review extensions of the standard stochastic volatility models in the time series literature to spatial data. Our starting point will be the stochastic volatility model considered in \citet{Robinson:2009} and \cite{tacspinar2021bayesian}, where the log-volatility terms are modelled through a first-order spatial autoregressive process. The resulting spatial stochastic volatility model shares similar properties with the standard stochastic volatility model in time series, and it is designed to capture volatility clustering in spatial data.    

\cite{tacspinar2021bayesian} specify the spatial stochastic volatility (SSV) model as 
\begin{align}
  &  Y(\xvec{s}_i) = e^{\frac{1}{2}h(\xvec{s}_i)} \varepsilon(\xvec{s}_i),\\
  &   h(\xvec{s}_i) - \mu_h  =  \phi \sum_{j = 1}^{n} w_{ij} \left(h(\xvec{s}_j) - \mu_h\right)  + u(\xvec{s}_i),
\end{align}
for $i=1,\hdots, n$, where the first equation is the outcome equation and the second equation is the log-volatility equation (or the state equation). In the outcome equation, $h(\xvec{s}_i)$ denotes the latent log-volatility term, and $\varepsilon(\xvec{s}_i)$ is an i.i.d. normal random variable with mean zero and unit variance. In the log-volatility equation, $\mu_h$ is the constant mean parameter, and $u(\xvec{s}_i)$ is an i.i.d. normal random variable with mean zero and variance $\sigma^2_u$. The $w_{ij}$'s are the non-stochastic spatial weights such that they are zero when $i=j$. These elements represent the degree of spatial association between the log-volatility terms. The scalar parameter $\phi$ is the spatial autoregressive parameter and provides a measure of spatial correlations among $h_i$'s. This model can be considered a spatial extension of the stochastic volatility model in \eqref{s1}-\eqref{s2}. 

Let $\xvec{h}=(h(\xvec{s}_1),\hdots,h(\xvec{s}_n))^{'}$ be the $n\times1$ vector of log-volatilities and $\xvec{u}=(u(\xvec{s}_1),\hdots,u(\xvec{s}_n))^{'}$ be the $n\times1$ vector of error terms. Also, let $\xmat{W} = (w_{ij})_{i,j = 1, \ldots, n}$ be the $n\times n$ non-stochastic matrix for the spatial weights. Then, the state equation can be written in vector form as
\begin{align}
\xvec{h} - \mu_h\xvec{1}_n &= \phi \xmat{W}(\xvec{h} - \mu_h\xvec{1}_n) + \xvec{u},
\end{align}
where $\xvec{1}_n$ is the $n\times1$ vector of ones. Define $\xmat{B}(\phi)=(\xmat{I}_n - \phi \xmat{W})$. Under some restrictions on the parameter space of $\phi$, $\xvec{h} = \mu_h\xvec{1}_n + \xmat{B}^{-1}(\phi)\xvec{u}$ exists, where $\xmat{I}_n$ is the $n\times n$ identity matrix.\footnote{The necessary and sufficient condition for the invertibility of $\xmat{B}(\phi)$ is that the spectral radius of $\phi \mathbf{W}$ must be less than $1$. See \citet{Lee:2004}, \citet{Lesage:2009}, \citet{KP:2010}, and \citet{Elhorst:2014} for a discussion on the parameter space of spatial parameters.}  

The conditional variance of $Y(\xvec{s}_i)$ varies over the relevant space, because $\text{Var}\left(Y(\xvec{s}_i)|h(\xvec{s}_i)\right)=e^{h(\xvec{s}_i)}$.
Furthermore, $ \E(Y(\xvec{s}_i)Y(\xvec{s}_j))=0$ for all $i\ne j$ implying that $Y(\xvec{s}_i)$'s are not spatially correlated. Let $\xvec{k}_i(\phi)$ denote the $i$th row vector of $\xmat{B}(\phi)$ placed in a column vector, and let $r\in\mathbb{N}$ be an even number. Then, it follows that 
\begin{align}
\E(Y(\xvec{s}_i)^r) &=\E\left(e^{\frac{1}{2}h(\xvec{s}_i) r}\right)\E\left(\varepsilon(\xvec{s}_i)^r\right)=e^{\frac{r \mu_h }{2}+\frac{r^2\sigma^2_u}{8}\Vert \xvec{k}_i(\phi)\Vert^2} \mu_{r},
\end{align}
where $\mu_{r}=\frac{r!}{2^{r/2}\times\left(r/2\right)!}$. Therefore, $\E(Y(\xvec{s}_i)^4)/\left[\E(Y(\xvec{s}_i)^2)\right]^2 - 3 = 3\left(e^{\sigma^2_u\Vert \xvec{k}_i(\phi)\Vert^2} - 1\right)>0$. Hence, $Y(\xvec{s}_i)$ has a leptokurtic symmetric distribution. Moreover, 
\begin{align}
\text{Cov}(Y(\xvec{s}_i)^r,\,Y(\xvec{s}_j)^r) &= \mu_{r}^2\left[e^{r\mu_h  + \frac{r^2\sigma^2_u}{8}\left(\Vert\xvec{k}_i(\phi)\Vert^2 + \Vert \xvec{k}_j(\phi)\Vert^2\right)}\left(e^{\frac{r^2\sigma^2_u }{8}2\xvec{k}_i^{'}(\phi)\xvec{k}_j(\phi)}-1\right)\right].
\end{align}
This covariance is generally not zero unless $\phi=0$. Thus, the higher moments of $Y(\xvec{s}_i)$'s are correlated, implying that $y(\xvec{s}_i)$'s are spatially dependent.

For the estimation, it is more convenient to turn the spatial stochastic volatility model into a linear state-space model, and to this end, both \citet{Robinson:2009} and \citet{tacspinar2021bayesian} transform the outcome equation by taking the square of both sides and then taking its natural logarithm (i.e., the log-squared transformation). Then, the outcome equation can be written as
\begin{align}
Y^{*}(\xvec{s}_i) = h(\xvec{s}_i) + \e^{*}(\xvec{s}_i),
\end{align}
as for spatial GARCH models. \citet{Robinson:2009} approximates the distribution of $\e^{*}(\xvec{s}_i)$ with the normal distribution and proposes a Gaussian pseudo maximum likelihood estimator for the estimation. The pseudo maximum likelihood estimators obtained in this manner may attain the standard large sample properties, but they tend to have poor finite sample properties, as mentioned above. Alternatively, \citet{tacspinar2021bayesian} propose to approximate the distribution of $\e^{*}(\xvec{s}_i)$ using a mixture of Gaussian distributions \blue{as in \eqref{10comp}} so that the resulting estimation system turns into a linear Gaussian state space model. They then use the data augmentation technique to facilitate the Bayesian estimation by treating $\xvec{h}$ as an additional parameter vector. The Bayesian MCMC algorithm is described in detail in Algorithm \ref{algo2} shown in the Appendix.

\section{Spatiotemporal volatility models}\label{sec:spatiotemporal}

In this section, we discuss spatiotemporal volatility specifications, which may allow for instantaneous spatial effects. In the context of these models, $Y_t(\xvec{s})$ is now repeatedly observed for $t = 1, \ldots, T$ and at all locations $\xvec{s} \in D$. In the case of  spatial econometrics models, the set of locations is assumed to be constant over time.  This means that we can observe the outcome variable's realisation in the same space across multiple time periods.  It is important to note that spatiotemporal models are naturally included in the purely spatial models, as time could be considered as one dimension of the points $\xvec{s}$. In such cases, the weight matrix $\xmat{W}$ has to be chosen accordingly so that future values do not influence past observations, as we will point out in Section \ref{sec:multiindex}.

Compared to multivariate time series described in Sections~\ref{sMARCH} and \ref{smSV}, spatiotemporal models account for spatial, temporal, and spatiotemporal dependence. Typically, a certain (geographical) structure of the effects is assumed to interpret the parameters in a geographical sense. Moreover, this implied structure makes the models suitable for cases when $n$ is larger than $T$.

\begin{figure}
    \centering
    \includegraphics[width=0.49\textwidth]{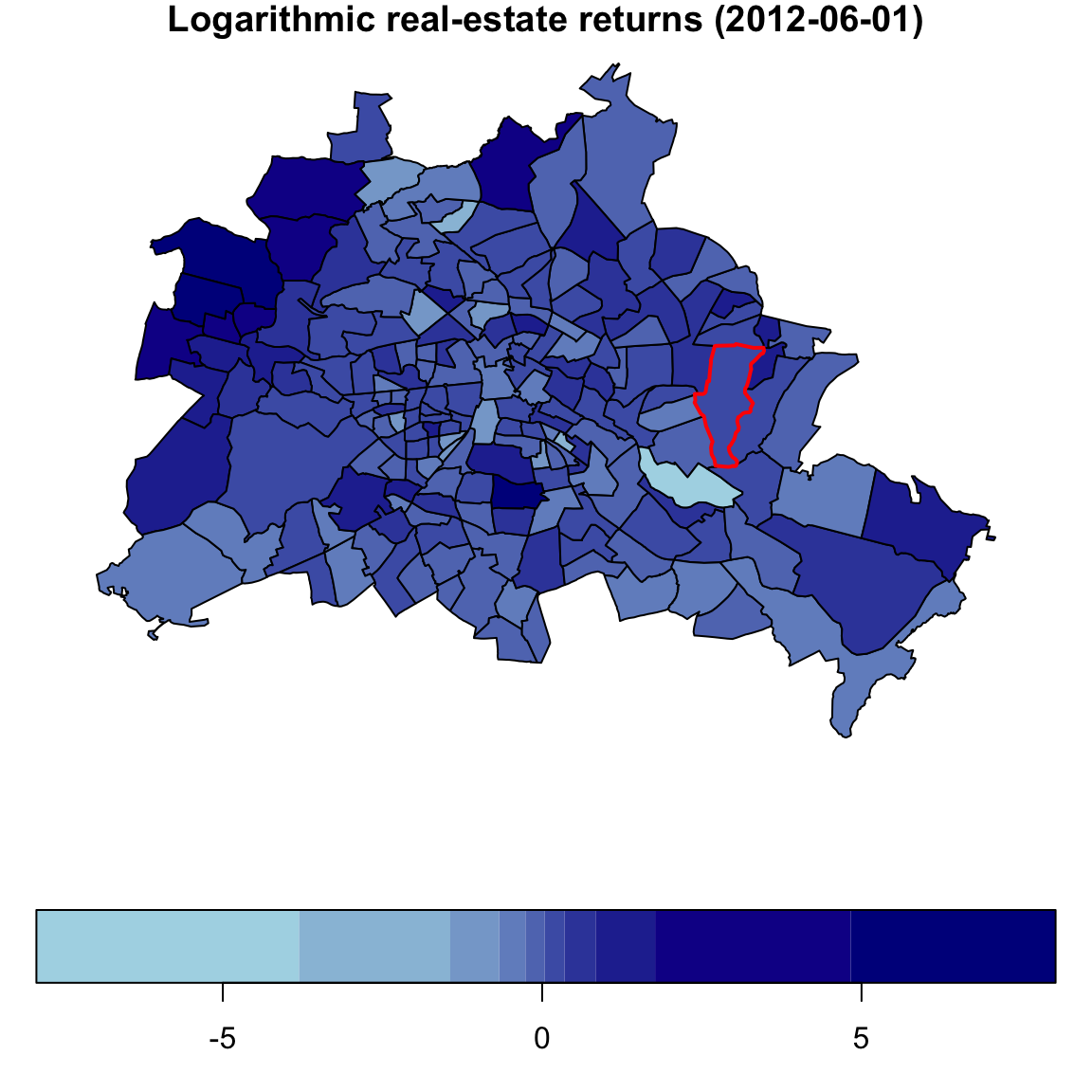}
    \includegraphics[width=0.49\textwidth]{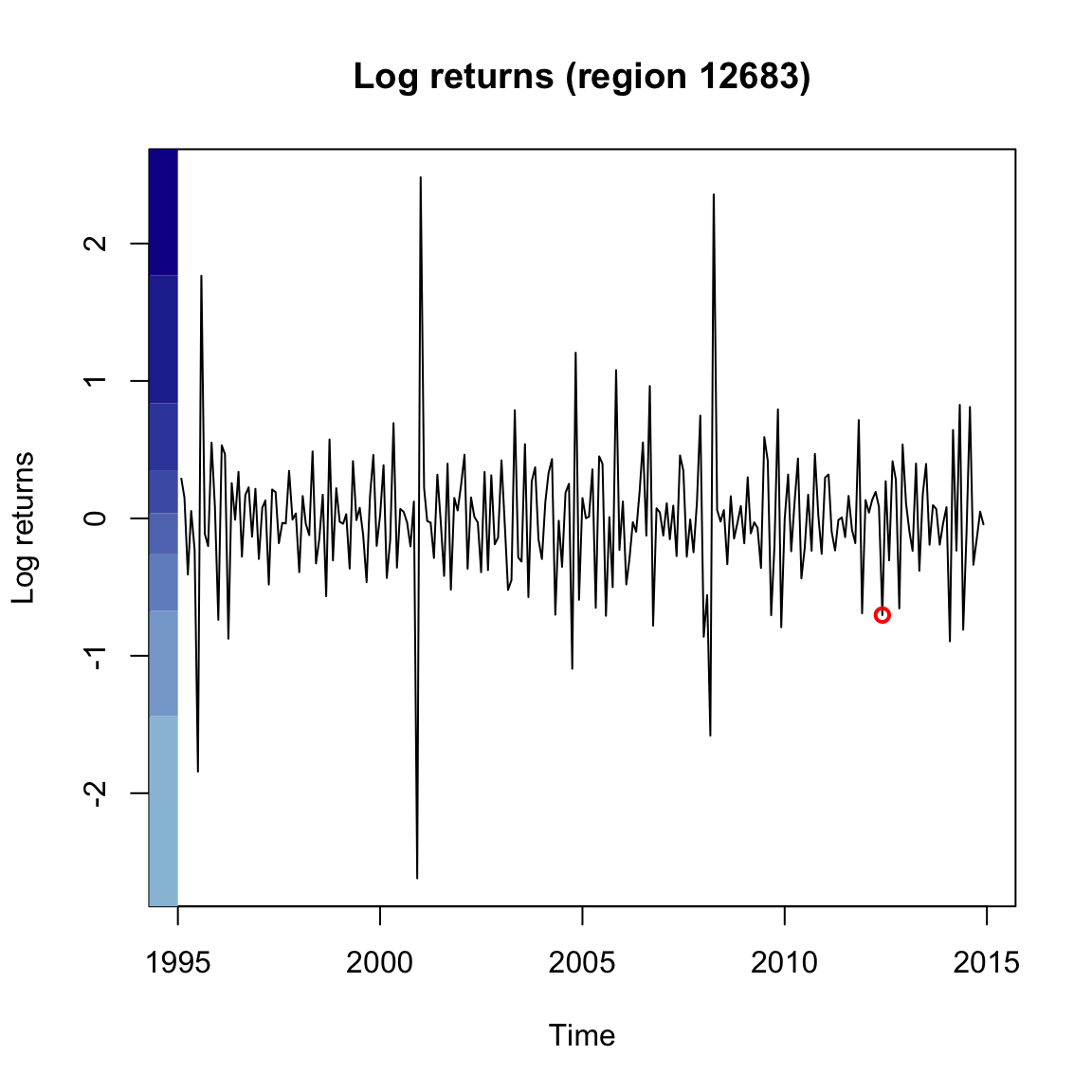}\\
    \includegraphics[width=0.49\textwidth]{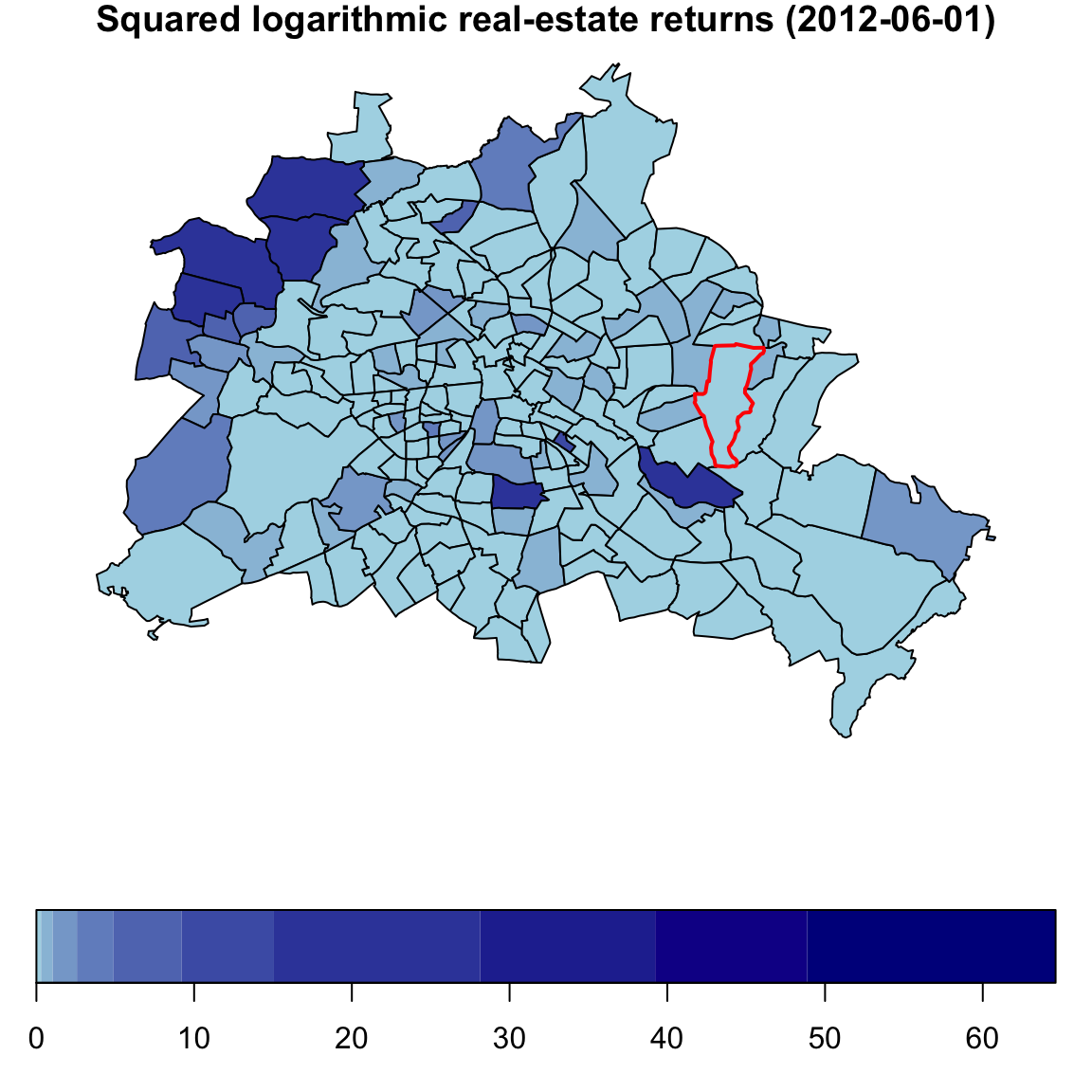}
    \includegraphics[width=0.49\textwidth]{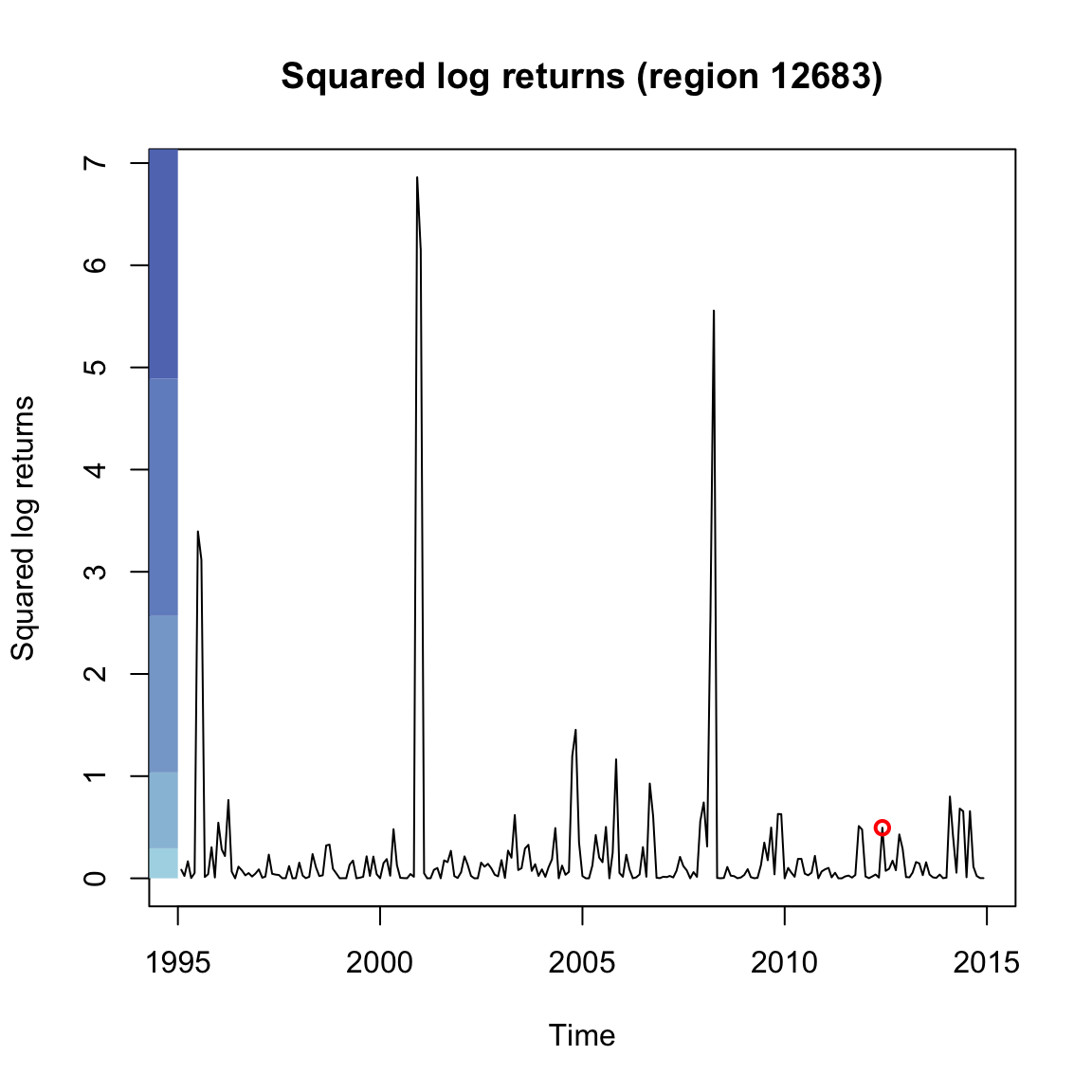}\\
    \caption{Example of a spatiotemporal process of real-estate returns. Top left: Monthly logarithmic returns on 01/06/2012 shown on a map of postcode regions of Berlin. Top right: Time-series plot of the monthly log returns of region 12683, which is highlighted with red borders on the map. The red circle corresponds to the observation shown in the map within the red borders. Bottom row: Squared log returns in the same representation as in the top panels.}
    % Right: Same data are shown in a network representation, where the directed edges point to the three nearest neighbours.}
    \label{fig:spatiotemporal_real_estate}
\end{figure}

As an example of a spatiotemporal process, we can consider the log-returns of the condominium sales in Berlin across time. Figure \ref{fig:spatiotemporal_real_estate} depicts the spatiotemporal process on a map for one selected time point, June 2012 (left panels), and as time series for one selected location, postcode region 12683 (right panels). Moreover, the observed monthly log returns are shown in the top panels, and the squared log returns are shown in the bottom panels. 

\begin{figure}
    \centering
    \includegraphics[width=0.49\textwidth]{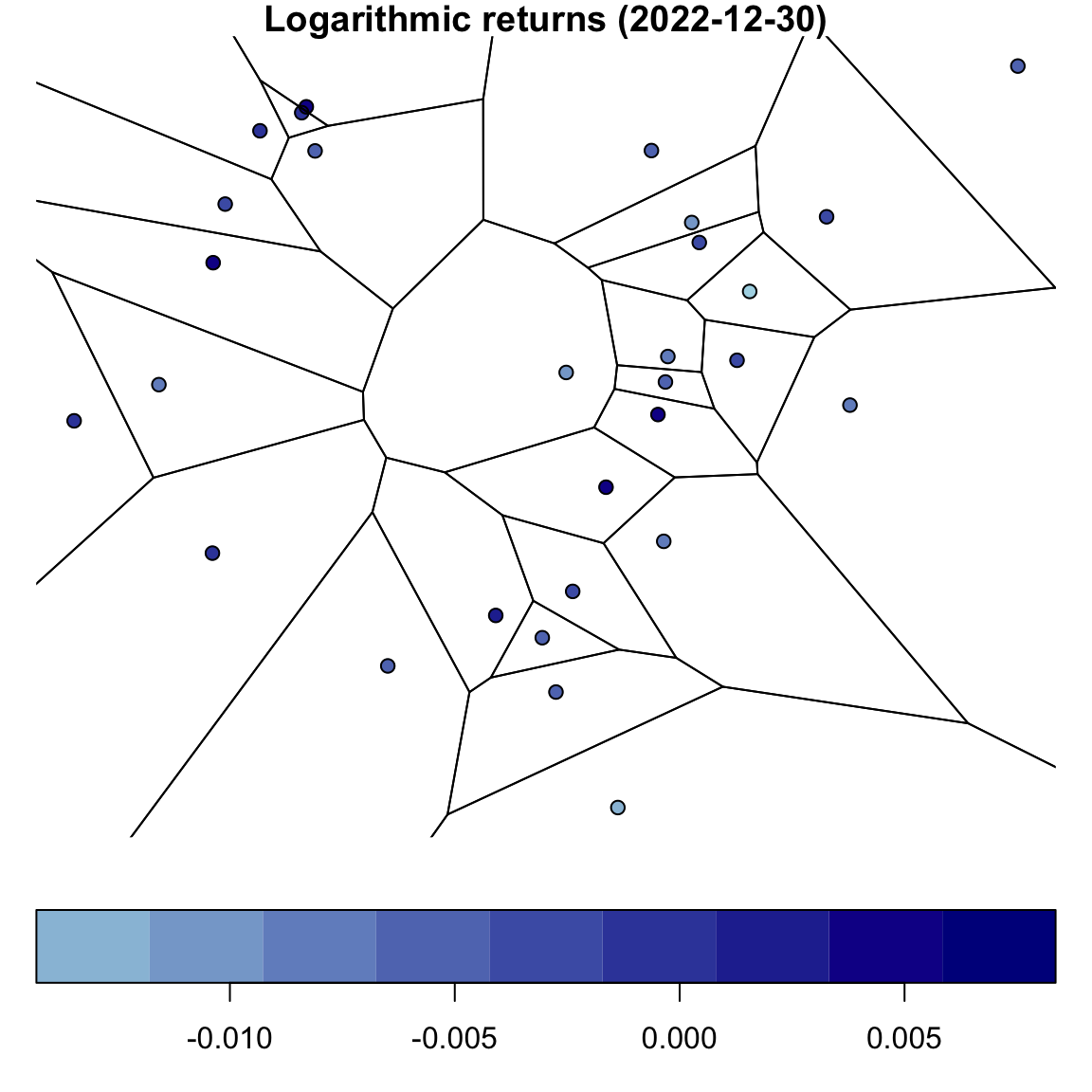}
    \includegraphics[width=0.49\textwidth]{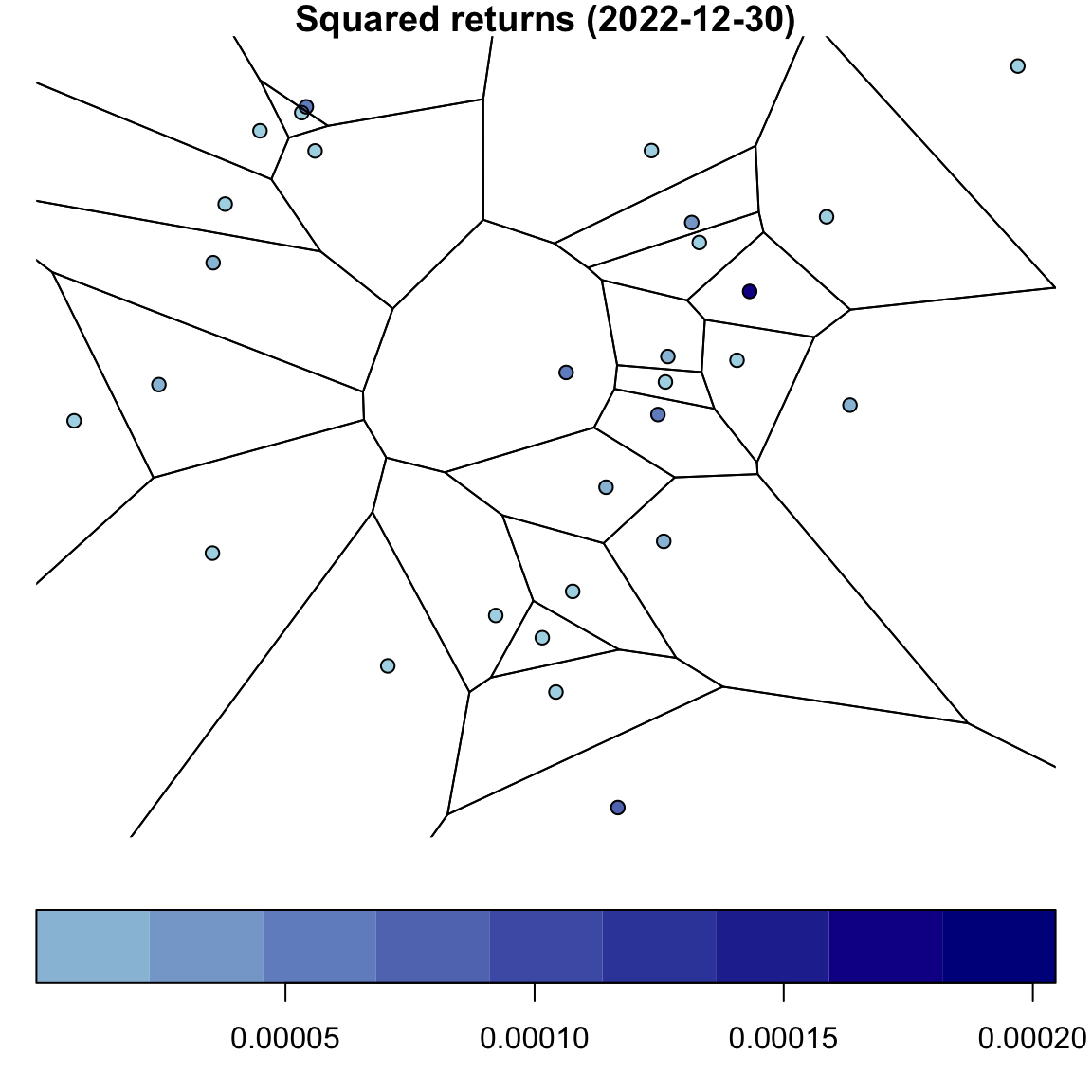}
    \caption{Example of a spatiotemporal process based on financial networks. Left: Daily logarithmic returns on 30/12/2022 of all stocks in the Dow Jones Index. The locations of the points are chosen according to distances between the estimated log-ARCH coefficients using the time series from 2011 to 2023 \citep{piccolo1990distance}. Right: Squared daily logarithmic returns on 30/12/2022 of all stocks in the Dow Jones Index.}
    % Right: Same data are shown in a network representation, where the directed edges point to the three nearest neighbours.}
    \label{fig:spatial_network}
\end{figure}

Another example of a spatiotemporal process from finance is the series of returns of all Dow Jones stocks across time. The similarity between the companies leads to interdependence between the series. For instance, \cite{fulle2022spatial} showed that the similarity of firms regarding their balance sheet data can be used to model interactions in the log-volatilities across financial networks. To represent the closeness of the stocks, we have plotted the returns in an artificial space reflecting the distances in the volatility behaviour according to \cite{piccolo1990distance}, so-called Piccolo distances. Figure \ref{fig:spatial_network} shows the log returns and squared returns for one selected time point, 30 December 2022, in a spatial representation on a map with Voronoi cells separating the stocks.

\subsection{Multi-index representation}\label{sec:multiindex}

In general, spatiotemporal models are already included in the spatial models when time is treated as one dimension of the locations $\xvec{s}$, and the spatial weight matrices are chosen appropriately. All the above-mentioned models assume an arbitrary (finite) dimension of the underlying spatial domain. For instance, consider a spatiotemporal process on the surface of the Earth with degrees latitude and longitude, $(\text{lat}, \text{long})'$. Then, the index of the $i$-th observation, $\xvec{s}_i$, would be composed of the spatial coordinate $(\text{lat}_i, \text{long}_i)'$ and the time point $t_i$ of the $i$-th observation, i.e., $\xvec{s}_i = (t_i, \text{lat}_i, \text{long}_i)'$. The key difference between the time index and the spatial coordinates is that future observation cannot influence past observations. That is, the weight matrix $\xmat{W}$ must account for the causal ordering in time, and $w_{ij}$ must be equal to zero if $t_i > t_j$. For example, let 
\begin{equation*}
    \xvec{Y} = (Y(1,\text{lat}_1, \text{long}_1), \ldots, Y(1,\text{lat}_n, \text{long}_n), Y(2,\text{lat}_1, \text{long}_1), \ldots, Y(T,\text{lat}_n, \text{long}_n))'
\end{equation*}
be an $nT$ dimensional vector of a spatiotemporal random process at $n$ locations and $T$ time points. With two spatial weight matrices
\begin{equation*}
    \xmat{W}_{\text{time}} =  \xmat{L}_1  \otimes \xmat{I}_n  \qquad \text{and} \qquad \xmat{W}_{\text{space}} = \xmat{I}_T \otimes \xmat{W} \, ,
\end{equation*}
where $\xmat{L}_1$ is a $T$-dimensional shift matrix (i.e., first subdiagonal equalling one), $\otimes$ is the Kronecker product, and $\xmat{W}$ is a regular $n$-dimensional spatial weights matrix (e.g., contiguity matrix), we obtain a spatiotemporal GARCH model with a first-order temporal lag implied by $\xmat{W}_{\text{time}}$ and a first-order spatial lag with a constant weight matrix $\xmat{W}$ directly from spatial GARCH models as proposed by \cite{otto2022general}. Similarly, the log-GARCH models of \cite{Takaki:2021} can be constructed for spatiotemporal data. In such a way, spatiotemporal processes can be modelled using spatial volatility models, and all theoretical results can be directly applied in the spatiotemporal setup.

Notice that the above-defined matrices are sparse. Current computational algorithms for sparse matrices are highly efficient from a time and memory perspective, as they only store the indices and values of the non-zero entries in the weight matrices. Thus, spatiotemporal models can often be estimated in this multi-index representation. However, when not using sparse-element objects and operations, the computational requirements can quickly explode with an increasing dimension $n$ and $T$. Thus, it is generally meaningful if the dimension of the spatial weight matrices is as small as possible, and we will return back to the representation with an index $t$ below, i.e., $Y_t(\xvec{s})$ is observed for $t = 1, \ldots, T$ at all locations $\xvec{s} \in D$.

\subsection{Spatiotemporal ARCH and GARCH models}\label{sec:st_arch}

The total number of coefficients in multivariate ARCH and GARCH models can increase faster than the cross-sectional dimension of these models. For example, in the BEKK specification considered by \cite{engle1995multivariate}, the number of parameters has an order of $O(n^2)$, where $n$ is the cross-sectional dimension of the model. Hence, multivariate models are often not applicable in realistic spatiotemporal settings. \cite{caporin2015proximity} consider parsimonious structured specifications using spatial econometrics tools. Consider the following BEKK specification:
\begin{align}
    &\Y_t=\xmat{\Sigma}_t\xvec{\e}_t,\\
    &\xmat{\Sigma}_t=\xmat{C}+\xmat{A}\Y_{t-1}\Y^{'}_{t-1}\xmat{A}^{'}+\xmat{B}\xmat{\Sigma}_{t-1}\xmat{B}^{'},
\end{align}
where $\Y_t=(Y_t(\xvec{s}_1,\hdots,\Y_t(\xvec{s}_n))^{'}$ is the $n\times1$ vector of the outcome variable, $\xmat{\Sigma}_t$ is the $n\times n$ matrix of covariances, $\xvec{\e}_t$ is the $n\times1$ vector of i.i.d. random variables that have $0$ mean and unit variance, and $\xmat{A}$, $\xmat{B}$ and $\xmat{C}$ are  $n\times n$ matrices of unrestricted parameters. \cite{caporin2015proximity} re-parametrise this model such that the number of parameters has an order of $O(n)$ by setting:
\begin{align*}
    &\xmat{C}=\xmat{S}^{-1}\xmat{V}\xmat{S}^{-1'},\quad \xmat{S}=\xmat{I}_n-\xmat{S}_1\xmat{W},\quad \xmat{S}_1=\Diag(\xmat{s}^{(1)}),\quad \xmat{V}=\Diag(\xvec{v}),\\
    &\xmat{A}=\xmat{A}_0+\xmat{A}_1\xmat{W},\quad \xmat{A}_j=\Diag(\xvec{\alpha}^{(j)}),\quad \xmat{B}=\xmat{B}_0+\xmat{B}_1\xmat{W},\quad \xmat{B}_j=\Diag(\xvec{\beta}^{(j)}),
\end{align*}
for $j=0,1$, where $\xmat{W}$ is an $n\times n$ spatial weights matrix, and $\xvec{s}^{(1)}$, $\xvec{v}$, $\xvec{\alpha}^{(j)}$ and $\xvec{\beta}^{(j)}$ are $n\times1$ vectors of unknown parameters. Note that a homogeneous parameters version can be obtained by assuming that the parameter vectors do not vary over the cross-sectional dimension. In this specification, the spillover effects arising through  $\xmat{A}\Y_{t-1}\Y^{'}_{t-1}\xmat{A}^{'}$ and $\xmat{B}\xmat{\Sigma}_{t-1}\xmat{B}^{'}$ depend on the specification adopted for $\W$. \cite{caporin2015proximity} discuss alternative specifications for $\W$ and consider a quasi-likelihood estimation approach for the model. \cite{Monica:2021} suggest an extended version of this model by assuming that $\W$ has a time-varying structure. However, notice that these models do not allow for instantaneous spatial spillovers in a GARCH sense. All spatial interactions enter the model at the first temporal lag, i.e., it must always take one time period for information to spill over to neighbouring locations.

\cite{otto2022dynamic} consider a spatiotemporal ARCH model with a logarithmic representation for the volatility equation. This model takes the following form:
\begin{align}
&Y_{t}(\xvec{s}_i)=h_{t}^{1/2}(\xvec{s}_i)\e_{t}(\xvec{s}_i),\label{2.1}\\
&\log h_{t}(\xvec{s}_i)=\sum_{l=1}^p\sum_{j=1}^n\rho_{l0}m_{l,ij}\log Y^2_{t}(\xvec{s}_j)+\gamma_0\log Y^2_{t-1}(\xvec{s}_i)+\sum_{l=1}^p\sum_{j=1}^n\delta_{l0}m_{l,ij}\log Y^2_{t-1}(\xvec{s}_i)\label{2.2}\nonumber\\ 
&\quad+\xvec{x}^{'}_{t}(\xvec{s}_i)\xvec{\beta}_0+\mu_{0}(\xvec{s}_i)+\alpha_{t0},
\end{align}
for $i=1,2,\hdots,n$ and $t=1,\hdots T$. Here, $\log h_{t}(\xvec{s}_i)$ is considered as the log volatility term in the region $\xvec{s}_i$ at time $t$, and $\e_{t}(\xvec{s}_i)$'s are i.i.d. random variables that have mean zero and unit variance. In \eqref{2.2},  $\{m_{l,ij}\}_{l=1}^p$, for $i,j=1,\hdots,n$, are the non-stochastic spatial weights, where $p$ is a finite positive integer, and $\{m_{l,ii}\}_{l=1}^p$ are zero for $i=1,\hdots,n$.  The spatial, temporal, and spatiotemporal effects of the log-squared outcome variable on the log-volatility term are measured by the unknown parameters $\gamma_0$, $\{\rho_{l0}\}_{l=1}^p$, and $\{\delta_{l0}\}_{l=1}^p$, respectively. In \eqref{2.2}, $\mf{x}_{t}(\xvec{s}_i)$ is a $k\times1$ vector of exogenous variables with the associated parameter vector $\xvec{\beta}_0$, and the regional and time fixed effects are denoted by $\xvec{\mu}_0=(\mu_{0}(\xvec{s}_1),\hdots,\mu_{0}(\xvec{s}_n))^{'}$ and $\xvec{\alpha}_0=(\alpha_{10},\hdots,\alpha_{T0})^{'}$. Both $\xvec{\mu}_0$ and $\xvec{\alpha}_0$ can be correlated with the exogenous variables in an arbitrary manner. 

\begin{figure}
	\begin{center}
	    \includegraphics[width = 0.45\textwidth]{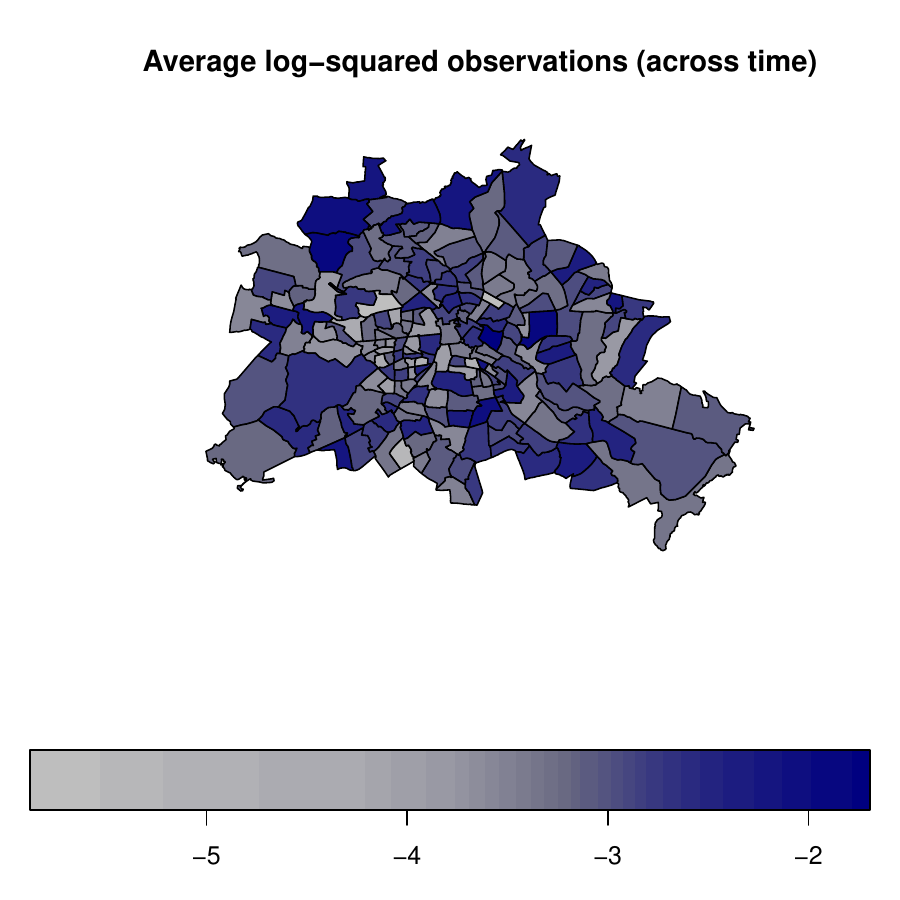}
		\includegraphics[width = 0.45\textwidth]{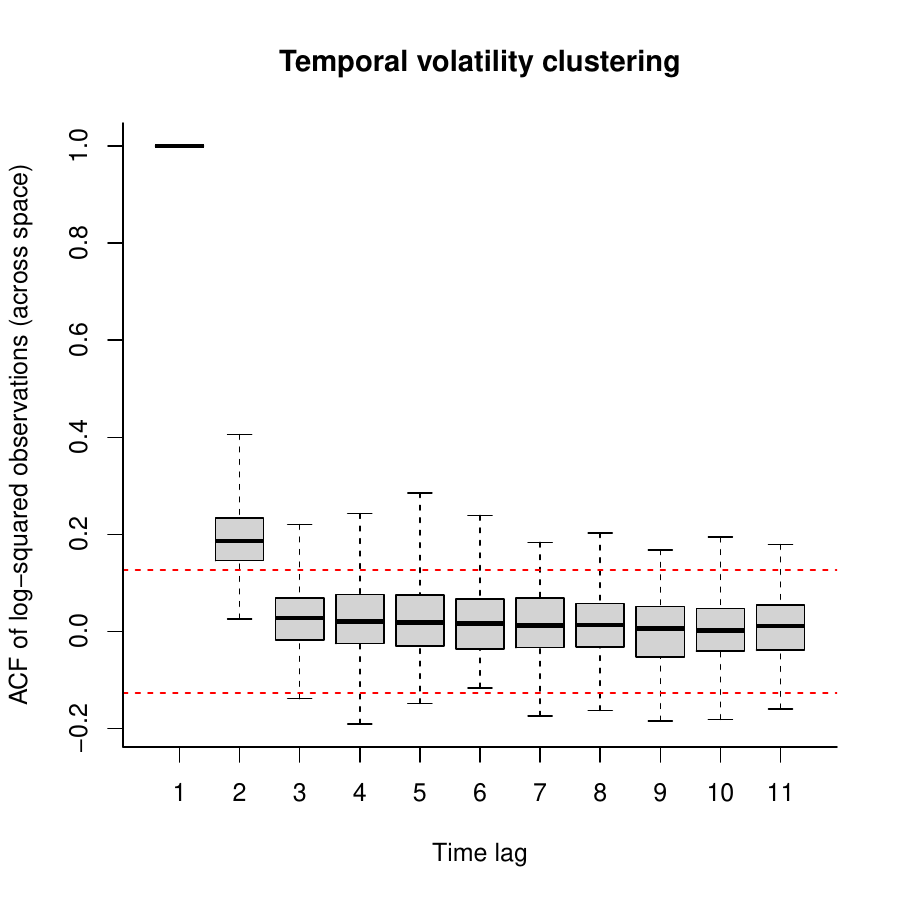}
		\includegraphics[width = 0.45\textwidth]{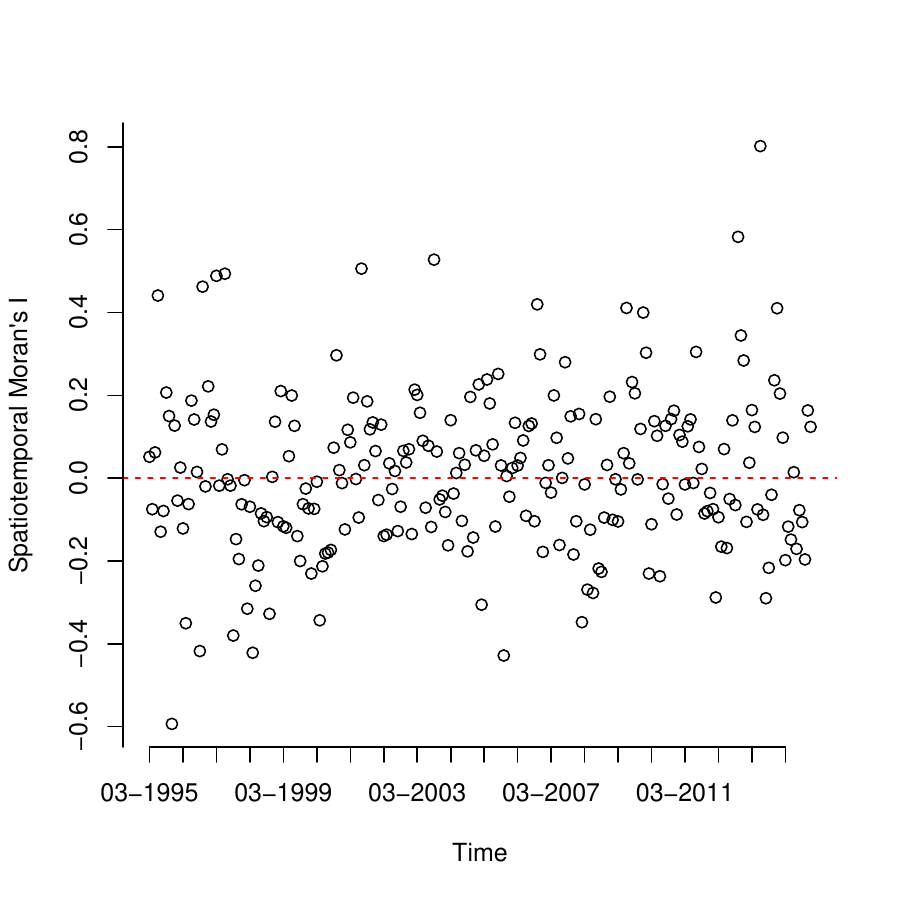}
	\end{center}
	\vspace{-0.80cm}
	\caption{Indication of spatial, temporal, and spatiotemporal correlation. Top left: Average log-squared house price return for each of the 190 postcode areas from February 1995 to December 2015. The map shows a weak clustering effect, especially for the outer regions (indicating a positive spatial dependence in the volatility). Top right: Temporal ACF is depicted as a series of boxplots showing the estimated temporal autocorrelation of all 190 locations. Bottom: Spatiotemporal correlation in terms of the slope of a regression line between the log squared returns and their temporally lagged neighbours (Moran's $I$ of first spatiotemporal lag). There is no clear pattern with varying coefficients around zero, which may indicate a weak spatiotemporal dependence \citep{otto2022dynamic}}\label{fig:acfs}
\end{figure}

To motivate the presence of spatial, temporal, and spatiotemporal effects in the log-volatility equation, \cite{otto2022dynamic} consider a monthly dataset of the real house price returns in Berlin at the postcode level over the period from January 1995 to December 2015. Figure \ref{fig:acfs} displays the average log-squared returns over Berlin's postcodes (the top left figure), the estimated temporal autocorrelation of the log-squared returns as a series of boxplots (the top right figure), and the estimated spatiotemporal autocorrelation in terms of Moran's $I$ across the time horizon (the bottom figure). The first figure shows a clustering pattern in the log-squared returns, indicating the presence of spatial dependence. From the ACF estimates, we can observe apparent temporal volatility clustering, while the spatiotemporal dependence is of a minor degree, irregularly fluctuating around zero.

Applying the log-squared transformation to the outcome equation in \eqref{2.1} yields
\begin{align}\label{2.3}
Y^{*}_{t}(\xvec{s}_i)=h^{*}_{t}(\xvec{s}_i)+\e^{*}_{t}(\xvec{s}_i),
\end{align}
where $Y^*_{t}(\xvec{s}_i)=\log Y^2_{t}(\xvec{s}_i)$, $h^{*}_{t}(\xvec{s}_i)=\log h_{t}(\xvec{s}_i)$ and $\e^*_{t}(\xvec{s}_i)=\log\e^2_{t}(\xvec{s}_i)$. Then, the model can be expressed in vector form as
\begin{align}
&\Y^{*}_t=\h^{*}_t+\bs{\e}^{*}_t,\label{2.4}\\
&\h^{*}_t=\sum_{l=1}^p\rho_{l0}\mf{M}_l\Y^{*}_t+\gamma_0 \Y^{*}_{t-1}+\sum_{l=1}^p\delta_{l0}\mf{M}_l \Y^{*}_{t-1}+\X_t\bs{\beta}_0+\bs{\mu}_0+\alpha_{t0}\mf{1}_n,\label{2.5}
\end{align}
where $\mf{M}_l=(m_{l,ij})$ is the $n\times n$ spatial weights matrices, $\Y^{*}_t=(Y^{*}_{t}(\xvec{s}_1),\hdots,Y^{*}_{t}(\xvec{s}_n))^{'}$, $\h^{*}_t=(h^{*}_{t}(\xvec{s}_1),\hdots,h^{*}_{t}(\xvec{s}_n))^{'}$, $\bs{\e}^{*}_t=(\e^{*}_{t}(\xvec{s}_1),\hdots,\e^{*}_{t}(\xvec{s}_n))^{'}$, $\X_t=(\mf{x}_{t}(\xvec{s}_1),\hdots,\mf{x}_{t}(\xvec{s}_n))^{'}$, and $\mf{1}_n$ is the $n\times1$ vector of ones.  The process in \eqref{2.5} indicates that $\mf{h}^{*}_t$ depends on the higher-order spatial lags of $\Y^{*}_{t}$ and $\Y^{*}_{t-1}$. Substituting \eqref{2.5} into \eqref{2.4}, we obtain
\begin{align}\label{2.7}
\Y^{*}_t=\sum_{l=1}^p\rho_{l0}\mf{M}_l\Y^{*}_t+\gamma_0 \Y^{*}_{t-1}+\sum_{l=1}^p\delta_{l0}\mf{M}_l\Y^{*}_{t-1}+\X_t\bs{\beta}_0+\bs{\mu}_0+\alpha_{t0}\mf{1}_n+\bs{\e}^{*}_t,
\end{align}
for $t=1,\hdots,T$. The elements of  $\bs{\e}^{*}_t$ in \eqref{2.7} are i.i.d across $i$ and $t$ but their mean may not be zero. \cite{otto2022dynamic} suggest adding and subtracting $\E\left(\bs{\e}^{*}_t\right)$ to obtain the following estimation equation:
\begin{align} % \label{3.1}
\Y^{*}_t&=\sum_{l=1}^p\rho_{l0}\mf{M}_l\Y^{*}_t+\gamma_0 \Y^{*}_{t-1}+\sum_{l=1}^p\delta_{l0}\mf{M}_l\Y^{*}_{t-1}+\X_t\bs{\beta}_0+\bs{\mu}_0 +\alpha_{t0}\mf{1}_n+\mu_{\e}\mf{1}_n+\boldsymbol{u}_t,
\end{align}
where $\boldsymbol{u}_t=(u_{t}(\xvec{s}_1),\hdots,u_{t}(\xvec{s}_n))^{'}=\bs{\e}^{*}_t-\E\left(\bs{\e}^{*}_t\right)$, and $\mu_{\e}=\E\left(\e^{*}_{t}(\xvec{s}_i)\right)$. Let $\sigma^2_0=\E(u^2_{t}(\xvec{s}_i))$. Then, it follows that the elements of $\boldsymbol{u}_t$ are i.i.d across $i$ and $t$ with mean zero and variance $\sigma^2_0$. To eliminate $\bs{\mu}_0$ and $\mu_{\e}\mf{1}_n$ from the model, they consider an orthonormal transformations. \cite{otto2022dynamic} suggest a GMM approach based on a set of linear and quadratic moment functions \citep{Yu:2014}. The Monte Carlo simulation results indicate that the suggested GMM estimator has good finite sample properties.  For details, we refer the interested reader to \cite{otto2022dynamic} directly.

When considering the same temporal ARCH parameter across spatial locations, one can see an additional benefit of spatiotemporal ARCH and GARCH models. Whereas their time-series counterparts often require long time series (without structural breaks) to reliably estimate the parameters, spatiotemporal models can also be estimated for shorter time frames because they can borrow information from all locations to estimate the temporal ARCH parameter $\gamma_0$. Simultaneously, spatial interactions between the series are included to account for the cross-sectional dependence.

We conclude this section by considering the circular spatiotemporal GARCH model considered by \cite{holleland2020stationary}. This model is a circular version of the following spatiotemporal GARCH model:
\begin{align}
    &Y_t(\xvec{s})=h_t(\xvec{s})V_t(\xvec{s}),\quad \xvec{s}\in\mathbb{Z}^d,\\
    &h^2_t(\xvec{s})=\mu+\sum_{i=1}^p\sum_{\mf{v}\in\Delta_{1i}}\alpha_i(\mf{v})Y^2_{t-i}(\mf{v}-\xvec{s})+\sum_{i=1}^q\sum_{\mf{v}\in\Delta_{2i}}\beta_i(\mf{v})h^2_{t-i}(\mf{v}-\xvec{s})
\end{align}
where $\{V_t(\xvec{s})\}$ is a sequence of i.i.d. random variables that have zero mean and unit variance, $\Delta_{1i}=\{\mf{v}\in\mathbb{Z}^d:\alpha_i(\mf{v})>0\}$, $\Delta_{2i}=\{\mf{v}\in\mathbb{Z}^d:\beta_i(\mf{v})>0\}$, $\mu$ is an unknown scalar parameter, $\bs{\alpha}=\{\alpha_i(\mf{v}), \mf{v}\in\Delta_{1i}, i=1,\hdots,p\}$ and $\bs{\beta}=\{\beta_i(\mf{v}), \mf{v}\in\Delta_{1i}, i=1,\hdots,p\}$ are unknown parameters. Let $\mathcal{R}=\mathbb{Z}^d/(\mf{m}\mathbb{Z}^d)$ be the quotient group of order $\mf{m}\in\mathbb{Z}^d_{+}$. The circular model considered in \cite{holleland2020stationary} is obtained by replacing $\mathbb{Z}^d$ with $\mathcal{R}$. In this way, $\{Y_t(\xvec{s})\}$ becomes a process indexed on $\mathbb{Z}\times\mathcal{R}$, and the difference $(\mf{v}-\xvec{s})$ and the sum $(\mf{v}+\xvec{s})$ are points in $\mathcal{R}$ with respect to modulus $\mf{m}$. Thus, the circular model takes the following form:
\begin{align}
    &Y_t(\xvec{s})=h_t(\xvec{s})V_t(\xvec{s}),\quad \xvec{s}\in\mathcal{R},\\
    &h^2_t(\xvec{s})=\mu+\sum_{i=1}^p\sum_{\mf{v}\in\Delta_{1i}}\alpha_i(\mf{v})Y^2_{t-i}(\mf{v}-\xvec{s}|\mf{m})+\sum_{i=1}^q\sum_{\mf{v}\in\Delta_{2i}}\beta_i(\mf{v})h^2_{t-i}(\mf{v}-\xvec{s}|\mf{m}).
\end{align}
\cite{holleland2020stationary} present the statistical properties of this circular model and suggest a quasi-maximum likelihood estimator for estimating the parameters. 

\subsection{Multivariate spatiotemporal ARCH models}

When we observe vectors of several features across space and time, we can model such data as a multivariate spatiotemporal process. \cite{otto2022multivariate} extended the spatiotemporal log-GARCH models for multivariate response variables. For that reason, they follow the idea of VEC-GARCH models, where the matrix of volatilities is transformed using the vech-operator \citep{bollerslev1988capital,engle1995multivariate}. More precisely, the multivariate spatiotemporal log-ARCH model is given by
\begin{equation} 
 	\Y_t  =  \xmat{H}_t^{(1/2)} \circ \xmat{\Xi}_t  \, ,
\end{equation}
where $\Y_t$ is an $n \times r$-dimensional matrix of all $r$ features of the response variable at all $n$ locations. Moreover, $\xmat{\Xi}_t$ is an $n \times r$-dimensional matrix of i.i.d random errors with mean zero and unit variance, and $\xmat{H}_t$ is the matrix of log-volatilities, which are defined as follows
\begin{equation} 
    \xmat{H}_t^{(\ln)} = \xmat{A} + \xmat{W}  \, \Y_t^{(\ln,2)} \,  \xmat{\Psi}  +  \Y_{t-1}^{(\ln,2)} \,  \xmat{\Pi} \,. 
\end{equation}
The operations $(\ln)$ and $(\ln, 2)$ should be understood as element-wise operations, i.e., $\xmat{H}^{(\ln)}_t = \left(\ln h_{j,t}(\xvec{s}_i)\right)_{i = 1, \ldots, n, j = 1, \ldots, r}$, and $\Y_t^{(\ln,2)} = \left(\ln Y_{j,t}^2(\xvec{s}_i)\right)_{i = 1, \ldots, n, j = 1, \ldots, r}$ denotes the $n \times r$-dimensional matrix of log-squared observations. The model has three coefficient matrices: (i) the location-specific intercepts $\xmat{A}$ ($n \times r$-dimensional), (ii) own and cross-variable spatial effects in $\xmat{\Psi}$ ($r \times r$-dimensional), and (iii) own and cross-variable temporal effects in $\xmat{\Pi}$ ($r \times r$-dimensional). Using the log-square transformation, the model can be transformed into a multivariate spatiotemporal autoregressive model. \cite{otto2022multivariate} propose a QML estimator based on the normal approximation of the transformed errors. The QML estimator is based on the estimation approach proposed by \cite{yang2017identification} for a multivariate spatial autoregressive model and \cite{Yu08} for the univariate spatiotemporal autoregressive model. Alternatively, the Gaussian mixture approximation could also be considered.

\subsection{Spatiotemporal stochastic volatility models}\label{sec:svm}

We can utilise two approaches to define a spatiotemporal stochastic volatility model. The first approach, known as the geostatistical approach \citep{Cressie:2015}, involves using known parametric covariance functions to model the correlation over time and space. This approach usually includes models obtained by extending a stationary and isotropic Gaussian process (GP), which can be defined in the following way:
\begin{align}\label{eq83}
    Y_t(\xvec{s})=\sigma V_t(\xvec{s}),\quad \xvec{s}\in D, t\in\mathbb{Z},
\end{align}
where $\sigma>0$ is a scalar unknown parameter and $V_t(\xvec{s})$ is a Gaussian process with mean $0$, variance $1$, and stationary covariance function $\text{Cov}\left(V_{t_1}(\xvec{s}_1),V_{t_2}(\xvec{s}_2)\right)=C(\mf{d},\lambda)$, where $\mf{d}=\xvec{s}_1-\xvec{s}_2$ and $\lambda=t_1-t_2$. In this approach, $C(\mf{d},\lambda)$ takes a known parametric function that satisfies certain properties such as stationarity, separability, and full symmetry. \citet{Gneiting:2007} review main covariance functions suggested in the literature. 

\citet{Peter:2011} consider an extension of \eqref{eq83} for temperature series, which takes the form of $Y_t(\xvec{s})=\sigma_t(\xvec{s}) V_t(\xvec{s})$, where $\sigma_t(\xvec{s})$ is modelled such that it can capture the spatially varying seasonality in the variance of $Y_t(\xvec{s})$.  \citet{huang2011class} suggest another extended version taking the following form:
\begin{align}
     Y_t(\xvec{s})=\sigma e^{\alpha h_t(\xvec{s})/2} V_t(\xvec{s}),
\end{align}
where $\sigma>0$ and $\alpha>0$ are scalar unknown parameters, $h_t(\xvec{s})$ is a Gaussian process with mean $0$, variance $1$, and correlation function $\rho_{h}$, and $V_t(\xvec{s})$ is another independent Gaussian process with mean $0$, variance $1$ and covariance function $\rho_V$. \citet{huang2011class} assume that the correlation functions $\rho_h$ and $\rho_V$ are isotropic and stationary in time in the following sense:
\begin{align}
    \text{corr}\left(h_{t_1}(\xvec{s}_1),h_{t_2}(\xvec{s}_2)\right)=\rho_h\left(\Vert\mf{d}\Vert,|\lambda|\right),\quad   \text{corr}\left(V_{t_1}(\xvec{s}_1),V_{t_2}(\xvec{s}_2)\right)=\rho_V\left(\Vert\mf{d}\Vert,|\lambda|\right).
\end{align}
This model can be considered as a spatiotemporal extension of the non-Gaussian geostatistical model proposed by \citet{Steel:2006}. For the estimation of the model, \citet{huang2011class} consider an approximation to the likelihood function of the model and show how an importance sampling approach and Monte Carlo integration can be used to evaluate and maximise the approximation. 

In the second approach, tools from spatial econometrics are used to specify spatial, temporal, and spatiotemporal effects in the log-volatility equation. These effects are incorporated into the model specification through spatial weights matrices that specify the degree of spatial dependence in the outcome variables. \citet{Otto:2022} suggested the following specification:
\begin{align}
&\Y_{t}=\mf{H}^{1/2}_t\mf{V}_{t}, \quad t=1,2,\hdots T, 
\end{align}
where $\Y_t=\left(Y_t(\xvec{s}_1),\hdots,Y_t(\xvec{s}_n)\right)^{'}$ is the $n\times1$ vector of outcome variable at time $t$, $\mf{H}^{1/2}_t=\text{diag}(e^{\frac{1}{2}h_{t}(\xvec{s}_1)},\hdots,e^{\frac{1}{2}h_{t}(\xvec{s}_n)})$ is the $n\times n$ diagonal matrix containing stochastic volatility terms $h_{t}(\xvec{s}_i)$'s, and $\mf{V}_{t}=\left(V_{t}(\xvec{s}_1),\hdots,V_{t}(\xvec{s}_n)\right)^{'}$ is the $n\times 1$ vector of disturbance terms whose elements are i.i.d standard normal random variables.  Let $\h_t=\left(h_{t}(\xvec{s}_1),\hdots,h_{t}(\xvec{s}_n)\right)^{'}$ be the $n\times1$ vector of stochastic volatility terms at time $t$. \citet{Otto:2022} consider the following process for $\h_t$:
\begin{align} % \label{2.2}
&\h_t-\bs{\mu}=\rho_1\W(\h_t-\bs{\mu})+\rho_2(\h_{t-1}-\bs{\mu})+
\rho_{3}\W(\h_{t-1}-\bs{\mu})+\U_t,
\end{align}
where $\bs{\mu}=\left(\mu(\xvec{s}_1),\hdots,\mu(\xvec{s}_n)\right)^{'}$ is the $n\times1$ vector of the time-invariant site-specific effects,  and $\U_{t}=\left(u_{t}(\xvec{s}_1),\hdots,u_{t}(\xvec{s}_n)\right)^{'}$ is the $n\times1$ vector of i.i.d. disturbance terms with $u_{t}(\xvec{s}_i)\sim N(0, \sigma^2)$ for all $i$ and $t$, where $\sigma^2$ is a scalar unknown parameter. The $n\times n$ spatial weights matrix $\W$ specifies the degree of linkages among the elements of $\h_t$. The scalar parameter $\rho_1$ captures contemporaneous spatial correlation, $\rho_2$ measures the temporal effect, i.e., the time dynamic effect, and $\rho_3$ represents the spatiotemporal effect, i.e., the spatial diffusion effect.  The reduced form of the volatility equation can be expressed as 
\begin{align}
&\h_t-\bs{\mu}=\Ss^{-1}(\rho_1)\A(\rho_2,\rho_3)(\h_{t-1}-\bs{\mu})+
\Ss^{-1}(\rho_1)\U_t,\label{eq88}
\end{align}
where $\Ss(\rho_1)=(\I_n-\rho_1\W)$ and $\A(\rho_2,\rho_3)=(\rho_2\I_n+\rho_3\W)$. When the cross-sectional dimension is fixed, the process for the log-volatility is stable if all eigenvalues of $\Ss^{-1}(\rho_1)\A(\rho_2,\rho_3)$ lie inside the unit ball. \citet{Otto:2022} show that an easy-to-check sufficient condition for the stability of the volatility equation is $|\rho_1|+|\rho_2|+|\rho_3|<1$ when $\W$ is row normalised. The statistical properties of this model are described in Section of  Appendix. 

\citet{Otto:2022} introduce a Bayesian estimation approach by assuming the following independent prior distributions: $\rho_1\sim\text{Uniform}(-1,1)$, $\rho_2\sim\text{Uniform}(-1,1)$, $\rho_3\sim\text{Uniform}(-1,1)$, $\bs{\mu}|\mf{b}_{\mu},\B_{\mu}\sim N(\mf{b}_{\mu},\B_{\mu})$, and $\sigma^2|a,b\sim\text{IG}(a,b)$. The estimation approach is based on the log-squared transformation and the finite Gaussian mixture approach described in Section~\ref{sec:spatial} for the SSV model. The log-squared transformation gives 
\begin{align}\label{4.2}
Y^*_{t}(\xvec{s}_i)=h_{t}(\xvec{s}_i)+V^{*}_{t}(\xvec{s}_i),
\end{align}
where $Y^*_{t}(\xvec{s}_i)=\log Y^2_{t}(\xvec{s}_i)$ and $V^*_{t}(\xvec{s}_i)=\log V^2_{t}(\xvec{s}_i)$. 
Define $\Y^*_t=\left(Y^*_{t}(\xvec{s}_1),Y^*_{t}(\xvec{s}_2),\hdots,Y^*_{t}(\xvec{s}_n)\right)^{'}$ and  $\V^*_t=\left(V^*_{t}(\xvec{s}_1),V^*_{t}(\xvec{s}_2),\hdots,V^*_{t}(\xvec{s}_n)\right)^{'}$. Then, in vector form, we have 
\begin{align}\label{4.4}
\Y^*_{t}=\h_{t}+\V^{*}_{t},
\end{align}
In order to convert \eqref{4.4} into a linear Gaussian state-space model, \citet{Otto:2022} approximate $p(V^*_{t}(\xvec{s}_i))$ with an $m$-component Gaussian mixture distribution:
\begin{align}\label{3.1}
f(V^*_{t}(\xvec{s}_i))\approx\sum_{j=1}^mp_j\times \phi(V^*_{t}(\xvec{s}_i)|\mu_j,\,\sigma^2_j),
\end{align}
where $\phi(V^*_{t}(\xvec{s}_i)|\mu_j,\,\sigma^2_j)$ denotes the Gaussian density function with mean $\mu_j$ and variance $\sigma^2_j$,  $p_j$ is the probability of $j$th mixture component and $m$ is the number of components. We can equivalently write \eqref{3.1} in terms of an auxiliary discrete random variable $z_{t}(\xvec{s}_i)\in\{1,2,\hdots,m\}$ that serves as the mixture component indicator:
\begin{align}\label{4.6}
V^*_{t}(\xvec{s}_i)|(z_{t}(\xvec{s}_i)=j)\sim N(\mu_j,\,\sigma^2_j),\quad\text{and}\quad \mathbb{P}(z_{t}(\xvec{s}_i)=j)=p_j,\quad j=1,2,\hdots,m,
\end{align}
where $\mathbb{P}(z_{t}(\xvec{s}_i)=j)=p_j$ is the probability that $z_{t}(\xvec{s}_i)$ takes the $j$th value.  Let $\mathbf{Z}_t=(z_{t}(\xvec{s}_1),\hdots,z_{t}(\xvec{s}_n))^{'}$, $\mathbf{d}_t=(\mu_{z_{t}(\xvec{s}_1)},\hdots,\mu_{z_{t}(\xvec{s}_n)})^{'}$ and $\boldsymbol{\Sigma}_t=\Diag(\sigma^2_{z_{t}(\xvec{s}_1)},\hdots,\sigma^2_{z_{t}(\xvec{s}_n)})$. Then, from \eqref{4.6}, we have $\V^*_t|\mf{Z}_t\sim N(\mathbf{d}_t,\,\boldsymbol{\Sigma}_t)$, which indicates that 
\begin{align}\label{4.7}
&\Y^*_t|\mf{Z}_t,\,\h_t\sim N\left(\h_t+\mathbf{d}_t,\,\boldsymbol{\Sigma}_t\right).
\end{align}
\citet{Otto:2022} suggest the Gibbs sampler described in Algorithm~\ref{a2} in the Appendix for estimating the model parameters.

\section{Further extensions and related models}\label{sec:furthermodels}

In this section, we consider alternative specifications for the spatial and spatiotemporal effects in the volatility equations. In Section \ref{sec5.1}, we consider some extensions of the basic SSV model introduced in Section \ref{sec3.2}.  In both spatial and spatiotemporal volatility models covered in Sections \ref{sec:spatial} and \ref{sec:spatiotemporal}, we usually consider spatial lag terms formulated with $\W$ to introduce spatial and spatiotemporal effects in the volatility equations. In Sections \ref{sec5.2} and \ref{sec5.3}, we consider alternative methods based on the matrix exponential and conditional autoregressive approaches to specify spatial dependence in the volatility equations.

\subsection{Extensions of spatial stochastic volatility models}\label{sec5.1}

% EXTENSIONS of the base SSV

Following the time series literature on the stochastic volatility models described in Section~\ref{suSV}, the SSV model described above can be extended in several directions. To this end, we consider several different spatial stochastic volatility models depending on the spatial specification adopted for the outcome and the log-volatility equations. Let $\x(\xvec{s}_i)=(x_1(\xvec{s}_i),\hdots,x_k(\xvec{s}_i))^{'}$ be the $k\times1$ vector of explanatory variables for $i=1,2,\hdots,n$, $\X=(\x(\xvec{s}_1),\hdots, \x(\xvec{s}_n))^{'}$ be the $n\times k$ matrix of explanatory variables, and $\xmat{W} = (w_{ij})_{i,j = 1, \ldots, n}$ and $\xmat{M} = (m_{ij})_{i,j = 1, \ldots, n}$ be two non-stochastic $n\times n$ spatial weights matrices that have zero diagonal elements. 

The first extension we consider allows for a spatial lag of the outcome variable as well as some exogenous explanatory variables in the outcome equation:
\begin{align}
&Y(\xvec{s}_i)  = \rho\sum_{j=1}^nm_{ij}Y(\xvec{s}_j) + \x(\xvec{s}_i)^{'}\xvec{\beta}+e^{h(\xvec{s}_i)/2}\e(\xvec{s}_i),\\
&h(\xvec{s}_i) - \mu_h = \phi\sum_{j=1}^n w_{ij}(h(\xvec{s}_j) - \mu_h) + u(\xvec{s}_i),
\end{align}
for $i=1,\hdots,n$. The first-order spatial autoregressive process for $Y(\xvec{s}_i)$'s introduces spatial correlations in the outcome variable. As before, $\e(\xvec{s}_i)$'s are i.i.d. standard normal random variables, and $u(\xvec{s}_i)$'s are an i.i.d. normal random variables with mean zero and variance $\sigma^2_u$. The scalar spatial autoregressive parameters are $\rho$ and $\phi$, and $\mu_h$ is the constant mean parameter. We will refer to this model as SAR-SSV.

The spatial autoregressive process allows for the global transmission of a shock. On the other hand, the spatial moving process transmits a shock locally \citep{anselin1988spatial, Fingleton:2008,Fingleton:2008b, Taspinar:2013, Dogan:2015}. An alternative specification to the SSV model is to use a spatial moving average process for the log-volatility terms:
\begin{align}
&Y(\xvec{s}_i)  = e^{h(\xvec{s}_i)/2}\e(\xvec{s}_i),\\
&h(\xvec{s}_i) - \mu_h = \phi\sum_{j=1}^n w_{ij}u(\xvec{s}_j) + u(\xvec{s}_i),
\end{align}
where $\phi$ is the scalar spatial moving average parameter and $u(\xvec{s}_i)$'s are an i.i.d normal random variables with mean zero and variance $\sigma^2_u$. We refer to this model as the SMA-SSV model.  

One can also allow for both a spatial autoregressive process and a spatial moving average process in the log-volatility equation to define the following extension of the SSV model:
\begin{align}
&Y(\xvec{s}_i)  =e^{h(\xvec{s}_i)/2}\e(\xvec{s}_i),\\
&h(\xvec{s}_i) - \mu_h = \phi_1\sum_{j=1}^n w_{1,ij}(h(\xvec{s}_j) - \mu_h) + \phi_2\sum_{j=1}^n w_{2,ij}u(\xvec{s}_j) + u(\xvec{s}_i),
\end{align}
where $\W_1 = (w_{1,ij})_{i,j = 1, \ldots, n}$ and $\W_2 = (w_{2,ij})_{i,j = 1, \ldots, n}$ are two non-stochastic $n\times n$ spatial weights matrices that have zero diagonal elements. We refer to this model as the SARMA-SSV model.  

Another variant can be defined by allowing the volatility feedback in the outcome (observation) equation, which can be considered as an analogous version of the model suggested by \citet{Koopman:2002}. This model is specified as 
\begin{align}
&Y(\xvec{s}_i)  = \alpha e^{h(\xvec{s}_i)}+e^{h(\xvec{s}_i)/2}\e(\xvec{s}_i),\\
&h(\xvec{s}_i) - \mu_h = \phi\sum_{j=1}^n w_{ij}(h(\xvec{s}_j) - \mu_h) + u(\xvec{s}_i),
\end{align}
where the scalar parameter $\alpha$ indicates the effect of the stochastic volatility on $Y(\xvec{s}_i)$. We refer to this model as the SARM-SSV model.  

An analogous version of the leverage effect model described in Section~\ref{suSV} can be defined by introducing correlation in the error terms of the outcome and log-volatility equations. This analogous takes the following form: 
\begin{align}
&Y(\xvec{s}_i)  = e^{h(\xvec{s}_i)/2}\e(\xvec{s}_i),\\
&h(\xvec{s}_i) - \mu_h = \phi\sum_{j=1}^n w_{ij}(h(\xvec{s}_j) - \mu_h) + u(\xvec{s}_i),
\end{align}
where $(\e(\xvec{s}_i),u(\xvec{s}_i))^{'}$ follows the following bivariate normal distribution 
\begin{align}\label{2.11}
\begin{pmatrix}
\e(\xvec{s}_i)\\
u(\xvec{s}_i)
\end{pmatrix}
\sim
N\left(0,\,
\begin{bmatrix}
1&\varrho\sigma_u\\
\varrho \sigma_u&\sigma^2_u
\end{bmatrix}
\right),
\end{align}
and $\varrho$ is the unknown correlation parameter. We refer to this model as the SAR-SSVL model. 

The next extension considers a scale mixture of normal distribution representation for the outcome variable. This is the analogous version described in Section~\ref{suSV} and takes the following form: 
\begin{align}
&Y(\xvec{s}_i)  = e^{h(\xvec{s}_i)/2}\omega(\xvec{s}_i)^{1/2}\e(\xvec{s}_i),\\
&h(\xvec{s}_i) - \mu_h = \phi\sum_{j=1}^n w_{ij}(h(\xvec{s}_j) - \mu_h) + u(\xvec{s}_i),
\end{align}
where the latent scale variables $\omega(\xvec{s}_i)$'s are i.i.d random variables having distribution $IG(\nu/2,\nu/2)$. It is well-known that this representation implies that the marginal distribution of $\omega(\xvec{s}_i)^{1/2}\e(\xvec{s}_i)$ (unconditional on $\omega(\xvec{s}_i)$) is the $t$ distribution with $\nu$ degrees of freedom \citep{Geweke:1993}. This specification can be considered as the spatial extension of the stochastic models considered in \citet{Harvey:1994}, \citet{Ruiz:1994} and \citet{Eric:2004}. We refer to this model as the SAR-SSVt model. 

The Bayesian estimation approach described in Algorithm~\ref{algo2} can also be considered for some of these alternative specifications. The main requirement of the estimation approach is that the model obtained through the log-squared transformation and the Gaussian mixture approximation should be in the form of a linear Gaussian state space model. Therefore, a similar approach can be adopted to estimate the SAR-SV, SMA-SSV, SARMA-SSV, SAR-SSVL, and SAR-SSVt models. However, the approach described for estimating the SSV model may not be extended to the SARM-SSV model.

\subsection{Matrix-exponential dependence structure}\label{sec5.2}

An alternative way to define spatial lag terms is through a matrix exponential term defined as $e^{\alpha \W}=\sum_{j=0}^\infty\frac{\alpha^j}{j!}\W^j$, where $\alpha$ is a scalar spatial parameter. The matrix exponential terms were first considered by \citet{Lesage:2007} to specify spatial dependence in an outcome variable as an alternative to the commonly used spatial autoregressive process. This specification type introduces an exponential decay rate for the cross-sectional dependence and can provide some computational advantages. \citet{su2023statistical} consider a matrix exponential term for spatial log-ARCH models, which is a special case of their logarithmic spatial heteroscedasticity model (log-SHE model). For further information, refer to \cite{Debarsy:2015} and \cite{Ye:2021, Ye:2022, Ye:2023} for the properties of spatial models defined in terms of matrix exponential terms. 

Using matrix exponential terms, we can alternatively define the spatial ARCH and GARCH processes mentioned in Section \ref{sec:spatial_arch_garch}, respectively, in the following way:
\begin{align}
& \h=\alpha_0\mf{1}_n+\left(\mf{I}_n-e^{\alpha_1\W_1}\right)\Y^{(2)},\\
& e^{\beta_1\W_2}\h=\alpha_0\mf{1}_n+\left(\mf{I}_n-e^{\alpha_1\W_1}\right)\Y^{(2)},\label{135}
\end{align}
where $\alpha_1$ and $\beta_1$ are scalar parameters. When $\alpha_1=\beta_1=0$, both processes reduce to $\h=\alpha_0\mf{1}_n$. Since a matrix exponential term is always invertible, the spatial GARCH process in \eqref{135} always has the reduced form given by
\begin{align}
\h=\alpha_0e^{-\beta_1\W_2}\mf{1}_n+e^{-\beta_1\W_2}\left(\mf{I}_n-e^{\alpha_1\W_1}\right)\Y^{(2)},
\end{align}
where $e^{-\beta_1\W_2}$ is the inverse of $e^{\beta_1\W_2}$. Importantly, this reduced form does not invoke any restriction on the parameter space of $\beta_1$, which was not the case for the model considered in Section \ref{sec:spatial_arch_garch}. Similarly, we can also consider the matrix exponential terms to define alternative versions of the spatial stochastic volatility models introduced in Section 3.2. For example, the log-volatility equation in the SSV model can be specified in the following way:
\begin{align}
e^{\phi\W}\left(\h - \mu_h\mf{1}_n\right)&= \boldsymbol{u},
\end{align}
where $\phi$ is the scalar spatial parameter. The reduced form of this process is $\h= \mu_he^{-\phi\W}\mf{l}_n+ e^{-\phi\W}\boldsymbol{u}$, which does not require any restrictions for $\phi$. Similarly, we can also introduce spatial and spatiotemporal effects in the models considered in Sections \ref{sec:st_arch} and \ref{sec:svm}. For example, the matrix exponential version of the spatiotemporal ARCH model considered by \cite{otto2022dynamic} can be formulated as
\begin{align}
&\h^{*}_t=\left(\mf{I}_n-e^{\sum_{l=1}^p\rho_{l0}\mf{M}_l}\right)\Y^{*}_t+\gamma_0 \Y^{*}_{t-1}+\left(\mf{I}_n-e^{\sum_{l=1}^p\delta_{l0}\mf{M}_l}\right)\Y^{*}_{t-1}+\mf{X}_t\bs{\beta}_0+\bs{\mu}_0+\alpha_{t0}\mf{1}_n,
\end{align}
where $e^{\sum_{l=1}^p\rho_{l0}\mf{M}_l}$ and $e^{\sum_{l=1}^p\delta_{l0}\mf{M}_l}$ are higher-order terms formulated by the sequence of weights matrices $\{\mf{M}_l\}$. Similarly, the log-volatility equation in the spatiotemporal model suggested by \citet{Otto:2022} can  alternatively be expressed as 
\begin{align}
&e^{\rho_1\W}\left(\h_t-\bs{\mu}\right)=\rho_2(\h_{t-1}-\bs{\mu})+
\left(\mf{I}_n-e^{\rho_{3}\W}\right)(\h_{t-1}-\bs{\mu})+\U_t,
\end{align}
where $\rho_1$ and $\rho_3$ are spatial and spatiotemporal parameters. 

\subsection{Conditional autoregressive dependence structure}\label{sec5.3}

Another alternative way that can be used to introduce spatial and spatiotemporal dependence in a volatility process is to use the conditional autoregressive (CAR) specification. We start with the following model considered by \citet{Besag:1991}:
\begin{align}
&Y(\xvec{s}_i)=\mu+\phi(\xvec{s}_i)+\e(\xvec{s}_i),\quad \e(\xvec{s}_i)\sim N(0,\sigma^2_{\e}),\label{eq.140}\\
&\phi(\xvec{s}_i)|\{\phi(\xvec{s}_{j}),j\ne i\}\sim N\left(\sum_{j=1}^n\frac{b_{ij}}{\sum_{k=1}^nb_{ik}}\phi(\xvec{s}_j),\frac{\sigma^2_{\phi}}{\sum_{k=1 }^nb_{ik}}\right),\label{eq.141}
\end{align}
where $\mu$ is a scalar unknown overall mean parameter, $\e(\xvec{s}_i)$ is a disturbance term that has $N(0,\sigma^2_{\e})$ distribution with the unknown variance parameter $\sigma^2_{\e}$, $\phi(\xvec{s}_i)$ is a spatially structured random effect term that has the CAR specification in \eqref{eq.141}. In the CAR process, we assume that $b_{ij}$'s are known constants with $b_{ij}=b_{ji}$ and $b_{ii}=0$, and $\sigma^2_{\phi}$ is an unknown scalar variance parameter. The constants $b_{ij}$ can be considered as spatial weights that determine the relationship between regions $\xvec{s}_i$ and $\xvec{s}j$. For example, $b_{ij}$ may be set to 1 if $\xvec{s}_i$ and $\xvec{s}_j$ are neighbours, and $0$ otherwise. In this specification, the conditional variance of $\phi(\xvec{s}_i)$ is spatially varying and depends on  $\sum_{j\ne i}b_{ij}$. \citet{Besag:1974} shows that the joint distribution of $\bs{\phi}=(\phi(\xvec{s}_1),\hdots,\phi(\xvec{s}_n))^{'}$ can be determined as
\begin{align*}
    \bs{\phi}|\sigma^2_{\phi}\sim N\left(\mf{0},\sigma^2_{\phi}(\mf{I}_n-\mf{B})^{-1}\mf{D}_{\phi}\right),
\end{align*}
where $\mf{B}$ is the $n\times n$ matrix with the $(i,j)$th element $B_{ij}=\frac{b_{ij}}{\sum_{k=1}^nb_{ik}}$ and $\mf{D}_{\phi}$ is the $n\times n$ diagonal matrix with the $i$th diagonal element $D_{ii}=\frac{1}{\sum_{k=1 }^nb_{ik}}$. Note that this result requires that $(\mf{I}_n-\mf{B})^{-1}\mf{D}_{\phi}$ is a positive definite matrix. \citet{Yan:2007} extends this model by assuming that $\e(\xvec{s}_i)$ has a stochastic volatility term as specified below:
\begin{align}
    &\e(\xvec{s}_i)|h(\xvec{s}_i)\sim N\left(0,e^{\mu_h+h(\xvec{s}_i)}\right),\\
    &h(\xvec{s}_i)|\{h(\xvec{s}_{j}),j\ne i\}\sim N\left(\sum_{i=1}^n\frac{c_{ij}}{\sum_{k=1}^nc_{ij}}h(\xvec{s}_j),\frac{\sigma^2_{h}}{\sum_{k=1}^nc_{ik}}\right),\label{eq.143}
\end{align}
where $\mu_h$ is a scalar mean parameter and $h(\xvec{s}_i)$ is the log-volatility term assumed to follow the CAR process specified in \eqref{eq.143}. In the CAR process, the weights $c_{ij}$ play the same role as $b_{ij}$ in the CAR process assumed for $\phi(\xvec{s}_i)$, and $\sigma^2_h$ is a scalar variance parameter. As in the case of $\bs{\phi}$, the joint distribution of $\h$ is $\h|\sigma^2_h\sim N\left(\mf{0},\sigma^2_h(\mf{I}_n-\mf{C})^{-1}\mf{D}_{h}\right)$, where we assume that $(\mf{I}_n-\mf{C})^{-1}\mf{D}_{h}$ is a positive definite matrix. In order to achieve identification for $\mu$ and $\mu_h$, \citet{Yan:2007} respectively requires that $\sum_{i=1}^n\phi(\xvec{s}_i) = 0$ and $\sum_{i=1}^nh(\xvec{s}_i) = 0$. The posterior distribution of the model can be expressed as
\begin{align*}
    f(\bs{\theta},\bs{\phi},\h|\Y)\propto f(\Y|\mu,\bs{\phi},\mu_h,\h)\times f(\bs{\phi}|\sigma^2_{\phi})\times f(\h|\sigma^2_h)\times f(\mu)\times f(\sigma^2_{\phi})\times f(\mu_h)\times f(\sigma^2_h)
\end{align*}
where $\bs{\theta}=(\mu,\bs{\phi},\mu_h,\sigma^2_h)^{'}$ and $f(\Y|\mu,\bs{\phi},\mu_h,\h)$ is the likelihood function given by
\begin{align}
    f(\Y|\mu,\bs{\phi},\mu_h,\h)\propto \exp\left(-\frac{1}{2}\sum_{i=1}^n(\mu_h+h(\xvec{s}_i))\right) \exp\left(-\frac{1}{2}\sum_{i=1}^n\frac{(Y(\xvec{s}_i)-\mu-\phi(\xvec{s}_i))^2}{\exp(\mu_h+h(\xvec{s}_i))}\right).
\end{align}
\citet{Yan:2007} assumes flat priors for the elements of $\bs{\theta}$ and shows that the conditional posterior distributions take standard forms, except for that of $\h$. \citet{Yan:2007} defines $\lambda(\xvec{s}_i)=\mu_h+h(\xvec{s}_i)$ and shows that the conditional posterior distribution of $\lambda(\xvec{s}_i)$ can be bounded, up to a scale, by a density of normal distribution. \citet{Yan:2007} suggests using this blanket distribution in an accept-reject algorithm to generate draws for $\lambda(\xvec{s}_i)$. Note that once we have draws for $\lambda(\xvec{s}_i)$, we can determined $\mu_h$ and $h(\xvec{s}_i)$ from the relation $\lambda_i(\xvec{s}_i)=\mu_h+h(\xvec{s}_i)$ via imposing the identification condition $\sum_{i=1}^nh(\xvec{s}_i) = 0$. 

\citet{Lesage:2008} suggested a version of the model in \eqref{eq.140} and \eqref{eq.141} that involves an alternative CAR process for the random effect term $\phi(\xvec{s}_i)$. In vector form, their version can be specified as 
\begin{align}
    &\Y=\mu\mf{1}_n+\bs{\phi}+\bs{\e},\\
    &\bs{\phi}|\sigma^2_{\phi},\rho\sim N\left(\mf{0},\,\sigma^2_{\phi}(\mf{I}_n-\rho\mf{W})^{-1}\mf{M}\right),
\end{align}
where $\bs{\e}=(\e(\xvec{s}_1),\hdots,\e(\xvec{s}_n))^{'}$ is the vector of disturbance terms, $\bs{\phi}=(\phi(\xvec{s}_1),\hdots,\phi(\xvec{s}_n))^{'}$ is the vector of random effects, $\sigma^2_{\phi}$ is a scalar variance parameter and $\rho$ is a scalar spatial parameter. \citet{Lesage:2008} consider this model for the knowledge spillovers arising from patent activity between European regions and specify the elements of the diagonal matrix $\mathbf{M}$ as the output gap and the elements of the symmetric matrix $\mathbf{W}$ as either based on a technological proximity index or based on an index of transport infrastructure. To ensure that $(\mf{I}_n-\rho\mf{W})^{-1}\mf{M}$ is positive definite, we should require that $\rho\in (1/\psi_{min},\,1/\psi_{max})$, where $\psi_{min}$ and $\psi_{max}$ are the minimum and maximum eigenvalues of $\mf{M}^{-1/2}\W\mf{M}^{1/2}$, respectively. In order to allow for outliers, \citet{Lesage:2008} assume that the elements of $\bs{\e}$ have a scale mixture of normal distribution, i.e., 
 $\e(\xvec{s}_i)|\omega(\xvec{s}_i),\sigma^2_{\e}\sim N\left(0,\sigma^2_{\e}\omega(\xvec{s}_i)\right)$. The scale mixture components $\mf{\omega}(\xvec{s}_i)$'s are independent with $\omega(\xvec{s}_i)|\nu\sim IG(\nu/2,\nu/2)$ and $\nu\sim Exp(\lambda_0)$, where $Exp(\lambda_0)$ is the exponential distribution with the rate parameter $\lambda_0$. For the remaining parameters, \citet{Lesage:2008} assume the following independent prior distributions: (i) $\sigma^2_{\phi}\sim IG(a_{\phi},b_{\phi})$, (ii) $\sigma^2_{\e}\sim IG(a_{\e},b_{\e})$, (iii) $\mu\sim N(\mu_0, V_0)$ and (iv) $\rho\sim Beta(a_0,a_0)$, where $Beta(a_0,a_0)$ is the beta distribution defined as 
 \begin{align*}
     f(\rho|a_0)=\frac{1}{B(a_0,a_0)}\frac{(\rho-1/\psi_{min})^{a_0-1}(1/\psi_{max}-\rho)^{a_0-1}}{(1/\psi_{min}-1/\psi_{max})^{2a_0-1}}.
 \end{align*}
\citet{Lesage:2008} show that setting $a_0=1.01$ gives a relatively uninformative prior for $\rho$ over the interval $(1/\psi_{min},\,1/\psi_{max})$. Under these priors, the conditional posterior distributions of $\bs{\phi}$, $\bs{\omega}$, $\mu$, $\sigma^2_{\phi}$ and $\sigma^2_{\e}$ are in standard forms while those of $\rho$ and $\nu$ take unknown forms. In the case of $\rho$, \citet{Lesage:2008} use the univariate numerical integration over the interval $(1/\psi_{min},\,1/\psi_{max})$ to produce the conditional posterior distribution. As for $\nu$, \citet{Lesage:2008} utilised the random walk Metropolis-Hastings algorithm described in \citet{Lesage:2009} to generate posterior draws.
 
%\section{Software for ARCH and GARCH models}

%Should we discuss the computational implementation??

%\cite{rugarch}, \cite{Otto19_RJournal} + spatial econometrics package MATLAB + fgarch R paket

\section{Conclusion and outlook}\label{sec:conclusion}

Spatial and spatiotemporal volatility models constitute a promising new class of models for modelling dependence among the volatility of neighbouring sites in spatial and spatiotemporal data. These models have been recently developed to account for the spatial and spatiotemporal effects in the volatility of an outcome variable. This paper has provided a comprehensive review of the recent literature on spatial and spatiotemporal volatility models. Compared to multivariate time-series GARCH models, which are typically not applicable in spatial settings because of the large number of cross-sectional locations, spatial and spatiotemporal volatility models incorporate a geographical structure to specify dependence in the volatilities. This structure allows modelling spatial and spatiotemporal spillovers across neighbouring locations in a GARCH-like sense. High volatilities may instantaneously spill over to neighbouring regions, thus forming spatial volatility clusters.

Besides motivating different alternative specifications and summarising estimation strategies, we discussed possible extensions and indicated future research directions. Notably, the strand of spatial and spatiotemporal volatility models can be extended in the following ways:
\begin{enumerate}
    \item Asymmetric and anisotropic spatial and spatiotemporal dependence in volatilities:  The study of asymmetric dependence in volatility is crucial for comprehending the dynamics of financial markets and the impact of shocks on volatility. Volatility exhibits a typical asymmetry known as the leverage effect. In time series analysis, exponential GARCH models have proven to be valuable in such scenarios. While exponential GARCH models for spatial data were briefly mentioned in \cite{OttoSchmid19_arxiv_unified}, they have not been thoroughly investigated. Considering housing prices, exploring asymmetric spatial spillovers presents an intriguing strand for future research. For instance, determining whether negative shocks on real-estate prices exert a stronger influence on local prices than positive shocks would be a compelling area to investigate.
    \item Matrix-exponential specification and conditional autoregressive heteroscedasticity models, and further extensions:  The effectiveness of spatial and spatiotemporal models heavily relies on the underlying dependence structure, which is typically unknown in practical applications. Therefore, exploring different dependence structures in future research would be of great interest. For example, matrix-exponential specifications for the spatial dependence in GARCH models and stochastic volatility models present intriguing possibilities, given their computational advantages.
    \item Spatial GARCH models for continuous spatial fields:  Presently, all spatial and spatiotemporal GARCH models have been designed for discrete spatial domains. Consequently, they cannot be directly utilised for predicting volatility at unknown locations, also known as kriging. This field would also relate to continuous-time GARCH models \citep[see, e.g.,][]{kluppelberg2004continuous}. As a notable exception, \cite{huang2011class} introduced a spatial stochastic volatility model within the geostatistical framework. However, in general, considering spatial (autoregressive) dependence in the volatility of a process has received relatively little attention and holds promise as a compelling direction for future research.
    \item Spatiotemporal GARCH model for financial networks:    Spatial weights matrices can be interpreted as adjacency matrices in networks, making spatiotemporal GARCH models akin to GARCH models for nodal attributes on networks. In this context, the spatial interactions describe the dependence across a (non-random) network, with $\xmat{W}$ representing the network structure. Consequently, the connections in $\xmat{W}$ should not be strictly understood in a geographical sense, rendering these models appealing for financial network data. For instance, \cite{mattera2023network} constructed financial networks based on Piccolo distances and utilised spatiotemporal log-ARCH models for volatility forecasting. Given the prevalence of GARCH and stochastic volatility models in finance, the application of spatiotemporal volatility models holds promise as a compelling pathway in finance, particularly when dealing with large financial networks.
\end{enumerate}

In summary, we believe that the class of spatial and spatiotemporal models offers intriguing opportunities for new theoretical developments and practical applications. Bearing in mind that a process' volatility is often interpreted as risk, identifying and predicting local areas of high volatilities -- risks -- is important in various fields, such as finance, economics, or environmental science. 

\section*{Glossary}

\begin{scriptsize}
\begin{itemize}
    \item[ARCH] Autoregressive Conditional Heteroscedasticity
    \item[GARCH] Generalised Autoregressive Conditional Heteroscedasticity\\[.1cm]
    \item[ARMA] Autoregressive Moving Average Process
    \item[BEKK-GARCH] Baba, Engle, Kraft, and Kroner GARCH
    \item[CAR] Conditional Autoregressive Model
    \item[CCC] Constant Conditional Correlations
    \item[DCC] Dynamic Conditional Correlations
    \item[DVEC-GARCH] Diagonal Vector GARCH
    \item[EGARCH] Exponential GARCH
    \item[FIGARCH] Fractionally integrated GARCH
    \item[GJR-GARCH] Glosten, Jaganathan and Runkle GARCH
    \item[GMM] Generalised Method of Moments
    \item[GP] Gaussian Process
    \item[i.i.d.] Independent and Identically Distributed
    \item[log-ARCH] Logarithmic ARCH
    \item[log-GARCH] Logarithmic GARCH
    \item[MCMC] Markov Chain Monte Carlo
    \item[MGARCH] Multivariate GARCH
    \item[NARCH] Non-linear ARCH
    \item[QML] Quasi-Maximum-Likelihood Estimator
    \item[SAR] Spatial/Simultaneous Autoregressive Model
    \item[SARMA] Spatial Autoregressive Moving Average
    \item[SARM-SSV] SAR in Mean Spatial Stochastic Volatility 
    \item[SAR-SSVL] SAR Spatial Stochastic Volatility with Leverage
    \item[SAR-SSVt] SAR Spatial Stochastic Volatility with Student's $t$ errors
    \item[SSV] Spatial Stochastic Volatility
    \item[SMA] Spatial Moving Average
    \item[TGARCH] Threshold GARCH
    \item[VARMA] Vector ARMA
    \item[VEC-GARCH] Vector GARCH 
\end{itemize}
\end{scriptsize}

\newpage
\section{Appendix}

\subsection{Bayesian MCMC algorithm for spatial GARCH models}
In this section, we describe the estimation approach suggested by \cite{Dogan2023bayesian} for the high-order spatial GARCH model in \eqref{eq56}. Note that the $10$-component Gaussian mixture distribution can be represented using an auxiliary random variable $a_i \in \{1,2,\hdots,10\}$ that serves as the mixture component indicator. In other words, $\e^*(\xvec{s}_i)|(a_i = j)\sim N(\mu_j,\,\sigma^2_j)$ and $P(a_i = j) = c_j$ for $j=1,2,\hdots,10$ and $i=1,2,\hdots,n$. 
Let $\xvec{a} = (a_1,\hdots,a_n)^{'}$, $\xvec{d}_a=(\mu_{a_1},\hdots,\mu_{a_n})^{'}$, and $\xmat{\Sigma}_a=\Diag(\sigma^2_{a_1},\hdots,\sigma^2_{a_n})$. Then, we have $\xvec{\e}^*|\xvec{a} \sim N(\xvec{d}_a,\,\xmat{\Sigma}_a)$. Then, from equation \eqref{reducedYs}, we can write $\y^*|\xvec{a},\xvec{\theta},\xvec{\delta} \sim N\left(\xvec{G}^{-1}(\xvec{\theta})\xmat{Z}\xvec{\delta} + \xmat{G}^{-1}(\xvec{\theta})\xmat{S}(\xvec{\beta}_1)\xvec{d}_a,\, \xmat{G}^{-1}(\xvec{\theta})\xmat{S}(\xvec{\beta}_1)\xmat{\Sigma}_a \xmat{S}^{'}(\xvec{\beta}_1)\xmat{G}^{-1'}(\xvec{\theta})\right)$. Furthermore, let us define  $\y^{**} = \xmat{S}^{-1}(\xvec{\beta}_1) \xmat{G}(\xvec{\theta})\y^{*} -  \xmat{S}^{-1}(\xvec{\beta}_1)\xmat{Z}\xvec{\delta}$. Then, it follows that $\y^{**}|\xvec{a} \sim N\left(\xvec{d}_a,\,\xmat{\Sigma}_a\right)$. Then, the joint posterior distribution $f(\xvec{\theta},\xvec{\delta},\xvec{a}|\y^*)$ can be stated as
\begin{align}
f(\xvec{\theta},\xvec{\delta},\xvec{a}|\y^*)\propto f(\y^*|\xvec{\theta},\xvec{\delta},\xvec{a})\times f(\xvec{\theta})\times f(\xvec{\delta})\times f(\xvec{a}),
\end{align}
where $f(\y^*|\xvec{\theta},\xvec{\delta},\xvec{a})$ denotes the conditional likelihood function of the transformed model. Let $\mf{b}^{(g)}$ be the draw generated at the $g$th iteration, where $\mf{b}\in\{\bs{\theta},\bs{\delta},\mf{a}\}$. Then, the following Gibbs sampler describes the steps for estimating the higher-order spatial GARCH model.

% Estimation algorithm
\begin{algorithm}[Estimation of the spatial GARCH model]\label{algo1}
\leavevmode   \normalfont
\begin{enumerate}
\item Sampling step for $\xvec{a}$:
\begin{align}\label{mixcid}
\mathbb{P}(a_i = j|\y^{**}) = \frac{c_j\times\varphi\left(Y^{**}(\xvec{s}_i)| \mu_j,\,\sigma^2_j\right)}{\sum_{k=1}^{10}c_k\times\varphi\left(Y^{**}(\xvec{s}_i)|\mu_k,\,\sigma^2_k\right)}
\end{align}
for $j=1,\hdots,10$ and $i=1,\hdots,n$, where the denominator is the normalizing constant.
\item Sampling step for $\xvec{\delta}$: 
\begin{align}
\xvec{\delta}|\y^{*},\xvec{\theta},\xvec{a} \sim N\left(\widehat{\xvec{\mu}}_\delta,\, \widehat{\xmat{V}}_\delta\right),
\end{align} 
where 
\begin{align*}
&\widehat{\xvec{\mu}}_\delta = \widehat{\xmat{V}}_\delta\left(\xmat{V}^{-1}_\delta\xvec{\mu}_{\delta} + \xmat{Z}^{'}\xmat{S}^{-1'}(\xvec{\beta}_1)\xmat{\Sigma}^{-1}_a\left(\xmat{S}^{-1}(\xvec{\beta}_1)\xmat{G}(\xvec{\theta})\y^{*} - \xmat{d}_a\right)\right)\\
& \widehat{\xmat{V}}_\delta = \left(\xmat{V}^{-1}_\delta +\xmat{Z}^{'}\xmat{S}^{-1'}(\xvec{\beta}_1)\xmat{\Sigma}^{-1}_a\xmat{S}^{-1}(\xvec{\beta}_1)\xmat{Z}\right)^{-1}.
\end{align*}
\item Sampling step for $\xvec{\theta}$: Use an adaptive Metropolis (AM) algorithm to sample $\xvec{\theta}$. At iteration $g$, generate a candidate value $\widetilde{\xvec{\theta}}$ according to $\widetilde{\xvec{\theta}}\sim f_g\left(\xvec{\theta}|\xvec{\theta}^{(0)},\hdots, \xvec{\theta}^{(g-1)}\right)$, where 
\begin{align*}
f_g\left(\xvec{\theta}|\xvec{\theta}^{(0)},\hdots, \xvec{\theta}^{(g-1)}\right)=
\begin{cases}
N\left(\xvec{\theta}^{(g-1)},\,\frac{(0.1)^2}{p+q} \xmat{I}_{p+q}\right),\quad\text{for}\quad g\leq2(p+q),\\
0.95\times N\left(\xvec{\theta}^{(g-1)},\,\frac{2.38^2}{p+q}\text{Cov}\left(\xvec{\theta}^{(0)},\hdots, \xvec{\theta}^{(g-1)}\right)\right)\\
\,\,+0.05\times N\left(\xvec{\theta}^{(g-1)},\,\frac{(0.1)^2}{p+q}\xmat{I}_{p+q}\right),\,\text{for}\, g>2(p+q).
\end{cases}
\end{align*}
Accept with the following probability
\begin{align}
\mathbb{P}(\xvec{\theta}^{(g-1)}, \widetilde{\xvec{\theta}})=\min\left(\frac{f(\y^{*}|\widetilde{\xvec{\theta}},\xvec{\delta}^{(g-1)},\xvec{a}^{(g-1)})}{f(\y^{*}|\xvec{\theta}^{(g-1)},\xvec{\delta}^{(g-1)},\xvec{a}^{(g-1)})},\,1\right).
\end{align}
\end{enumerate}
\end{algorithm}

In Step 1, note that the mixture component indicators are conditionally independent given $\y^{**}$. Thus, each component is a discrete random variable that takes values between $1$ and $10$ with the conditional posterior probability given in \eqref{mixcid}. The AM algorithm in Step 3 uses the historical draws of  $\xvec{\theta}$ to form the covariance matrix of the proposal distribution $f_g\left(\xvec{\theta}|\xvec{\theta}^{(0)},\hdots, \xvec{\theta}^{(g-1)}\right)$ \citep{Haario:2001,Roberts:2009}. 
The candidate value $\widetilde{\xvec{\theta}}$ generated in this step is subject to the stability condition $\left\Vert \sum_{r=1}^p\alpha_{1r} \xmat{W}_{1r} +  \sum_{l=1}^q\beta_{1l}\xmat{W}_{2l}\right\Vert<1$. If the candidate does not meet this condition, a new candidate $\widetilde{\xvec{\theta}}$ is drawn until the stability condition is met.

\subsection{Estimation of the spatial stochastic volatility models}
In this section, we describe the estimation approach suggested by \citet{tacspinar2021bayesian} for the estimation of the SSV model. \citet{tacspinar2021bayesian} assume that the distribution of $\e^{*}(\xvec{s}_i)$ can be approximated by the $10$-component Gaussian mixture distribution stated in \eqref{10comp}. The parameters of the $10$-component Gaussian mixture distribution were provided in Table~\ref{mixt}. Note again that  the $10$-component Gaussian mixture distribution approximates the distribution of $\e^{}(\xvec{s}_i)$ closely, as shown in Figure~\ref{comp}. 

We will again represent the $10$-component Gaussian mixture distribution using an auxiliary random variable $a_i \in \{1,2,\hdots,10\}$ that serves as the mixture component indicator. In other words, $\e^*(\xvec{s}_i)|(a_i = j)\sim N(\mu_j,\,\sigma^2_j)$ and $\mathbb{P}(a_i = j) = c_j$ for $j=1,2,\hdots,10$ and $i=1,2,\hdots,n$. Hence, the SSV model becomes linear and Gaussian given the component indicator variable $\mathbf{a} = (a_1,\hdots,a_n)^{'}$.

\citet{tacspinar2021bayesian} assume the following independent prior distributions for $\sigma^2_u$, $\phi$ and $\mu_h$: $\sigma^2_u\sim IG(a_0,b_0)$, $\lambda\sim \text{Uniform}(-1/\tau,\,1/\tau)$, $\mu_h\sim N(\mu_0,V_{\mu})$, where $IG(a_0,b_0)$ is the inverse gamma distribution with shape parameter $a_0$ and scale parameter $b_0$, $\text{Uniform}(-1/\tau,\,1/\tau)$ is the uniform distribution over the interval $(-1/\tau,\,1/\tau)$, and $\tau$ is the spectral radius of $\xmat{W}$. Let $\y=(Y(\xvec{s}_1),\hdots,Y(\xvec{s}_n))^{'}$ be the $n\times1$ vector of observations on the outcome variable. Let $\y^*$ and $\bs{\e}^*$ denote the $n\times1$ vector of observations on the transformed outcome variable and the transformed error terms, respectively, using the log-squared transformation. Then, the joint posterior distribution $f(\xvec{h},\xvec{a},\sigma^2_u,\mu_h,\phi|\y^*)$ can be stated as
\begin{align}
f(\xvec{h},\xvec{a},\sigma^2_u,\mu_h,\lambda|\y^*)\propto f(\y^*|\xvec{h},\xvec{a})\times f(\xvec{h}|\sigma^2_u,\mu_h,\phi)\times f(\xvec{a})\times f(\sigma^2_u)\times f(\mu_h)\times f(\phi),
\end{align}
where $f(\y^*|\xvec{h},\xvec{a})$ denotes the conditional likelihood function of the transformed SSV. Let $\xvec{d}_a=(\mu_{a_1},\hdots,\mu_{a_n})^{'}$ and $\xmat{\Sigma}_a=\Diag(\sigma^2_{a_1},\hdots,\sigma^2_{a_n})$. Then, we have $\xvec{\e}^*|\xvec{a} \sim N(\xvec{d}_a,\,\xmat{\Sigma}_a)$. Hence, $\y^*|\xvec{a},\,\xvec{h}\sim N\left(\xvec{h}+\xvec{d}_a,\,\xmat{\Sigma}_a\right)$ and
$\xvec{h}|\sigma^2_u,\mu_h,\phi \sim N\left(\mu_h\mf{1}_n, \, \sigma^2_u\xmat{B}^{-1}(\phi)\xmat{B}^{-1'}(\phi)\right)$. Let $\mf{a}^{(g)}$ be the draw generated at the $g$th iteration, where $\mf{a}\in\{\bs{h},\mu_h,\sigma^2_u\}$. Then, \citet{tacspinar2021bayesian} suggested the following Gibbs sampler for the estimation of the SSV model. 
% Estimation algorithm
\begin{algorithm}[Estimation of the SSV model]\label{algo2}
\leavevmode   \normalfont
\begin{enumerate}
\item Sampling step for $\xvec{a}$:
\begin{align}
\mathbb{P}(a_i = j|y^*(\xvec{s}_i), h(\xvec{s}_i)) = \frac{c_j\times\varphi\left(y^*(\xvec{s}_i)|h(\xvec{s}_i) + \mu_j,\,\sigma^2_j\right)}{\sum_{k=1}^{10}c_k\times\varphi\left(y^*(\xvec{s}_i)|h(\xvec{s}_i) + \mu_k,\,\sigma^2_k\right)}
\end{align}
for $j=1,\hdots,10$ and $i=1,\hdots,n$, where the denominator is the normalizing constant.
\item Sampling step for $\xvec{h}$:
\begin{align}
\xvec{h}|\y^*, \xvec{a},\sigma^2_u,\mu_h,\phi \sim N\left(\widehat{\h},\, \widehat{\xmat{H}}^{-1}_h\right),
\end{align} 
where $\widehat{\xmat{H}}_h=\xmat{\Sigma}_a^{-1} + \frac{1}{\sigma^2_u}\xmat{B}^{'}(\phi)\xmat{B}(\phi)$ and $\widehat{\h}=\widehat{\xmat{H}}^{-1}_h\left(\xmat{\Sigma}_a^{-1}(\y^* - \xvec{d}_a)+\frac{\mu_h}{\sigma_u^2}\xmat{B}^{'}(\phi)\xmat{B}(\phi)\xvec{1}_n\right)$. 

\item Sampling step for $\sigma^2_u$:
\begin{align}
\sigma^2_u|\h,\mu_h,\phi \sim IG\left(a_0+\frac{n}{2},\,b_0+\frac{1}{2}(\h-\mu_h\xvec{1}_n)^{'}\xmat{B}^{'}(\phi)\xmat{B}(\phi)(\h-\mu_h\xvec{1}_n)\right).
\end{align}
\item Sampling step for $\mu_h$: 
\begin{align}
\mu_h|\h,\sigma^2_u,\phi\sim N\left(\widehat{\mu}_0,\widehat{V}_{\mu}^{-1}\right),
\end{align}
where $\widehat{V}_\mu=V^{-1}_\mu+\frac{1}{\sigma^2_u}\xvec{l}^{'}_n\xmat{B}^{'}(\phi)\xmat{B}(\phi)\xvec{1}_n$ and $\widehat{\mu}_0=\widehat{V}_\mu^{-1}\left(\frac{1}{\sigma^2_u}\xvec{1}^{'}_n\xmat{B}^{'}(\phi)\xmat{B}(\phi)\h+V^{-1}_\mu\mu_0\right)$.
\item Sampling step for $\phi$:
\begin{align}
f(\phi|\h,\mu_h,\sigma^2_u)\propto\left|\mf{B}(\phi)\right|\times\exp\left(-\frac{1}{2\sigma^2_u}(\h-\mu_h\xvec{1}_n)^{'}\xmat{B}^{'}(\phi)\xmat{B}(\phi)(\h-\mu_h\xvec{1}_n)\right),
\end{align}
which does not correspond to any known density function. A random-walk Metropolis-Hastings algorithm can be used to sample from this distribution \citep{Lesage:2009}. At iteration $g$, a candidate value $\widetilde{\phi}$ is generated according to 
\begin{align}
\widetilde{\phi}=\phi^{(g-1)} + z_\phi \times N(0,1),
\end{align}
where $z_\phi$ is the tuning parameter. The candidate value $\widetilde{\phi}$ is accepted with probability  $\mathbb{P}(\widetilde{\phi},\,\phi^{(g-1)})=\min\left(\frac{f(\widetilde{\phi}|\h^{(g-1)},\sigma^{2(g-1)}_u,\mu_h^{(g-1)})}{f(\phi^{(g-1)}|\h^{(g-1)},\sigma^{2(g-1)}_u,\mu_h^{(g-1)})},1\right)$. 
\end{enumerate}
\end{algorithm}

\subsection{Statistical properties of the spatiotemporal stochastic volatility model}\label{apen7.3}
In this section, we provide the statistical properties of the model considered in \citet{Otto:2022}. Consider the following expression:
\begin{align}\label{4.1}
Y_{t}(\xvec{s}_i)=e^{\frac{1}{2}h_{t}(\xvec{s}_i)}V_{t}(\xvec{s}_i) \qquad \text{for $i = 1, \ldots, n$ and $t = 1, \ldots, T$.}
\end{align}
Then, the conditional variance of $Y_{t}(\xvec{s}_i)$ given $h_{t}(\xvec{s}_i)$ is $\text{Var}\left(Y_{t}(\xvec{s}_i)|h_{t}(\xvec{s}_i)\right)=e^{h_{t}(\xvec{s}_i)}$, indicating that the conditional variance is both time and space varying. This equation also indicates that $h_{t}(\xvec{s}_i)$ can be called the log-volatility since $h_{t}(\xvec{s}_i)=\log\left(\text{Var}\left(Y_{t}(\xvec{s}_i)|h_{t}(\xvec{s}_i)\right)\right)$. In order to determine the unconditional moments of $Y_{t}(\xvec{s}_i)$, we need to determine the distribution of $\h=(\h^{'}_1,\hdots,\h^{'}_T)^{'}$. Let $\bs{\rho}=(\rho_1,\rho_2,\rho_3)^{'}$, and define the $nT\times nT$ matrix $\J(\rh)$ as
\begin{align}\label{4.9}
\J(\rh)=
\begin{pmatrix}
\Ss(\rho_1)&\0&\hdots&\0&\0\\
-\A(\rho_2,\rho_3)&\Ss(\rho_1)&\hdots&\0&\0\\
\vdots&\ddots&\ddots&\vdots&\vdots\\
\0&\0&\hdots&-\A(\rho_2,\rho_3)&\Ss(\rho_1)
\end{pmatrix}.
\end{align}
Then, we can express the log-volatility equation as  
\begin{align}\label{4.10}
\J(\rh)(\h-\bs{l}_T\otimes\bs{\mu})=
\begin{pmatrix}
\Ss(\rho_1)(\h_1-\bs{\mu})\\
\U_2\\
\vdots\\
\U_T\\
\end{pmatrix},
\end{align}
where $l_T$ is the $T\times1$ vector of ones. Assume that the spatial dynamic process for $\mathbf{h}_t$ has been operating for a long time so that it is possible to express $\Ss(\rho_1)(\h_1-\bs{\mu})$ in the following way:
\begin{align}\label{recursive}
\Ss(\rho_1)(\h_1-\bs{\mu})=\sum_{j=0}^{\infty}\left(\A(\rho_2,\rho_3)\Ss^{-1}(\rho_1)\right)^{j}\U_{1-j},
\end{align}
which implies that 
\begin{align*}
\text{Var}(\Ss(\rho_1)(\h_1-\bs{\mu}))&=\sigma^2\sum_{j=0}^{\infty}\left(\A(\rho_2,\rho_3)\Ss^{-1}(\rho_1)\right)^{j}\left(\A(\rho_2,\rho_3)\Ss^{-1}(\rho_1)\right)^{'j}=\sigma^2\K(\rh),
\end{align*}
where $\K(\rh)=\sum_{j=0}^{\infty}\left(\A(\rho_2,\rho_3)\Ss^{-1}(\rho_1)\right)^{j}\left(\A(\rho_2,\rho_3)\Ss^{-1}(\rho_1)\right)^{'j}$. Then, from \eqref{4.10}, we obtain
\begin{align}
\text{Var}\left(\h-\bs{l}_T\otimes\bs{\mu}\right)=\sigma^2\J^{-1}(\rh)\Pp(\rh)\J^{'-1}(\rh),
\end{align}
where
\begin{align} % \label{4.13}
\Pp(\rh)=
\begin{pmatrix}
\K(\rh)&\0&\hdots&\0&\0\\
\0&\I_n&\hdots&\0&\0\\
\vdots&\vdots&\ddots&\vdots&\vdots\\
\0&\0&\hdots&\0&\I_n
\end{pmatrix}.
\end{align}
Let $\bs{\Omega}=\sigma^2\J^{-1}(\rh)\Pp(\rh)\J^{'-1}(\rh)$. Then, the distribution of $\h$ is
\begin{align}\label{4.14}
\h|\rh,\bs{\mu},\sigma^2\sim N(\bs{l}_T\otimes\bs{\mu},\,\bs{\Omega}).
\end{align}
Note that when $\bs{\rho}=\mf{0}$, $\bs{\Omega}$ reduces to $\sigma^2\mf{I}_{nT}$, and thus the result in \eqref{4.14} becomes $\h|\bs{\mu},\sigma^2\sim N(\bs{l}_T\otimes\bs{\mu},\,\sigma^2\mf{I}_{nT})$. Consider the following partition of $\bs{\Omega}$:
\begin{align} \label{4.13}
\bs{\Omega}=
\begin{pmatrix}
\bs{\Omega}_{11}&\bs{\Omega}_{12}&\hdots&\bs{\Omega}_{1,T-1}&\bs{\Omega}_{1T}\\
\bs{\Omega}_{21}&\bs{\Omega}_{22}&\hdots&\bs{\Omega}_{2,T-1}&\bs{\Omega}_{2T}\\
\vdots&\vdots&\ddots&\vdots&\vdots\\
\bs{\Omega}_{T1}&\bs{\Omega}_{T2}&\hdots&\bs{\Omega}_{T,T-1}&\bs{\Omega}_{TT}\\
\end{pmatrix},
\end{align}
where each $\bs{\Omega}_{st}$ for $s,t=1,2\hdots,T$ is an $n\times n$ sub-matrix of $\bs{\Omega}$. Let $\Omega_{ij,st}$ be the $(i,j)$th element of $\bs{\Omega}_{st}$ for $i,j=1,2,\hdots,n$. Let $r\in\mathbb{N}$ be a natural even number. Then, the even moments of $Y_{t}(\xvec{s}_i)$ can be expressed as
\begin{align}
\E\left(Y^r_{t}(\xvec{s}_i)\right)=\E\left(e^{\frac{r}{2}h_{t}(\xvec{s}_i)}\right)\E(V^r_{t}(\xvec{s}_i))=\exp\left(\frac{\mu(\xvec{s}_i)r}{2}+\frac{r^2}{8}\Omega_{ii,tt}\right)\gamma(r)
\end{align}
where $\gamma(r)=\frac{r!}{2^{r/2}(r/2)!}$. Then, it follows that $\E\left(Y^4_{t}(\xvec{s}_i)\right)/\left(\E\left(Y^2_{t}(\xvec{s}_i)\right)\right)^2-3=3\left(\exp(\Omega_{ii,tt})-1\right)>0$, which indicates that $Y_{t}(\xvec{s}_i)$ has a leptokurtic symmetric distribution. The covariance between $Y^r_{t}(\xvec{s}_i)$ and $Y^r_{s}(\xvec{s}_j)$:
\begin{align}
&\text{Cov}\left(Y^r_{t}(\xvec{s}_i),Y^r_{s}(\xvec{s}_j)\right)\\
&=\E\left(e^{\frac{r}{2}(h_{t}(\xvec{s}_i)+h_{s}(\xvec{s}_j))}V^r_{t}(\xvec{s}_i)V^r_{s}(\xvec{s}_j)\right)-\E\left(e^{\frac{r}{2}h_{t}(\xvec{s}_i)}V^r_{t}(\xvec{s}_i)\right)\E\left(e^{\frac{r}{2}h_{s}(\xvec{s}_j)}V^r_{s}(\xvec{s}_j)\right)\nonumber\\
&=\gamma^2(r)\exp\left(\frac{r\left(\mu(\xvec{s}_i)+\mu(\xvec{s}_j)\right)}{2}+\frac{r^2}{8}\left(\Omega_{ii,tt}+\Omega_{jj,ss}+2\Omega_{ij,ts}\right)\right)\nonumber\\
&\quad-\gamma^2(r)\exp\left(\frac{\mu(\xvec{s}_i)r}{2}+\frac{r^2}{8}\Omega_{ii,tt}\right)\exp\left(\frac{\mu(\xvec{s}_j)r}{2}+\frac{r^2}{8}\Omega_{jj,ss}\right)\nonumber\\
&=\gamma^2(r)\exp\left(\frac{r\left(\mu(\xvec{s}_i)+\mu(\xvec{s}_j)\right)}{2}+\frac{r^2}{8}\left(\Omega_{ii,tt}+\Omega_{jj,ss}\right)\right)\left(\exp\left(\frac{r^2}{4}\Omega_{ij,ts}\right)-1\right).\nonumber
\end{align}
This result indicates that $\text{Cov}\left(Y^r_{t}(\xvec{s}_i),Y^r_{s}(\xvec{s}_j)\right)=0$ when $\bs{\rho}=\mf{0}$ holds, because  $\left(\exp\left(\frac{r^2}{4}\Omega_{ij,ts}\right)-1\right)=0$.

\subsection{Bayesian MCMC algorithm for the spatiotemporal stochastic volatility model}
In this section, we describe the Gibbs sampler considered in \citet{Otto:2022} for the estimation of their suggested model. Let $\mf{a}^{(g)}$ be the draw generated at the $g$th iteration, where $\mf{a}\in\{\h_t,\bs{\mu},\rh,\sigma^2\}$. 
\begin{algorithm}[Estimation Algorithm]\label{a2}
\leavevmode   \normalfont
\begin{enumerate}
\item Sampling step for $\Z$:  Note that $\Z_t$ is a discrete random variable, and its conditional posterior probability mass function is
\begin{align}
f(\Z_t|\Y_t,\h_t,\bs{\mu},\rh,\sigma^2)\propto f(\Z_t)f\left(\Y^{*}_t|\Z_t,\h_t\right)=\prod_{i=1}^nf\left(Y^{*}_{t}(\xvec{s}_i)|z_{t}(\xvec{s}_i),h_{t}(\xvec{s}_i)\right)f(z_{t}(\xvec{s}_i)),
\end{align}
for $t=1,\hdots,T$. Thus,
\begin{align}\label{4.17}
\mathbb{P}\left(z_{t}(\xvec{s}_i)=j|Y^{*}_{t}(\xvec{s}_i)\right)=\frac{p_j\phi\left(Y^{*}_{t}(\xvec{s}_i)|h_{t}(\xvec{s}_i)+\mu_j,\,\sigma^2_j\right)}{\sum_{k=1}^{10}p_k\phi\left(Y^{*}_{t}(\xvec{s}_i)|h_{t}(\xvec{s}_i)+\mu_k,\,\sigma^2_k\right)},\,j=1,\hdots,10,\,i=1,\hdots,n,
\end{align} 
for $t=1,\hdots,T$, where the denominator is the normalization constant.
\item Sampling step for $\h$: Let $\Si=\Diag\left(\Si_1,\hdots,\Si_T\right)$ and $\mf{d}=(\mf{d}^{'}_1,\hdots,\mf{d}^{'}_T)^{'}$. Using standard regression results on $f(\h|\Y,\Z,\bs{\mu},\rh,\sigma^2)\propto \prod_{t=1}^Tf(\Y^{*}_t|\Z,\h_t)f(\h|\rh,\bs{\mu},\sigma^2)$, we obtain
\begin{align}
\h|\Y,\Z,\bs{\mu},\rh,\sigma^2\sim N(\widehat{\mf{b}}_h,\widehat{\B}_h),
\end{align}
where 
\begin{align*}
\widehat{\B}_h=\left(\Om^{-1}+\Si^{-1}\right)^{-1},\quad\widehat{\bs{b}}_h=\widehat{\B}_h\left(\Om^{-1}(\bs{l}_T\otimes\bs{\mu})+\Si^{-1}(\Y^{*}-\mf{d})\right).
\end{align*}
\item Sampling step for $\bs{\mu}$: Using \eqref{4.9} and \eqref{4.13}, it can be shown that
\begin{align*}
\Om^{-1}=
\begin{pmatrix}
\Om^{*}_{11} & \Om^{*}_{12} &\0& \ldots&\0& \0 \\
\Om^{*}_{21}&\Om^{*}_{22}&\Om^{*}_{23}&\hdots&\0&\0\\
\vdots & \ddots &\ddots &\ddots&\vdots& \vdots \\
\0&\0& \0 &\ldots&\Om^{*}_{T-1,T-1}&\Om^{*}_{T-1, T}\\
\0&\0& \0 &\ldots&\Om^{*}_{T,T-1}&\Om^{*}_{TT}
\end{pmatrix},
\end{align*}
where
\begin{align*}
&\Om^{*}_{11}= \sigma^{-2}\Ss^{'}(\rho_1)\K^{-1}(\rh)\Ss(\rho_1)+\sigma^{-2}\A^{'}(\rho_2,\rho_3)\A(\rho_2,\rho_3),\quad \Om^{*}_{TT}=\sigma^{-2}\Ss^{'}(\rho_1)\Ss(\rho_1),\\
&\Om^{*}_{ii}=\sigma^{-2}\Ss^{'}(\rho_1)\Ss(\rho_1)+\sigma^{-2}\A^{'}(\rho_2,\rho_3)\A(\rho_2,\rho_3),\quad i=2,\hdots,T-1,\\
&\Om^{*}_{i,i+1}=\Om^{*'}_{i+1,i}=-\sigma^{-2}\A^{'}(\rho_2,\rho_3)\Ss(\rho_1),\quad i=1,\hdots,T-1.
\end{align*}
Then, from $f(\bs{\mu}|\Y,\h,\Z,\bs{\mu},\rh,\sigma^2)\propto f(\h|\rh,\bs{\mu},\sigma^2)f(\bs{\mu})$, we obtain
\begin{align}
\bs{\mu}|\Y,\h,\Z,\bs{\mu},\rh,\sigma^2\sim N(\widehat{\mf{b}}_{\mu},\widehat{\B}_{\mu}),
\end{align}
where
$$
\widehat{\B}_{\mu}=\left(\B^{-1}_{\mu}+\sum_{j=1}^T\sum_{i=1}^T\Om^{*}_{ij}\right)^{-1},\quad
\widehat{\mf{b}}_{\mu}=\widehat{\B}_{\mu}\left(\B^{-1}_\mu\bs{b}_\mu+\sum_{j=1}^T\sum_{i=1}^T\Om^{*}_{ij}\h_i\right)
$$
\item Sampling step for $\sigma^2$: From $f(\sigma^2|\Y,\h,\bs{\mu},\rh)\propto f(\h|\rh,\bs{\mu},\sigma^2)f(\sigma^2)$, we obtain
\begin{align*}
\sigma^2|\Y,\h,\Z,\bs{\mu},\rh\sim\text{IG}(\widehat{a},\,\widehat{b}),
\end{align*}
where 
$$
\widehat{a}=a+nT/2,\quad\widehat{b}=b+\frac{1}{2}\left(\h-(\bs{l}_T\otimes\bs{\mu})\right)^{'}\left(\J^{'}(\rh)\Pp^{-1}(\rh)\J^{}(\rh)\right)\left(\h-(\bs{l}_T\otimes\bs{\mu})\right).
$$
\item Sampling step for $\rh$: The conditional posterior density of $\rh$ is non-standard. \citet{Otto:2022} suggest using the adaptive Metropolis (AM) algorithm suggested in \citet{Haario:2001} and \citet{Roberts:2009} to generate draws from $f(\rh|\Y,\h,\Z,\sigma^2)$. At the iteration $g$, a candidate value $\tilde{\bs{\rho}}$ is generated from the following proposal distribution: 
\begin{align*}
f_g\left(\rh|\rh^{(0)},\hdots, \rh^{(g-1)}\right)=
\begin{cases}
N\left(\rh^{(g-1)},\,\frac{(0.1)^2}{3}\times \I_3\right),\quad\text{for}\quad g\leq g_0,\\
0.95\times N\left(\rh^{(g-1)},\,\frac{c(2.38)^2}{3}\times\text{Cov}\left(\rh^{(0)},\hdots, \rh^{(g-1)}\right)\right)\\
\quad\quad +0.05\times N\left(\rh^{(g-1)},\,\frac{(0.1)^2}{3}\times \I_3\right),\quad\text{for}\quad g>g_0,
\end{cases}
\end{align*}
where $g_0$ is the length of the initial sampling period, $\text{Cov}\left(\rh^{(0)},\hdots, \rh^{(g-1)}\right)$ is the empirical covariance matrix of historical draws defined by $\text{Cov}\left(\rh^{(0)},\hdots, \rh^{(g-1)}\right)=\frac{1}{g}\sum_{j=0}^{g-1}\bs{\rho}^{(j)}\bs{\rho}^{(j)'}-\bar{\bs{\rho}}^{(g-1)}\bar{\bs{\rho}}^{(g-1)'}$ with $\bar{\bs{\rho}}^{(g-1)}=\frac{1}{g}\sum_{j=0}^{g-1}\bs{\rho}^{(j)}$, and $c$ is a scalar tuning parameter used to achieve a reasonable acceptance rate. Then, the acceptance probability is:
\begin{align*}
\mathbb{P}(\rh^{(g-1)}, \tilde{\rh})=\min\left(\frac{f\left(\h|\tilde{\rh},\bs{\mu}^{(g)},\sigma^{2(g)}\right)}{f\left(\h|\rh^{(g-1)},\bs{\mu}^{(g)}, \sigma^{2(g)}\right)},\,1\right),
\end{align*}
where $f(\h|\rh,\bs{\mu},\sigma^{2})$ is given in \eqref{4.14}. Then, $\tilde{\rh}$ is accepted with the probability $\mathbb{P}(\rh^{(g-1)}, \tilde{\rh})$.
\end{enumerate}
\end{algorithm}
The simulation results presented in \citet{Otto:2022} demonstrate that this Gibbs sampler performs satisfactorily.

\newpage
\bibliography{references}\label{sec:bib}

\end{document}